\documentclass[preprint]{aastex61}
\usepackage{graphicx}

\def\Ms{$\textrm{M}_{\odot}$}

\def\H2{H$_{2}$}

\def\roH2{$\rho_{\textrm{H}_2}$}

\def\MH2{M$_{\textrm{H}_2}$}

\submitjournal{AJ}

\shorttitle{Rings in S0 galaxies} \shortauthors{Proshina et al.}

\begin{document}

\title{Star-Forming rings in lenticular
galaxies: origin of the gas}\footnotemark[0]\thanks{Based on observations
made with the Southern African Large Telescope (SALT)}

\correspondingauthor{Olga Sil'chenko}
\email{olga@sai.msu.su,olgasil.astro@gmail.com}

\author{Irina S. Proshina}
\affil{Sternberg Astronomical Institute, M.V. Lomonosov Moscow State University, Universitetsky pr., 13, Moscow, 119991 Russia}
\email{ii.pro@mail.ru}

\author{Alexei Yu. Kniazev}
\affil{South African Astronomical Observatory, PO Box 9, 7935 Observatory, Cape Town, South Africa}
\affil{Southern African Large Telescope Foundation, PO Box 9, 7935 Observatory, Cape Town, South Africa}
\affil{Sternberg Astronomical Institute, M.V. Lomonosov Moscow State University, Universitetsky pr., 13, Moscow, 119991 Russia}
\email{akniazev@saao.ac.za}

\author[0000-0003-4946-794X]{Olga K. Sil'chenko}
\affil{Sternberg Astronomical Institute, M.V. Lomonosov Moscow State University, Universitetsky pr., 13, Moscow, 119991 Russia}
\affil{Physics Department, M.V. Lomonosov Moscow State University}
\email{olga@sai.msu.su}

\begin{abstract}
Rings in S0s are enigmatic features which can however betray the evolutionary
paths of particular galaxies. We have undertaken long-slit spectroscopy of  five lenticular galaxies
with UV-bright outer rings. The observations have been made with the Southern African Large Telescope (SALT) to reveal the kinematics, chemistry, and
the ages of the stellar populations and the gas characteristics in the rings and surrounding disks. 
Four of the five rings are also bright in the H$\alpha$ emission line,  and the spectra of the gaseous rings extracted
around the maxima of the H$\alpha$ equivalent width reveal excitation by young stars betraying current star
formation in the rings. The integrated level of this star formation is 0.1--0.2~\Ms\ per year, with
the outstanding value of 1~\Ms\ per year in NGC~7808. The difference of chemical composition between the ionized gas 
of the rings which demonstrate nearly solar metallicity and the underlying stellar disks which are metal poor
implies recent accretion of the gas and star formation ignition; the star formation history estimated by using different 
star formation indicators implies that the star formation rate decreases with e-folding time of less than 1~Gyr. In NGC~809
where the UV-ring is well visible but the H$\alpha$ emission line excited by massive stars is absent, the star formation 
has already ceased.

\end{abstract}


\section{Introduction}

Lenticular galaxies are disk galaxies which differ from spirals by the
smooth appearance of their reddish large-scale stellar disks and by
visible absence of starforming sites organized as spiral arms. As a class, they belong to the
red sequence and are usually characterized as `quiescent' galaxies.
However, despite the absence of visible star formation, the disks
of S0s possess often noticeable amounts of cold gas \citep{welchsage03,welchsage06},
and to understand what conditions for star formation are lacking 
in the massive gaseous disks of S0 galaxies is a separate problem \citep{pogge_eskridge}.
The frequent decoupling between the gaseous and stellar kinematics in the S0 disks
provoked a suggestion that the cold gas in S0s was mostly accreted from
outside \citep{bertola92,kuijken,davis_2011,ilg_gas}, and perhaps the
conditions of such accretion -- off-plane direction of infall and/or transmittent
regime of inflow -- were often unfavourable for star formation ignition 
\citep{katkov_salt}. However, when the ultraviolet space telescope GALEX
had surveyed a large sample of nearby galaxies with a spatial resolution
of a few arcsec, numerous UV-rings were found in optically red, `quiescent'
lenticular galaxies \citep{gildepaz,marino11}. After the old claims by \citet{pogge_eskridge}
that current star formation in S0s, if present, is concentrated in rings,
this feature has been confirmed with the new observational data \citep{salim12}.

Outer {\it stellar} rings are often present in S0 galaxies. According to the statistics
of the recent catalogue of stellar rings ARRAKIS \citep{arrakis} based on the
data of the NIR survey of nearby galaxies S4G \citep{sheth10}, the occurence of
outer stellar rings is maximal just among lenticular galaxies: up to 50\%\ of
S0 galaxies ($T=-1$) have outer stellar rings \citep{arrakis}. Interestingly,
while the fraction of outer rings rises toward the morphological type of --1,
the fraction of strong bars falls \citep{buta2010,laurikainen_2013}. This fact
puts into doubt the theory that all the rings have the resonance nature and
are produced by bars \citep{ss1976,atha_82,buta_combes}, at least as concerning
the rings in lenticular galaxies. Though the rings found by \citet{arrakis} in 
the data of the Spitzer infrared space telescope consist certainly of old stars,
the subsequent mining in the GALEX data has shown that among the outer NIR rings
of lenticular galaxies more than the half are also seen in the ultraviolet \citep{kostuk15}.
It means that they contain stars younger than 200~Myr \citep{kennrev}, and that
recent star formation is common in the outer rings of lenticular galaxies. The next step
in the study of the outer rings in S0 galaxies is to look for gas in these rings, to see if it is
excited by young stars, and to deduce something about the origin of the rings based on the gas
properties. It is the topic of the present paper.

\begin{figure*}[t]
\centering
\begin{tabular}{c c c c}
 \includegraphics[width=3.5cm]{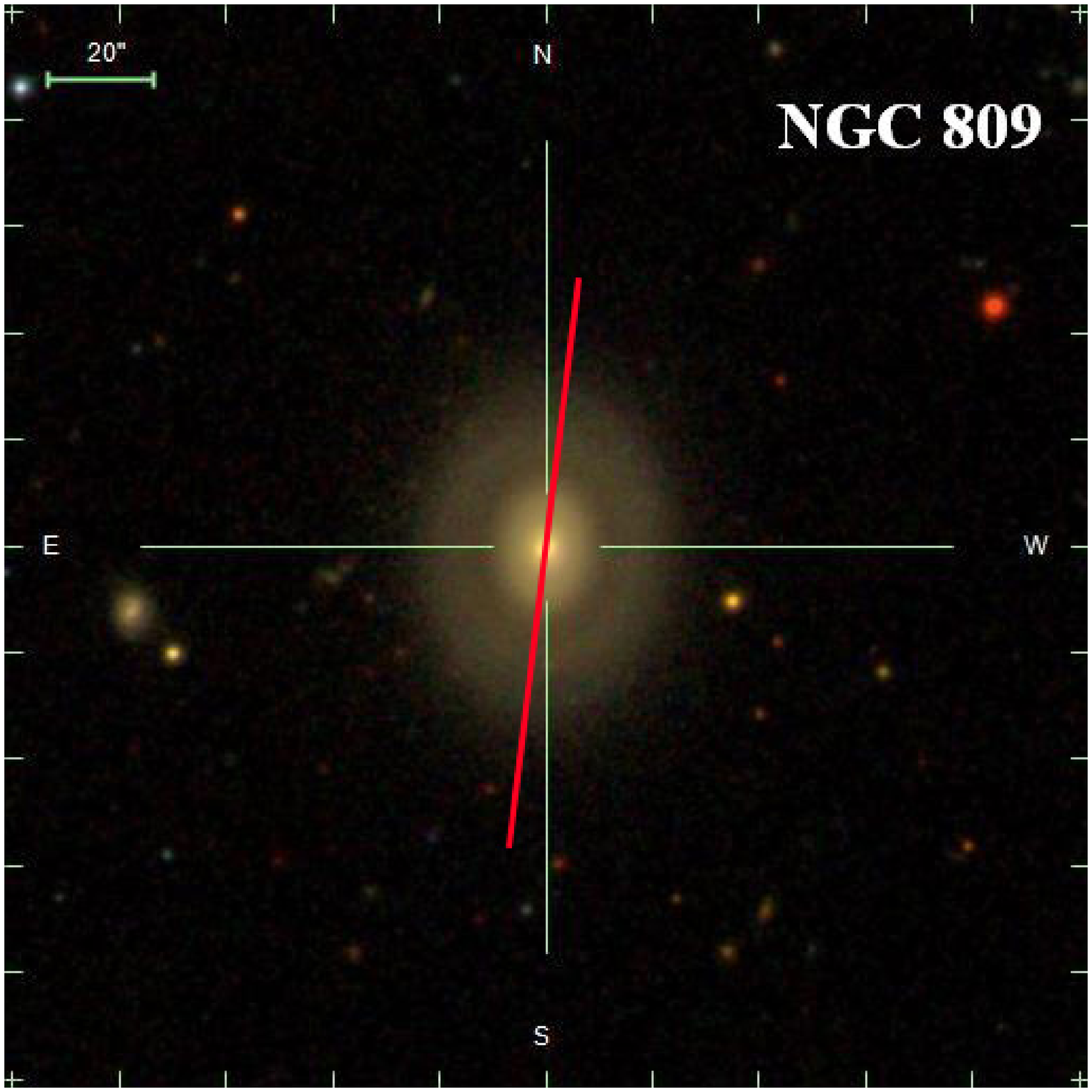} & \includegraphics[width=4cm]{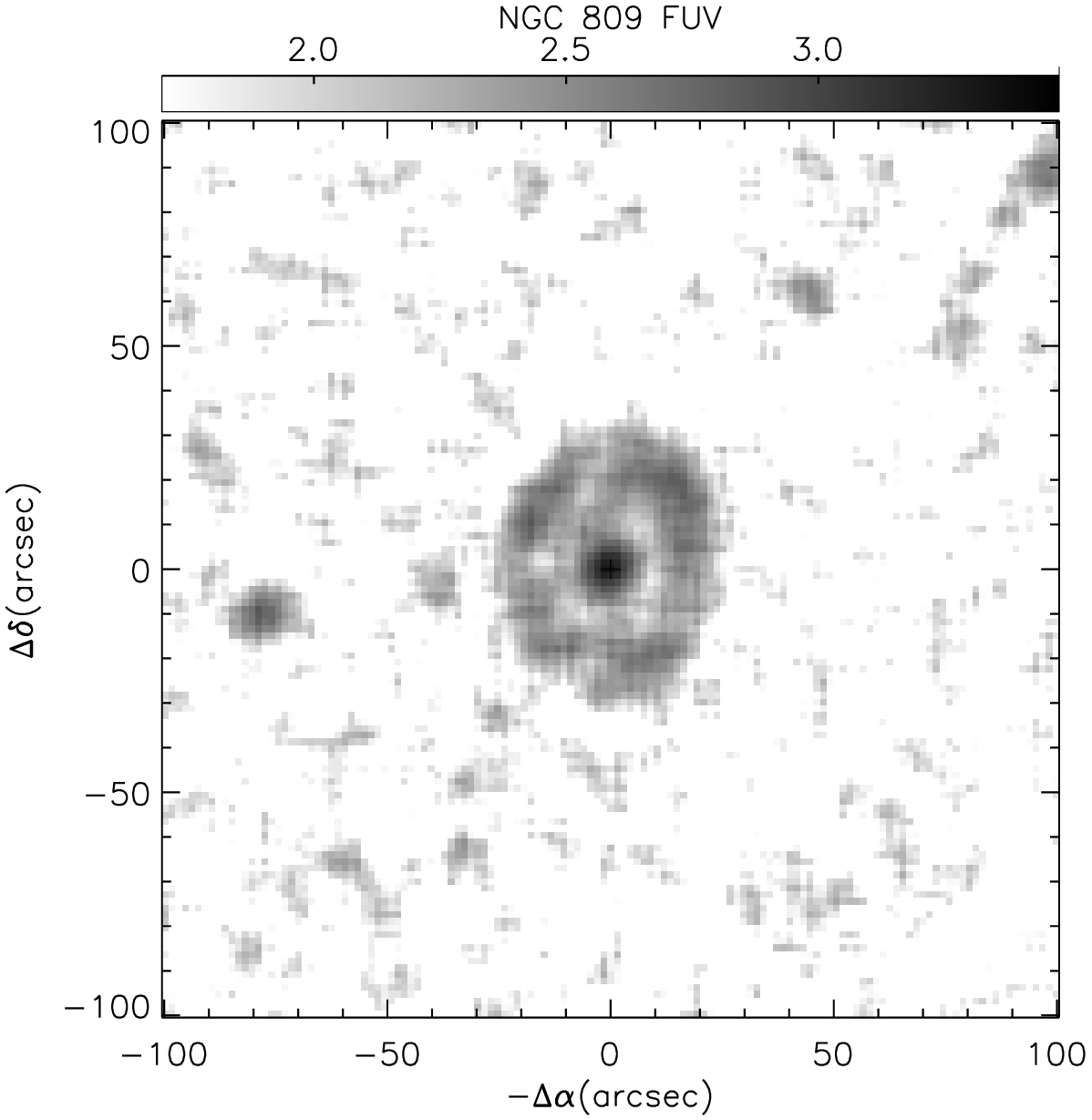} &
 \includegraphics[width=3.5cm]{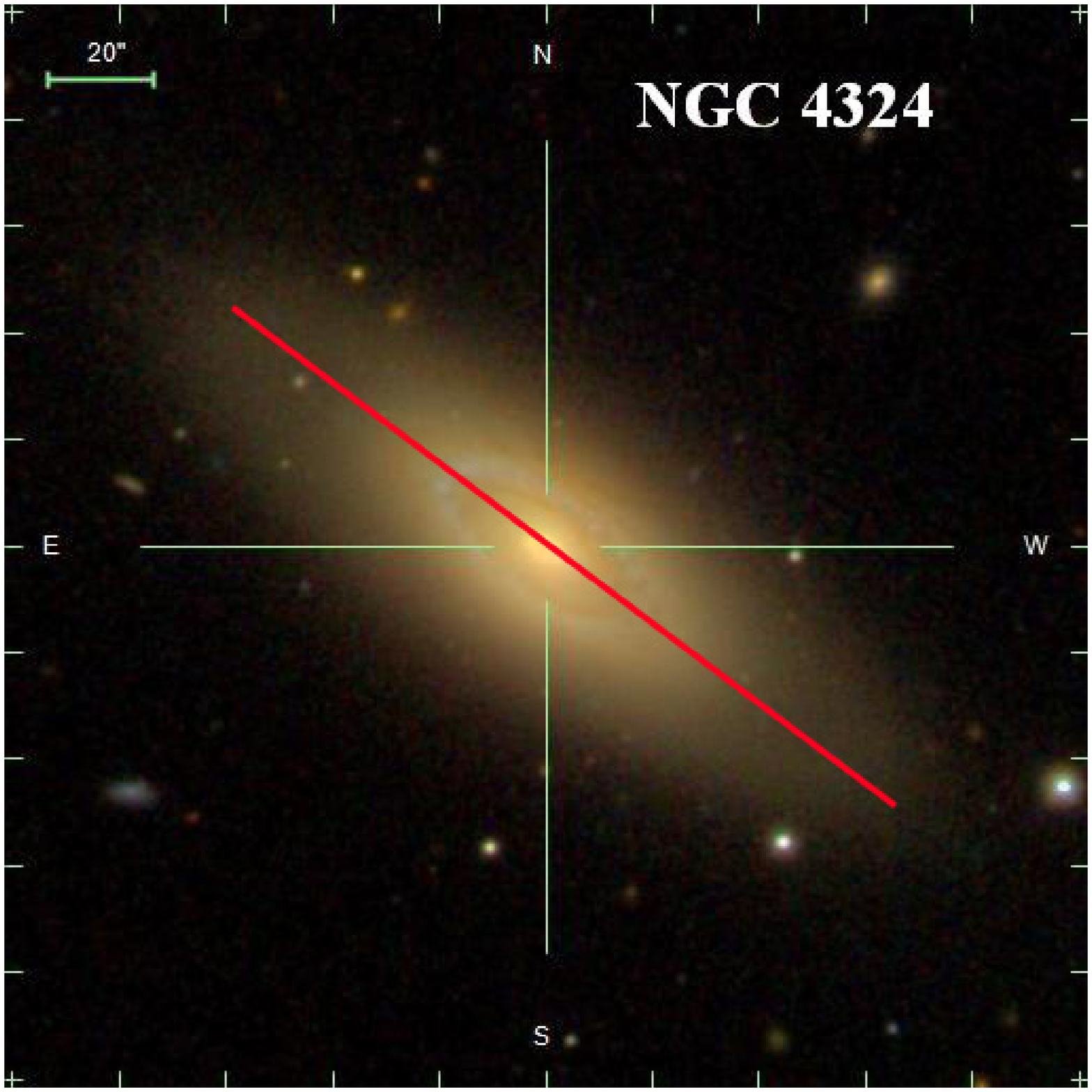} & \includegraphics[width=4cm]{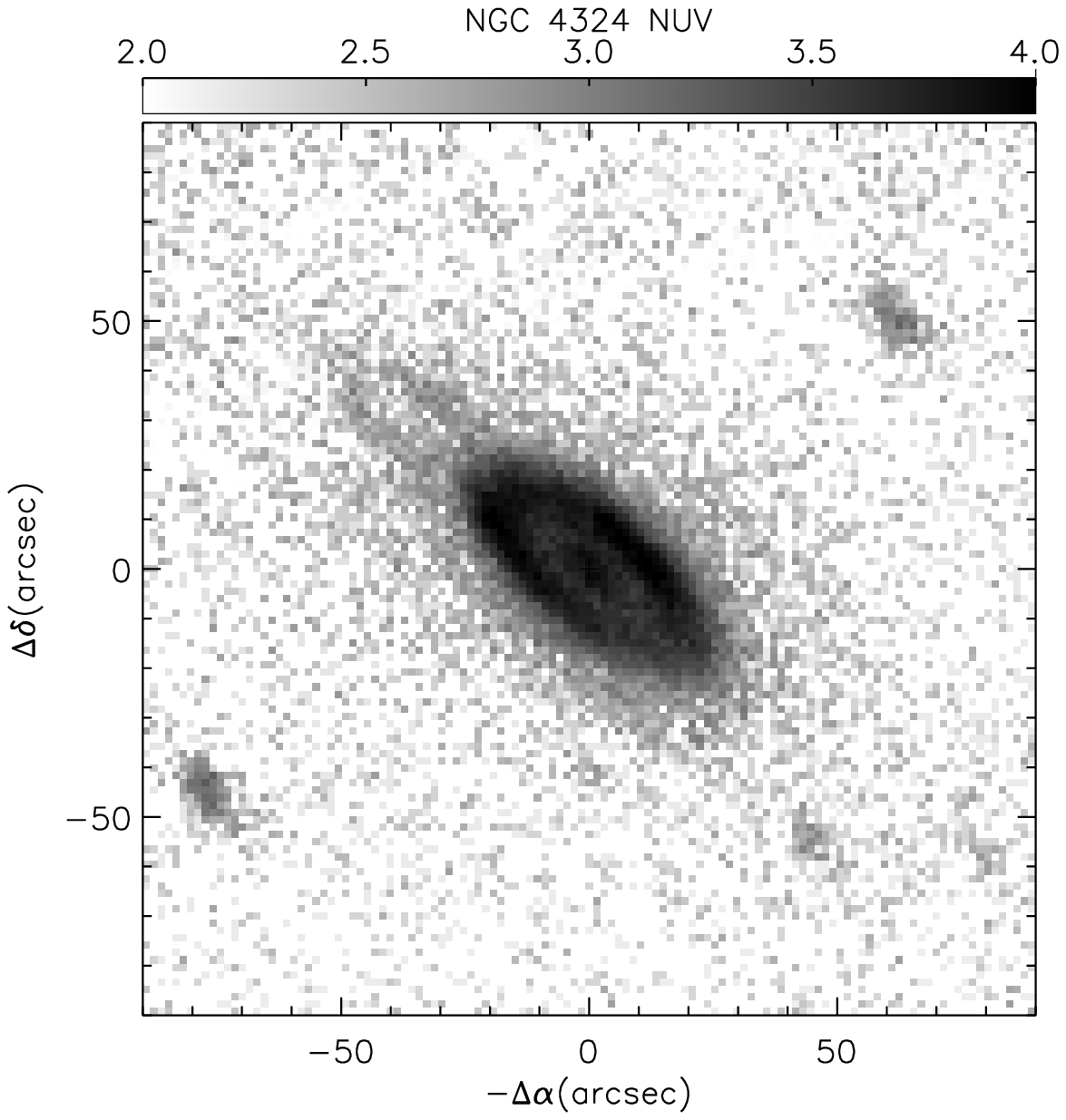} \\
 \includegraphics[width=3.5cm]{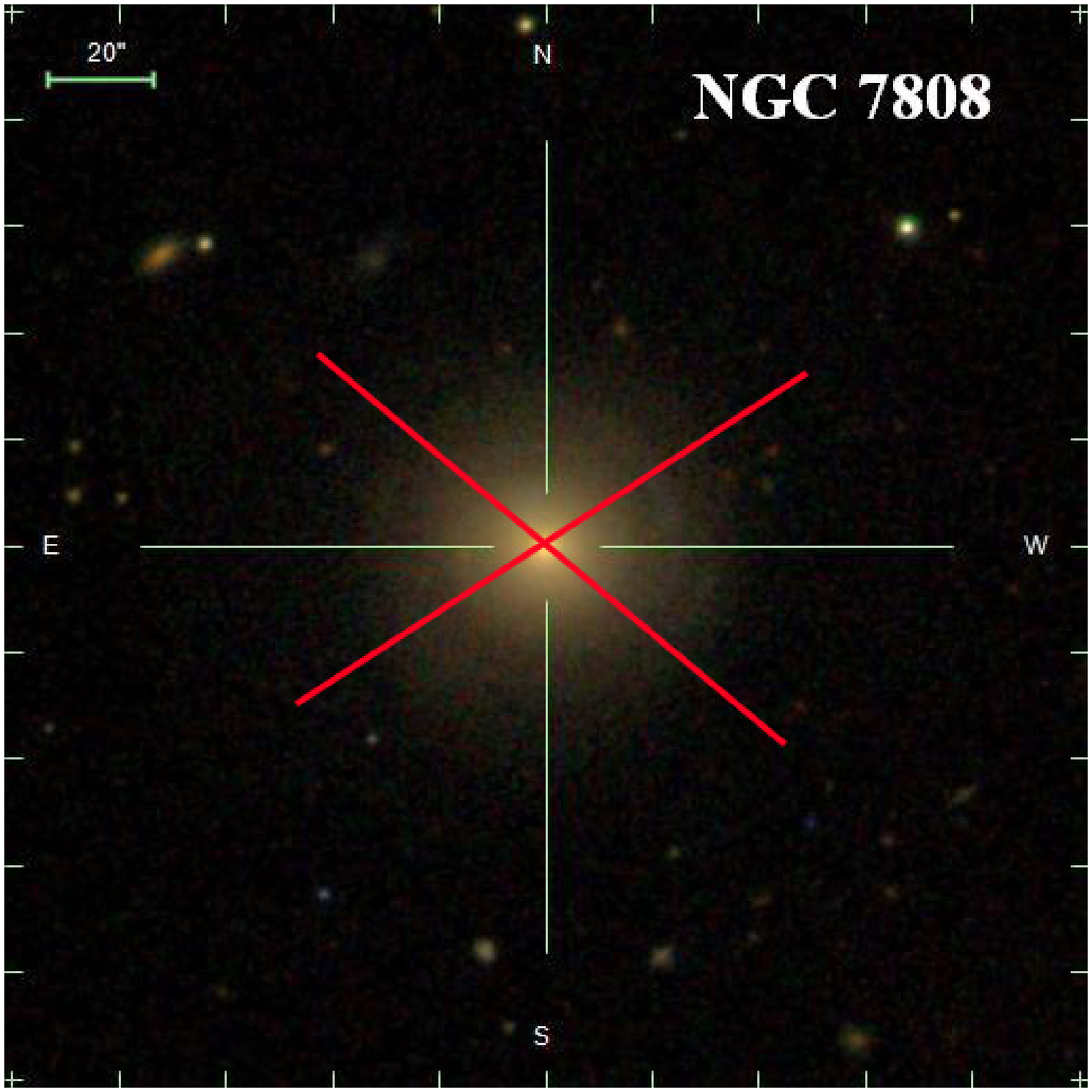} & \includegraphics[width=4cm]{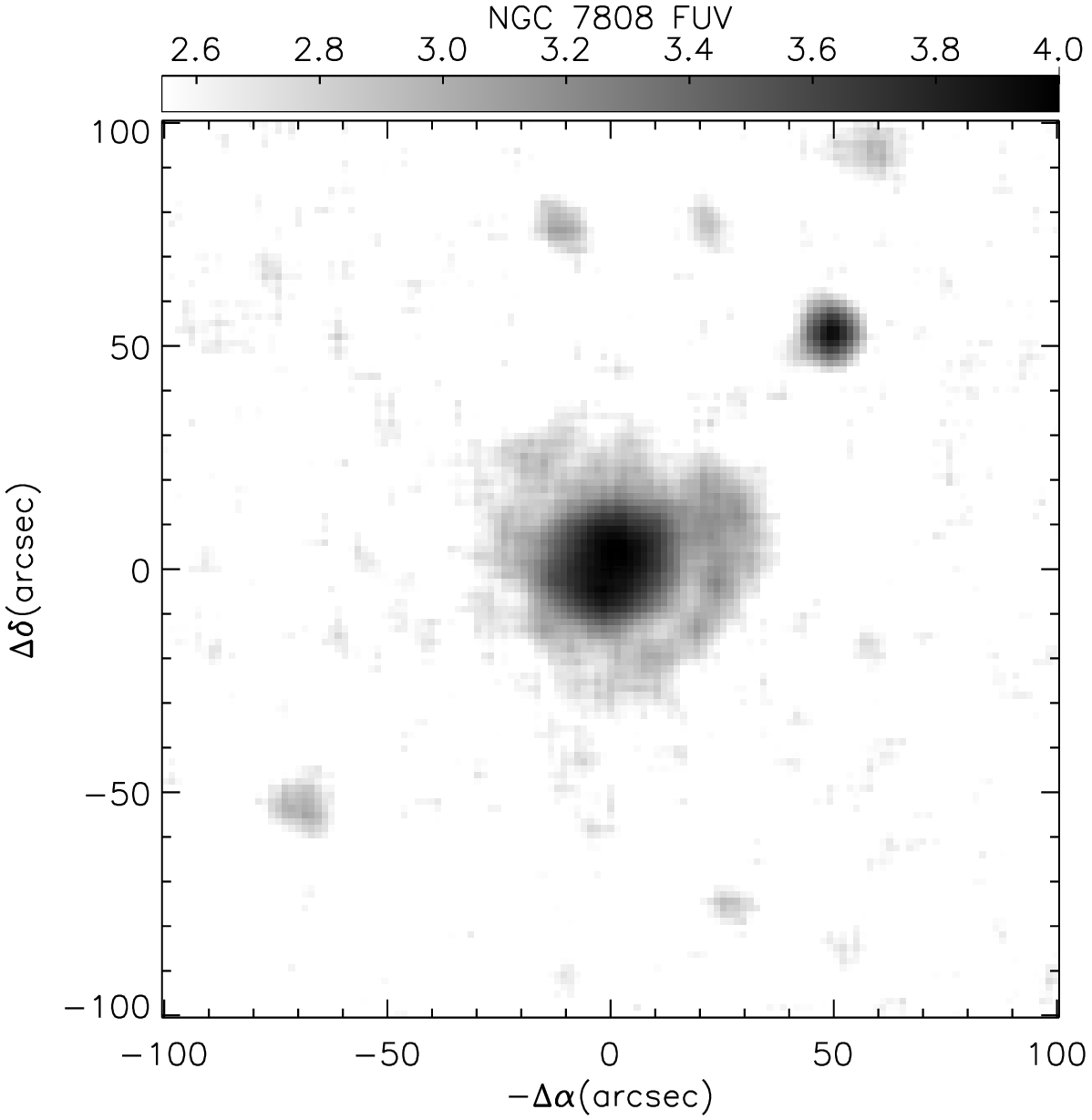} &
 \includegraphics[width=3.5cm]{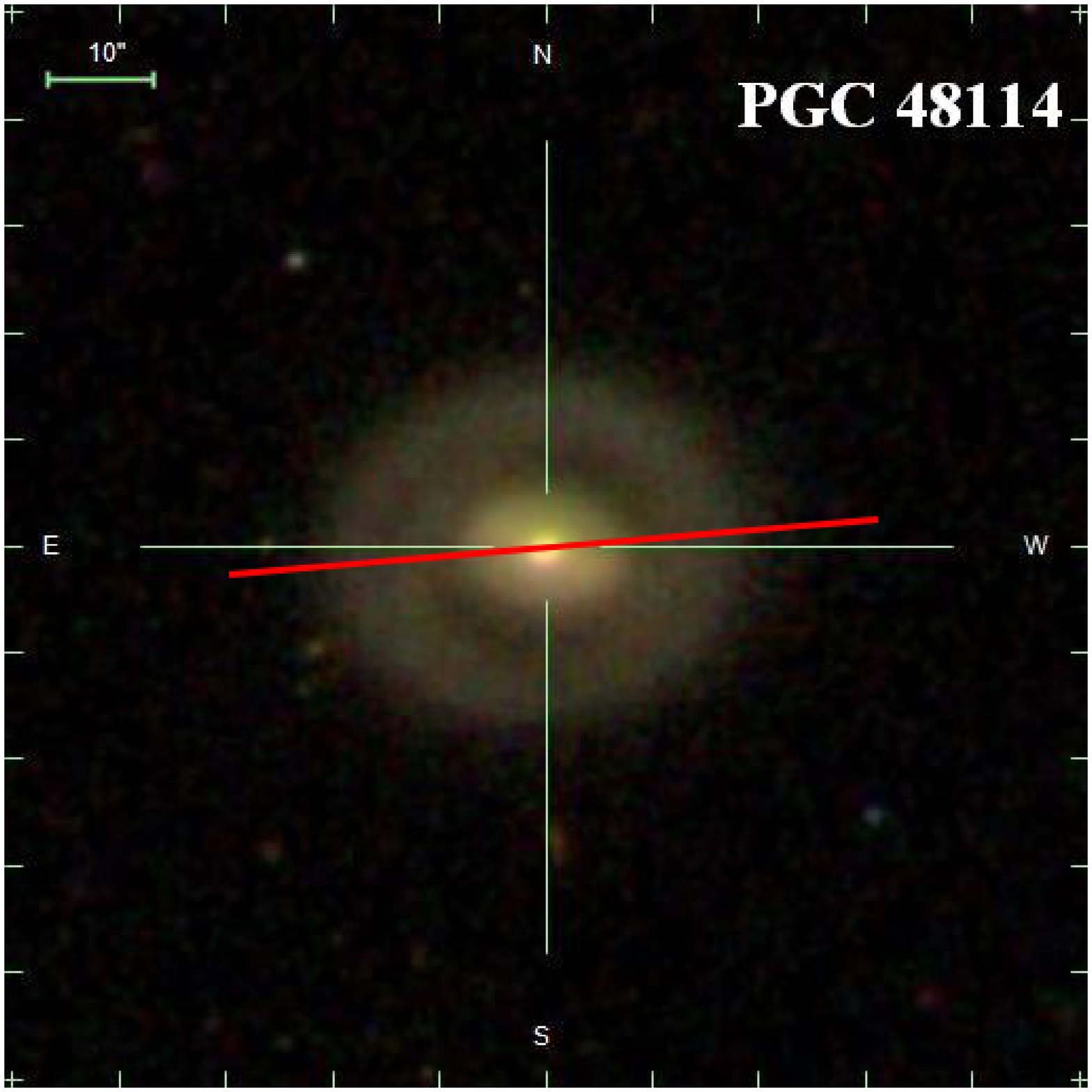} & \includegraphics[width=4cm]{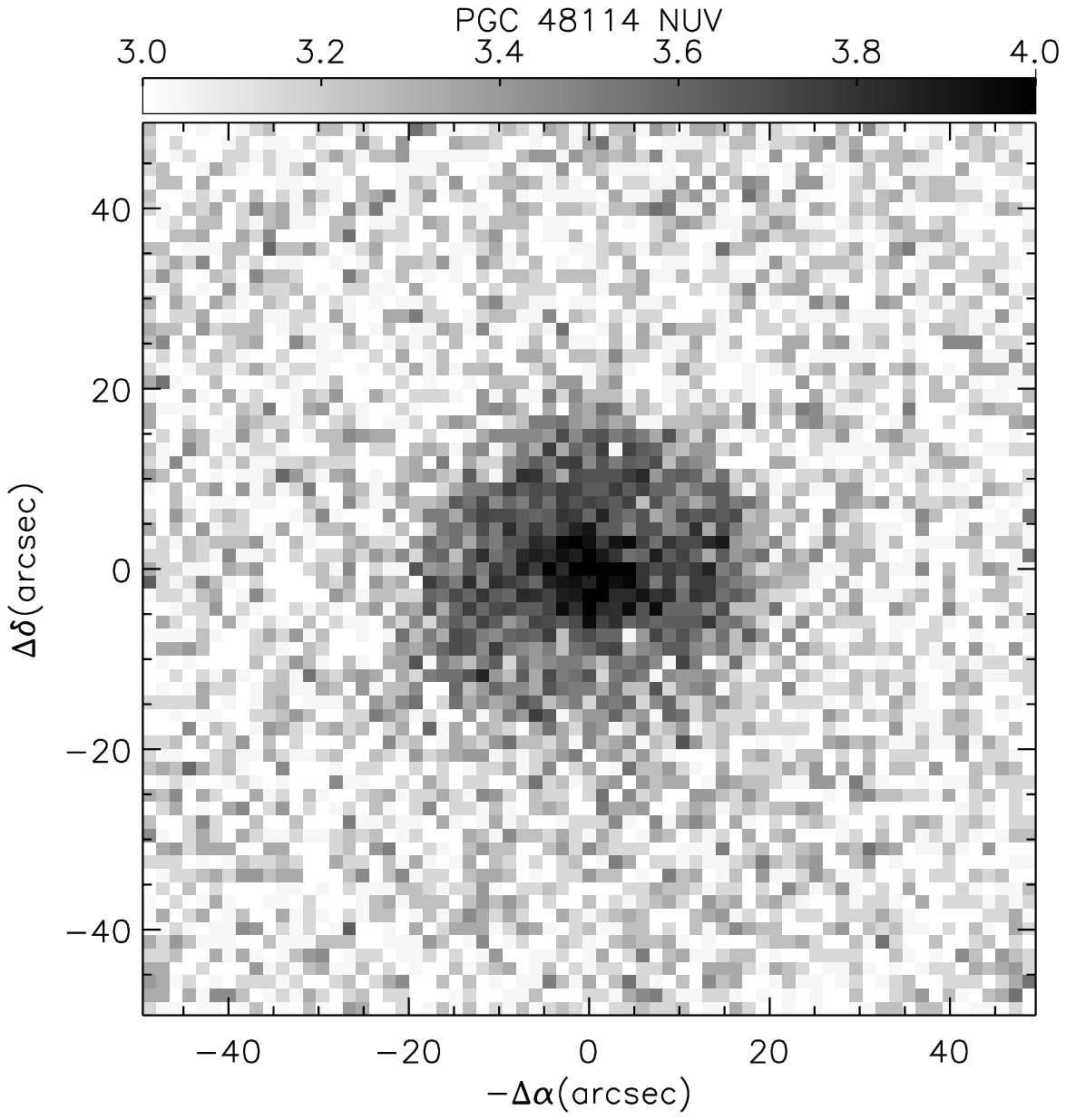} \\
\end{tabular}
\caption{False-coloured SDSS images of the galaxies under
consideration paired with the GALEX maps where the intensities are scaled logarithmically:
 NGC~809 (upper left), NGC~4324 (upper right),
NGC~7808 (bottom left), and PGC~48114 (bottom right). The positions
of the spectrograph slit are superposed as red straight lines.
}
\label{sdssview}
\end{figure*}

Here we study spectrally a small sample of five lenticular galaxies where the 
UV-bright rings are present. In NGC~809 and NGC~7808 the presence of
ultraviolet rings was reported by \citet{gildepaz} and measured by
\citet{ilyina_sil}; for NGC~4324 a nice picture of its UV-ring is presented
by \citet{cortese09}. The distant S0 galaxy PGC~48114 was noted as a
ring galaxy by \citet{kostyuk75}, and our inspection of the corresponding
GALEX image has confirmed the presence of an extended UV-signal in this object.
Finally, NGC~2697 has been found by us  serendipitously and is confirmed to
have a ring morphology in the GALEX maps by visual inspection. The literature
data on the global properties of the galaxies under consideration are given
in the Table~\ref{tbl_lit}, and the combined-colour images of four galaxies taken from the
SDSS survey database are presented in Fig.~\ref{sdssview}. All the galaxies look
like bona fide early-type disk galaxies -- without spirals, with smooth reddish appearance;
all of them are classified as unbarred galaxies.

\begin{table*}
\caption{Global parameters of the sample galaxies}
\label{tbl_lit}
%\begin{center}
\begin{flushleft}
\begin{tabular}{l|ccccc}
\hline\noalign{\smallskip}
Galaxy & NGC~809 & NGC~2697 & NGC~4324 & NGC~7808 & PGC~48114 \\
\hline
Galaxy type (NED$^1$) & (R)S0$^+$: & SA0$^+$(s): & SA(r)0$^+$ & (R')SA0$^0$ & S? (S0$^5$) \\
Type of a ring & RL..+$^6$ & & (L)SA(r)0$^+$/E?$^7$ & PLA.0$^6$ & R$_1$R$_2$SAB(l)0$^+$$^7$ \\
$V_r $ (NED), $\mbox{km} \cdot \mbox{s}^{-1}$ &  5367 & 1824 & 1665 & 8787 & 6984 \\
Distance$^4$, Mpc  & 71 & 29 & 27.5 & 115 & 99 \\
$R_{25}^{\prime \prime}$(RC3$^2$) & 44 & 55 & 83 & 38 & 20 \\
$R_{25}$, kpc (RC3$+$NED) & 14.7 & 7.6 & 11.1 & 20.5 & 9.3 \\
$M_B$(LEDA)  & --20.03 & --18.67 & --19.75 & --21.39 & --19.73 \\
$M_H$(NED)  & --23.74 & --22.26 & --23.64 & --24.71 & --23.34 \\
$(g-r)$ (SDSS$^3$) & 0.77 & -- & 0.74 & 0.77 & 0.80 \\
{\it PA}$_{phot}$ (LEDA$^5$) & $173.5^{\circ}$ & $123^{\circ}$ & $54.5^{\circ}$ & -- & $89^{\circ}$  \\
$M(\mbox{HI})$, $10^8\,M_{\odot}^8$ &  -- & 6 & 17 &  -- & -- \\
$M(\mbox{H}_2)$, $10^8\,M_{\odot}^9$ &  -- & 4.1 & 0.9 &  -- & -- \\
Environment$^{10}$ & 1  & 2 & 2 &  2 & 1 \\
 & (625, $+1.8$) & (10, $-0.8$, 237) & (23, $+2.0$, 100) & (7, $+0.8$, 94) & (159, $+1.5$) \\
\hline
\multicolumn{6}{l}{$^1$\rule{0pt}{11pt}\footnotesize
NASA/IPAC Extragalactic Database (http://ned.ipac.caltech.edu).}\\
\multicolumn{6}{l}{$^2$\rule{0pt}{11pt}\footnotesize
Third Reference Catalogue of Bright Galaxies \citep{rc3} .}\\
\multicolumn{6}{l}{$^3$\rule{0pt}{11pt}\footnotesize
Sloan Digital Sky Survey, Data Release 9 \citep{dr9} .}\\
\multicolumn{6}{l}{$^4$\rule{0pt}{11pt}\footnotesize
from NED, `Cosmology corrected' option} \\
\multicolumn{6}{l}{$^5$\rule{0pt}{11pt}\footnotesize
Lyon-Meudon Extragalactic Database (http://leda.univ-lyon1.fr) }\\
\multicolumn{6}{l}{$^6$\rule{0pt}{11pt}\footnotesize
according to \citet{nair_abraham} } \\
\multicolumn{6}{l}{$^7$\rule{0pt}{11pt}\footnotesize
according to \citet{buta17} } \\
\multicolumn{6}{l}{$^8$\rule{0pt}{11pt}\footnotesize
The source of the HI data -- Extragalactic Distance Database (http://edd.ifa.hawaii.edu).}\\
\multicolumn{6}{l}{$^9$\rule{0pt}{11pt}\footnotesize
The source of the H$_2$ data -- \citet{atlas3d_18} }\\
\multicolumn{6}{l}{$^{10}$\rule{0pt}{11pt}\footnotesize
The environment types derived from the HYPERLEDA and NED searching are coded:}\\ 
\multicolumn{6}{l}{\footnotesize
1  -- a pair member (numbers in parentheses: separation in kpc, magnitude difference),}\\
\multicolumn{6}{l}{\footnotesize
2 -- a group member (numbers in parentheses: N gal in the group, magnitude difference}\\
\multicolumn{6}{l}{\footnotesize
with the second(first-)ranked galaxy, and separation with it in kpc).}\\
\end{tabular}
\end{flushleft}
\end{table*}

\section{Long-slit spectroscopy}

\subsection{Observations and data reduction}

The spectral observations were made in a long-slit mode with the Robert Stobie Spectrograph
\citep[RSS;][]{Burgh03,Kobul03} operating at the Southern African Large Telescope (SALT)
\citep{Buck06,Dono06}; a slit width was 1.25 arcsec.  All observational details are given
in Table~\ref{tbl_logobs}.  The slit was aligned along the major axis
for every galaxy except NGC~7808 which looks face-on.  
The volume-phase grating PG0900 was used for our program to cover 
the spectral range of 3760$-$6860~\AA\ with a final reciprocal dispersion of
$\approx0.97$~\AA\ pixel$^{-1}$ and FWHM spectral resolution of 5.5~\AA.  The
seeing during the observations was in the range of 1.5$-$3.0~arcsec.  The RSS pixel
scale is 0\farcs1267, and the effective field of view is 8\arcmin\ along the slit.
We used a binning factor of 2 or 4 to give final spatial sampling of
0\farcs253 pixel$^{-1}$ and 0\farcs507 pixel$^{-1}$ respectively.  Every night the spectrum of
an Ar comparison arc was exposed to calibrate wavelength scale after each galaxy observation,
and spectral flats were observed regularly to correct for
pixel-to-pixel variations.  Spectrophotometric standard stars were observed
during twilights, after the observations of objects, that allowed relative flux
calibration of the SALT/RSS spectra.

% Table 2 %%%%%%%%%%%%%%%%%%%%%%%%%%%%%%%%%%%%%%%%%%%%%%%%%%%%%%%%%%%%%%%%%%%%%%%%%%%%%%%%%%%%
\begin{table*}
\centering
\caption{Long-slit spectroscopy of the studied galaxies}
%\begin{flushleft}
\label{tbl_logobs}
\begin{tabular}{cclccccc}
\hline\noalign{\smallskip}
Galaxy       & Date   &  Exp.      & Binning    &Slit    & PA(slit) & Seeing          \\
	       &           & [sec]      &            &[arcsec]&  [deg]   & [FWHM, arcsec]  \\ \hline\noalign{\smallskip}

NGC\,809                    & 07.10.2011 & 1020$\times$3      & 2$\times$2 &  1.25  & 173 & 2.5 \\ \hline
NGC\,2697                   & 22.11.2012 & 900$\times$3      & 2$\times$4 &  1.25   & 305 & 2.2--2.7 \\
                            & 16.02.2013 &  900$\times$3      & 2$\times$4 &  1.25   & 305 & 2.7 \\ \hline
NGC\,4324                   & 10.03.2013 &  800$\times$3     & 2$\times$4 &  1.25   & 233 & 3.5 \\ \hline
NGC\,7808                  & 02.12.2012 & 600$\times$3      & 2$\times$4 &  1.25   & 50 & 1.5 \\
                            & 04.12.2012 & 600$\times$3      & 2$\times$4 &  1.25   & 123 & 3.5 \\ \hline
PGC\,48114                  & 20.02.2013 & 600$\times$3      & 2$\times$4 &  1.25   & 275 & 2.7--3.2 \\ \hline
\end{tabular} 
%\end{flushleft}
\end{table*}

Primary data reduction was done with the SALT science pipeline \citep{Cr2010}.
After that, the bias and gain corrected and mosaicked long-slit data were
reduced in the way described in \citet{Kn08}.  The accuracy of the spectral
linearisation was checked using the night-sky lines [OI]~$\lambda$5577 and [OI]~$\lambda$6300; 
the RMS scatter of their wavelengths measured along the slit is 0.04~\AA.  Since the
diameters of the galaxies do not exceed 3 arcmin, the sky spectra from the slit edges were 
used to estimate the background during the galaxy exposures.

\subsection{Data analysis}

The spectral data have been used to calculate line-of-sight velocities of the ionized gas
in the galaxies along the slit and to estimate equivalent widths of the emission lines by means of Gaussian 
multi-component fitting of the [NII]$\lambda$6548+6583+H$\alpha$(emission)+H$\alpha$(absorption) line blend; 
also the Gaussian fitting of the individual emission lines  H$\beta$, [OIII]$\lambda$5007 (and of [SII]$\lambda$6716.4, 6730.6 
where they are in the spectral range covered by observations, i.e. in NGC 2697 and NGC 4324) was made.
For all the galaxies we have also determined the line-of-sight velocities of the stellar component 
by cross-correlating the blue-green part of the spectra (4050-5590 \AA) with the spectrum
of HD~58972 -- a bright K-giant star. We summed the spectra over radially increasing bins in the peripheral
parts of the galaxies to improve the S/N ratio by taking into account the radial galaxy structure (see the
next Section).

After that the Lick indices H$\beta$, Mgb, Fe5270, and Fe5335 \citep{Faber_1985, licksystem1, Worthey_Ottaviani_1997} 
were calculated in the same bins. The Lick-index system includes equivalent widths of a dozen of strong absorption lines
in integrated spectra of stellar populations which allow to determine stellar-population properties, in particular,
mean metallicities and ages, by confronting Balmer-line measurements against the metal-line measurements.
The calibration of the Lick indices measured with our instrumental setup (spectrograph, grating, and slit width)
into the standard index system is described by \citet{katkov_salt}. A separate problem is correction of the Lick index
H$\beta$  for the emission which is prominent in all of our sample galaxies. This was accomplished through the measurements
of the H$\alpha$ emission-line intensity and equivalent widths \citep{sil06}. After deriving the emission-line equivalent
widths by means of Gaussian fitting the blend of [NII]$\lambda$6548+6583+H$\alpha$(emission)+H$\alpha$(absorption),
we diagnosed the gas excitation mechanism by calculating the [NII]/H$\alpha$ ratio.
If $\log (\mbox{[NII]}\lambda6583/\mbox{H}\alpha )$ was less or equal to --0.3, we concluded that the gas was excited
by young stars \citep{kauffman_2003}; if it was larger, then the excitation mechanism in the galactic 
disc was thought to be by shock waves \citep{allen_shock} and/or by old hot stars \citep{oldstar_excitation}. 
In the former case we calculated the emission-line
flux of the H$\beta$ emission as the H$\alpha$ emission flux divided by 2.85, in the latter case --
as one fourth \citep{stasodre}.

\section{Structure of the galaxies}

\begin{figure*}[p]
\centering
\begin{tabular}{c c}
\begin{tabular}{c}
\vspace{0.2cm}
 \includegraphics[width=3.5cm]{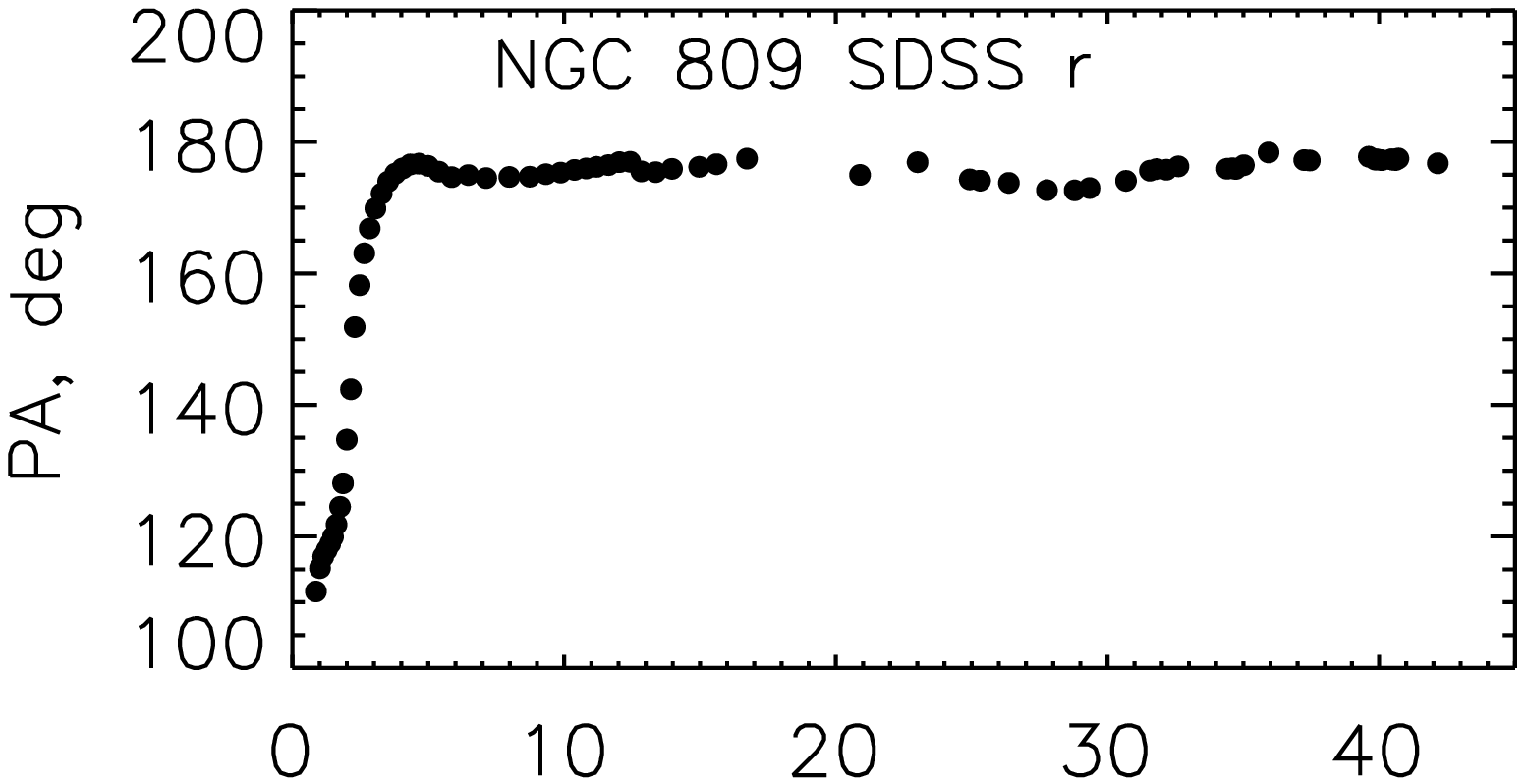} \\
 \includegraphics[width=3.5cm]{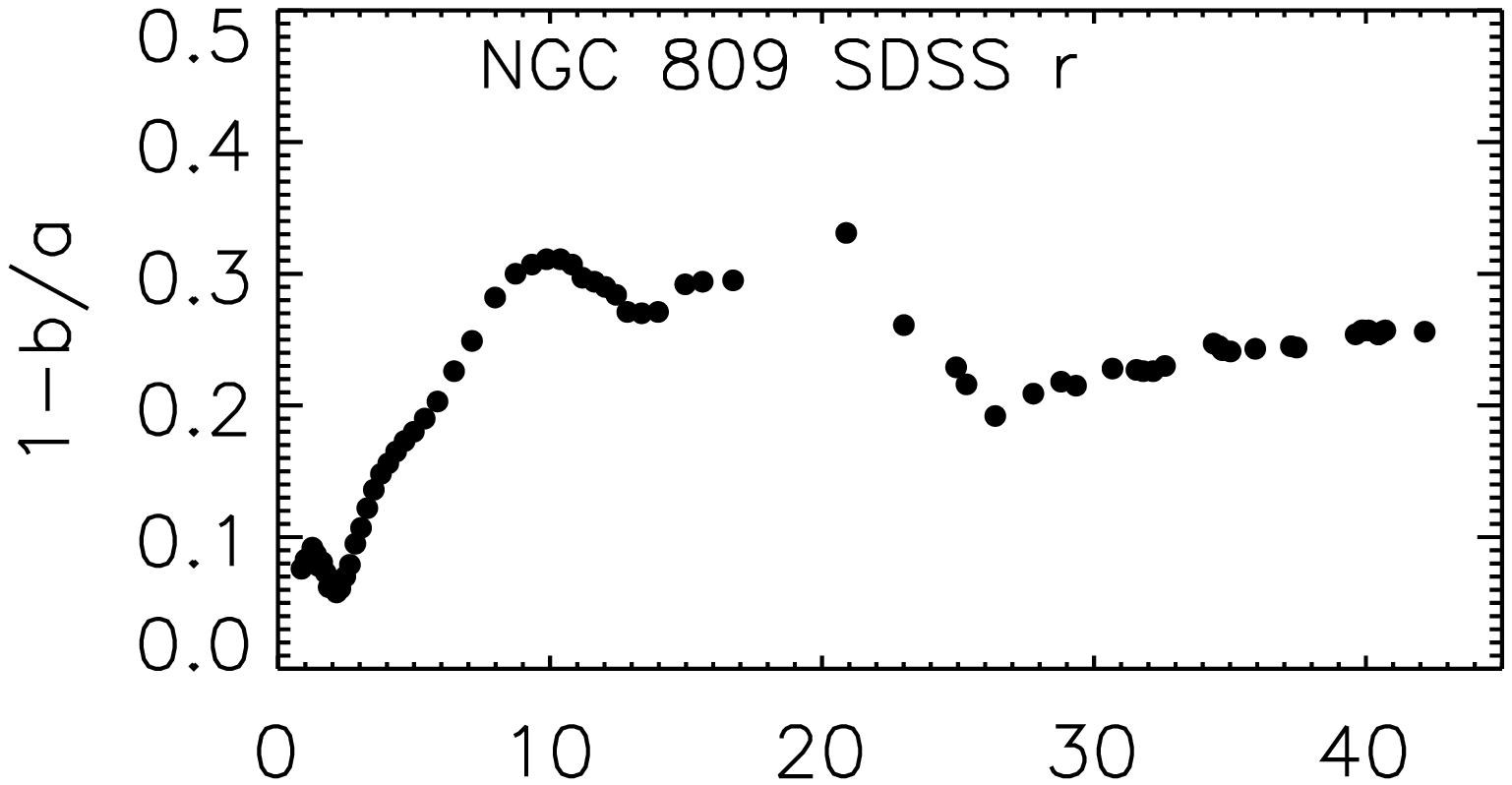}\\
\end{tabular}
 &
\begin{tabular}{c}
 \includegraphics[width=4.3cm]{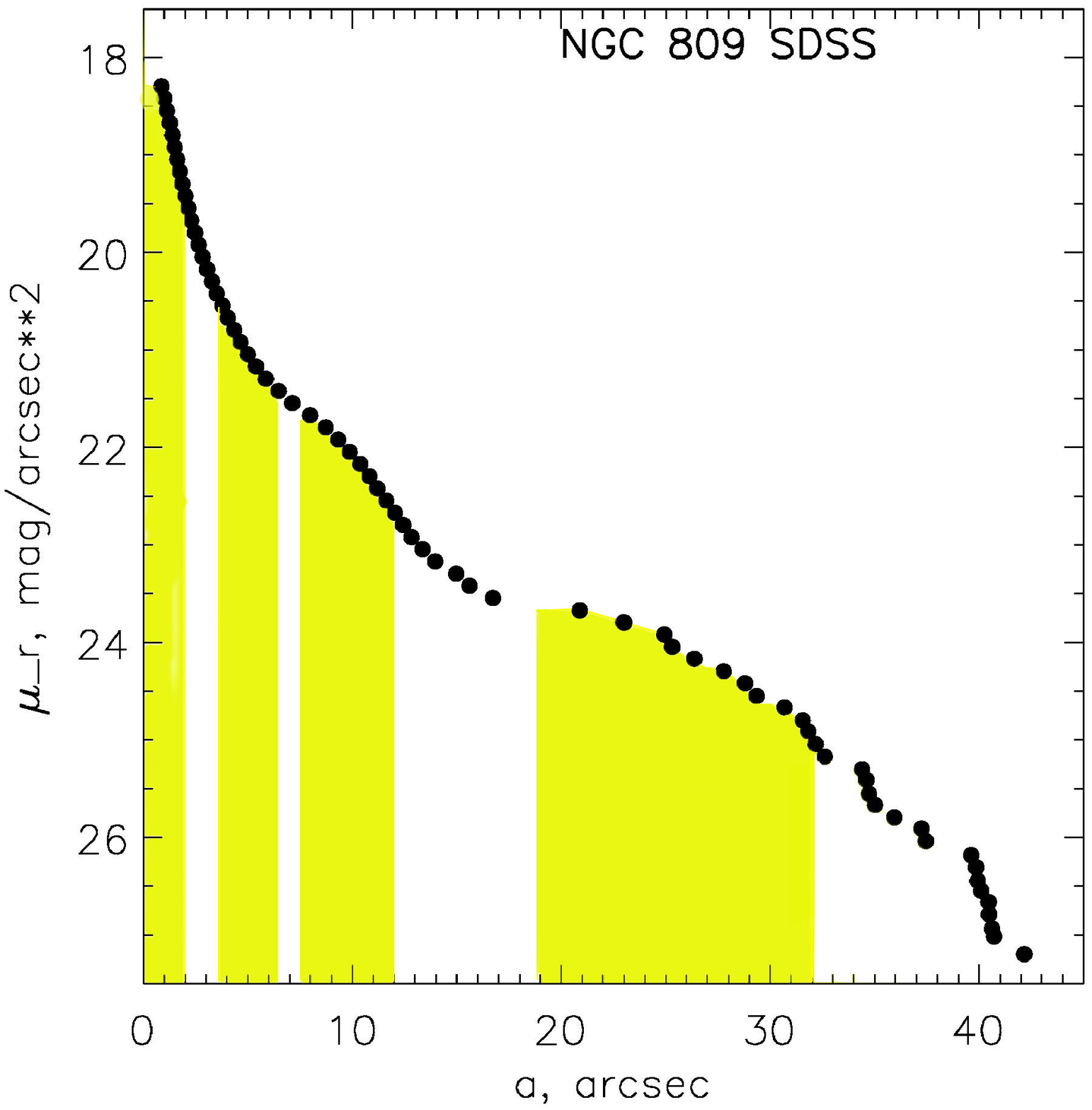} \\
\end{tabular} \\
\begin{tabular}{c}
\vspace{0.2cm}
 \includegraphics[width=3.5cm]{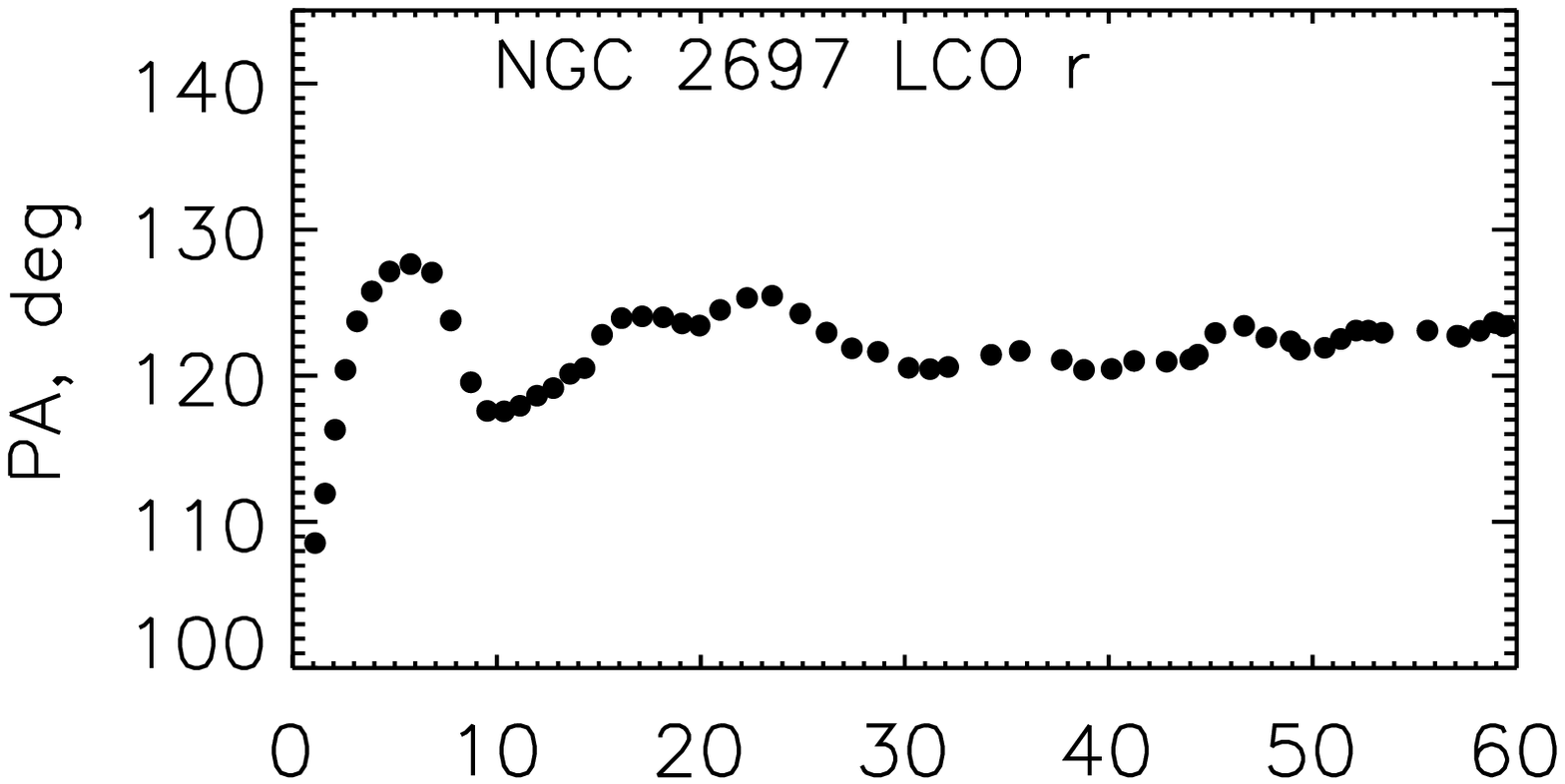} \\
 \includegraphics[width=3.5cm]{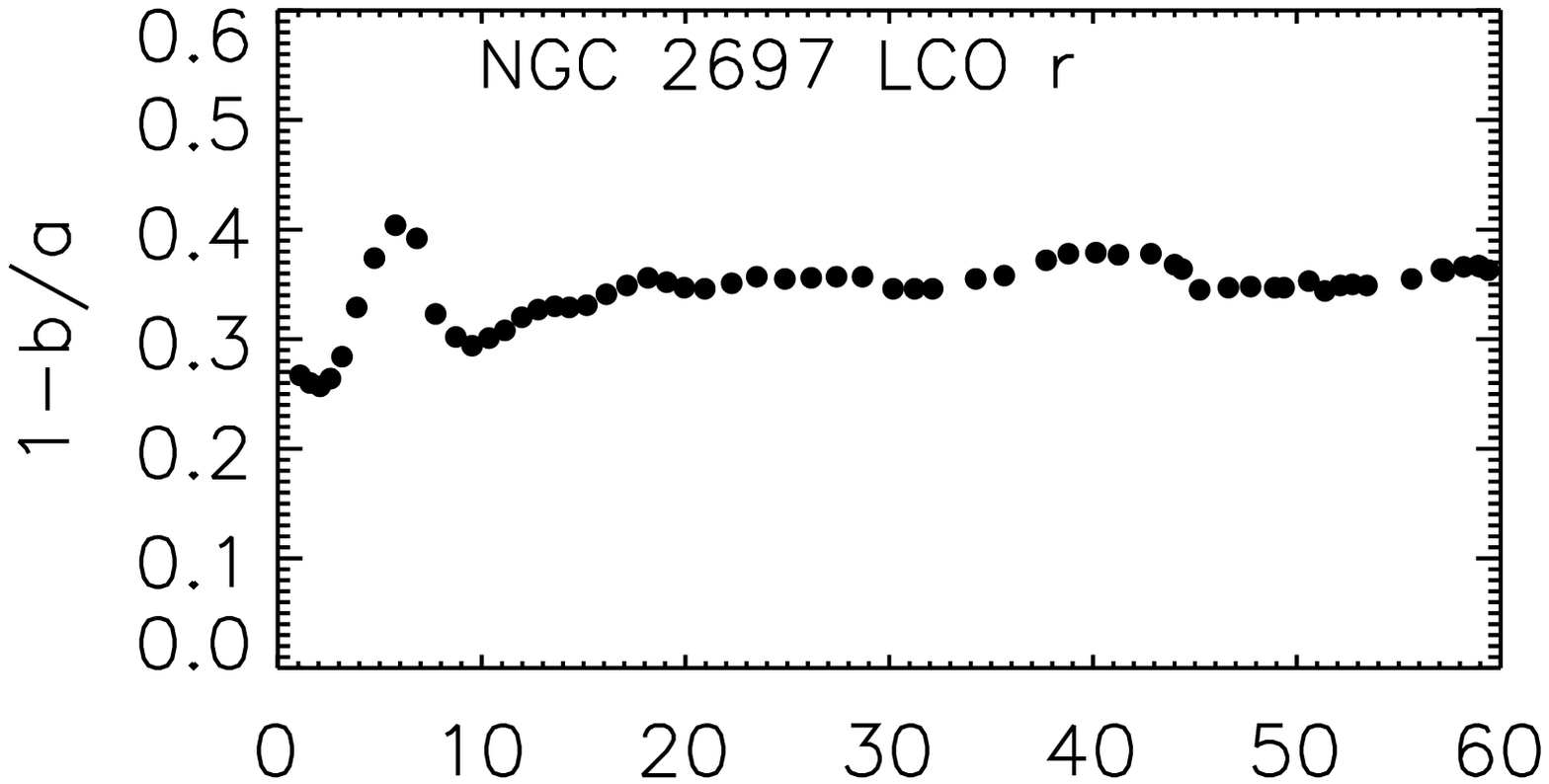} \\
\end{tabular}
 &
\begin{tabular}{c}
 \includegraphics[width=5.0cm]{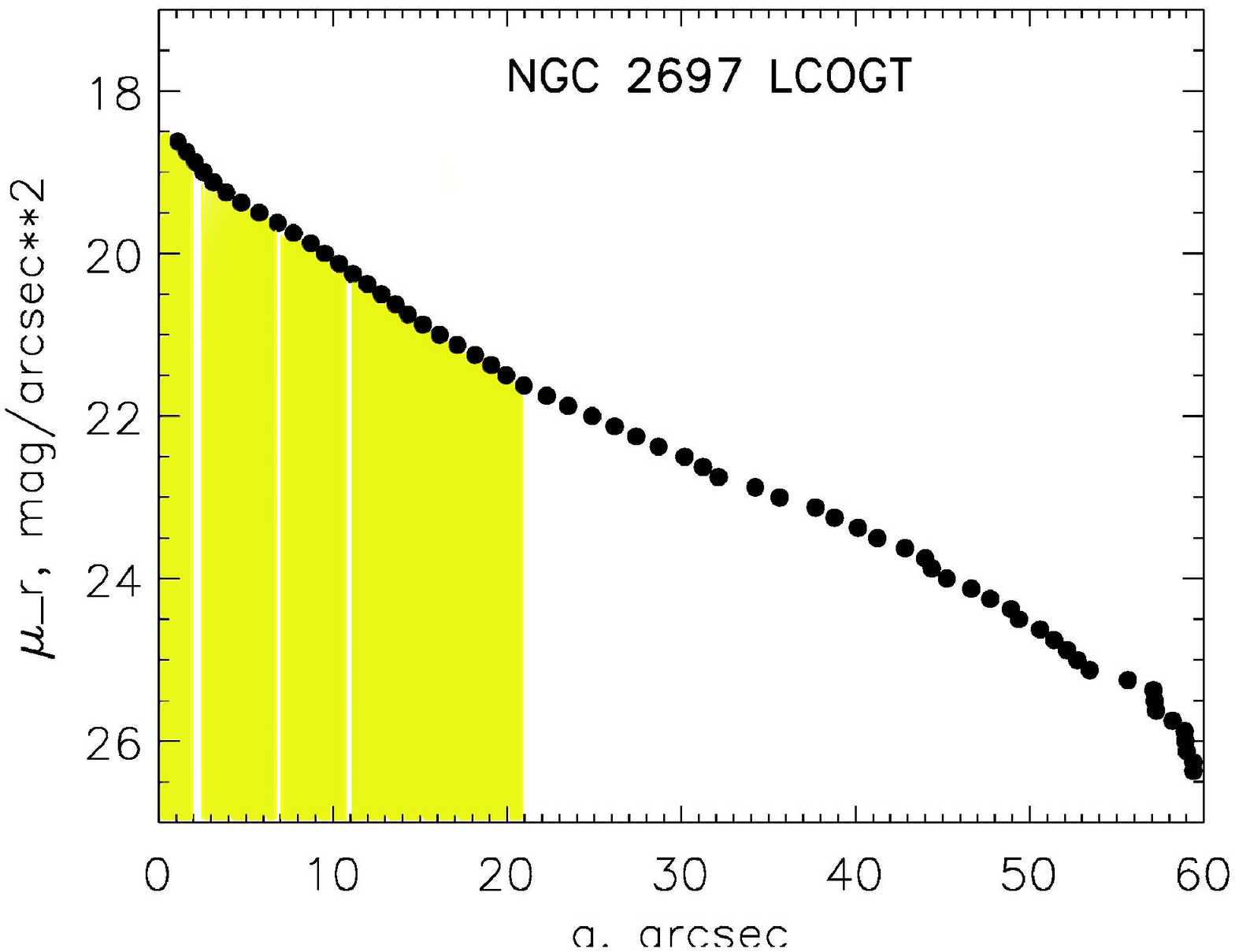} \\
\end{tabular} \\
\begin{tabular}{c}
\vspace{0.2cm}
 \includegraphics[width=3.5cm]{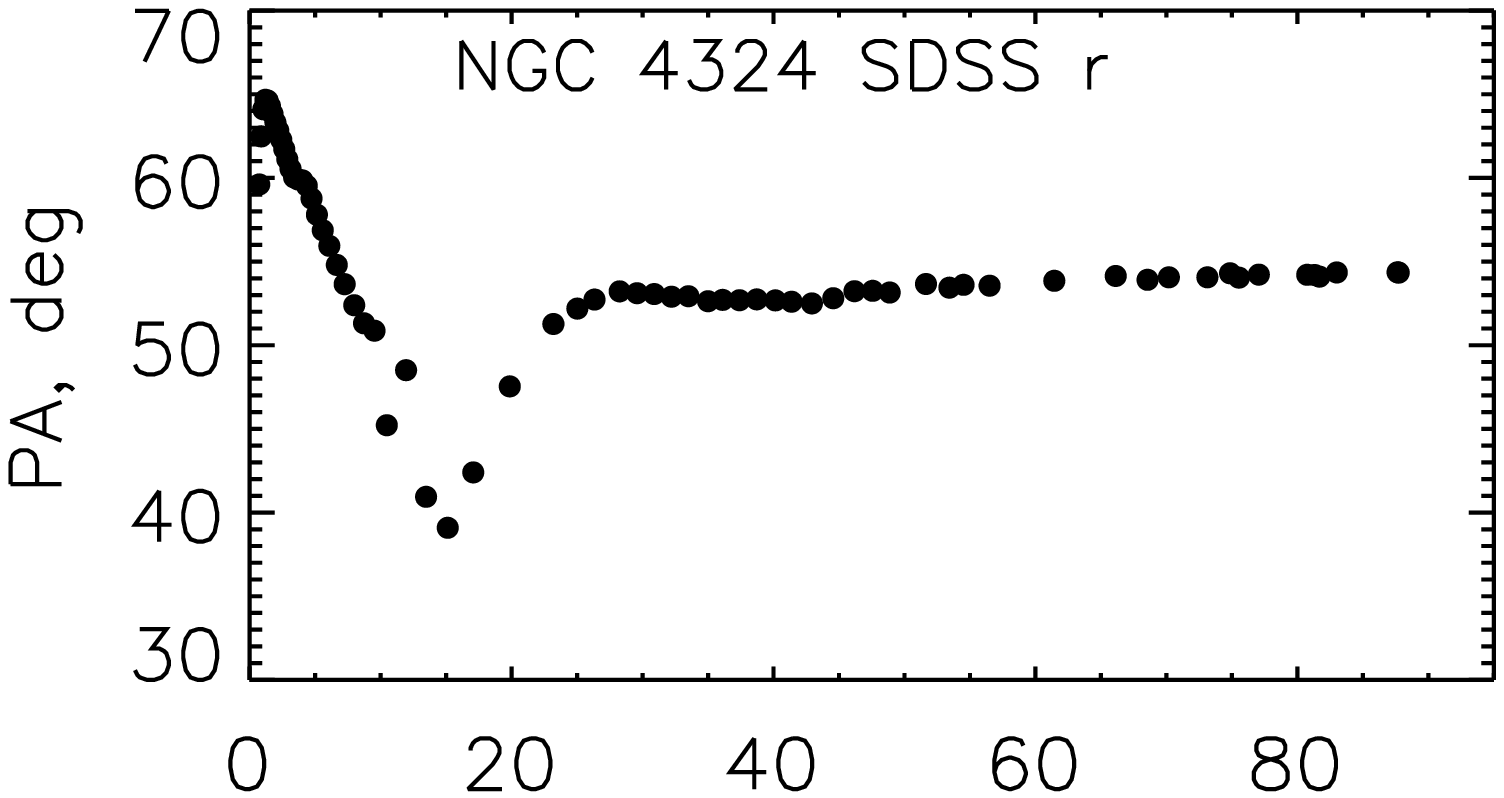} \\
 \includegraphics[width=3.5cm]{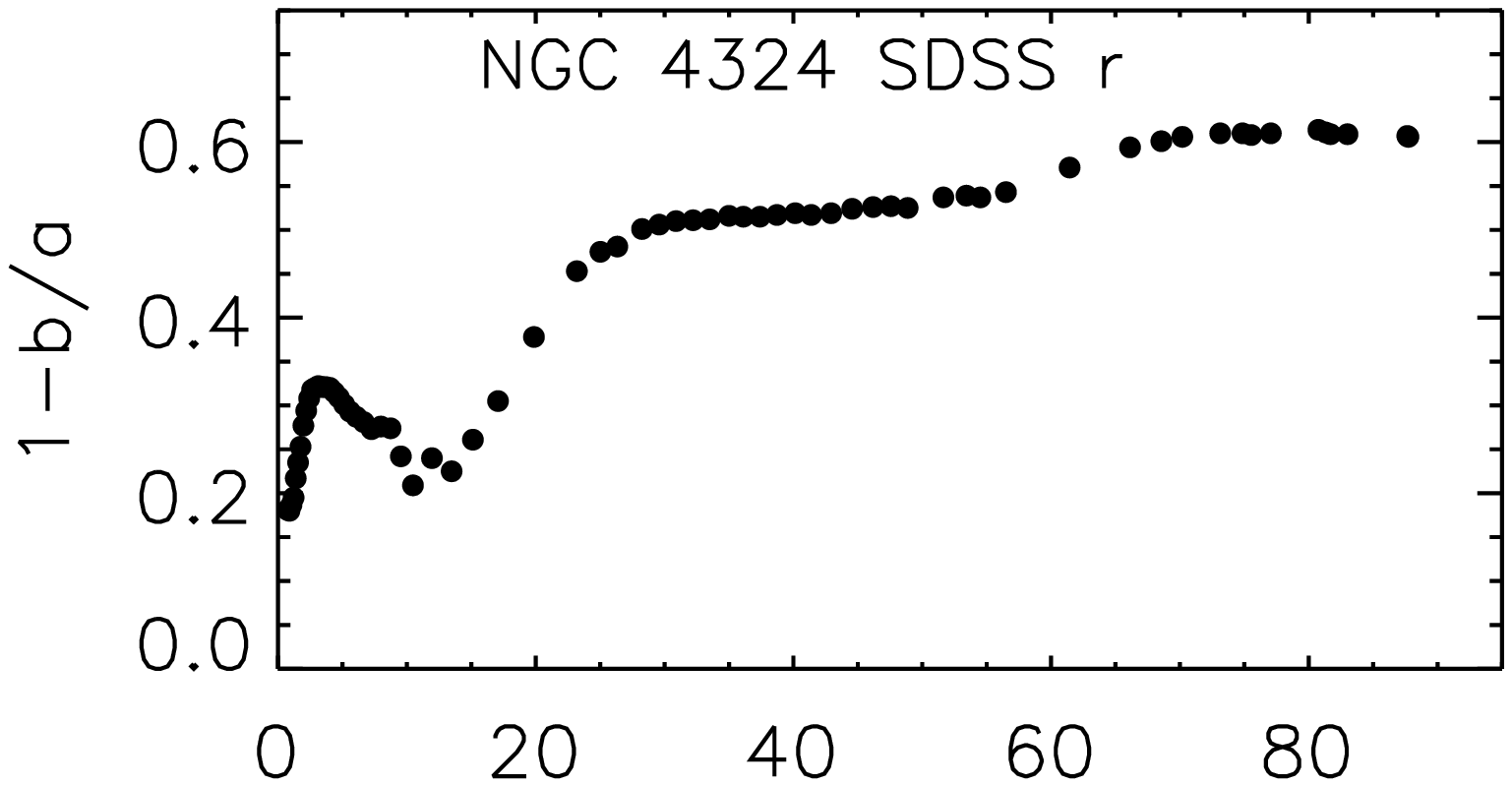} \\
\end{tabular}
 &
\begin{tabular}{c}
 \includegraphics[width=4.0cm]{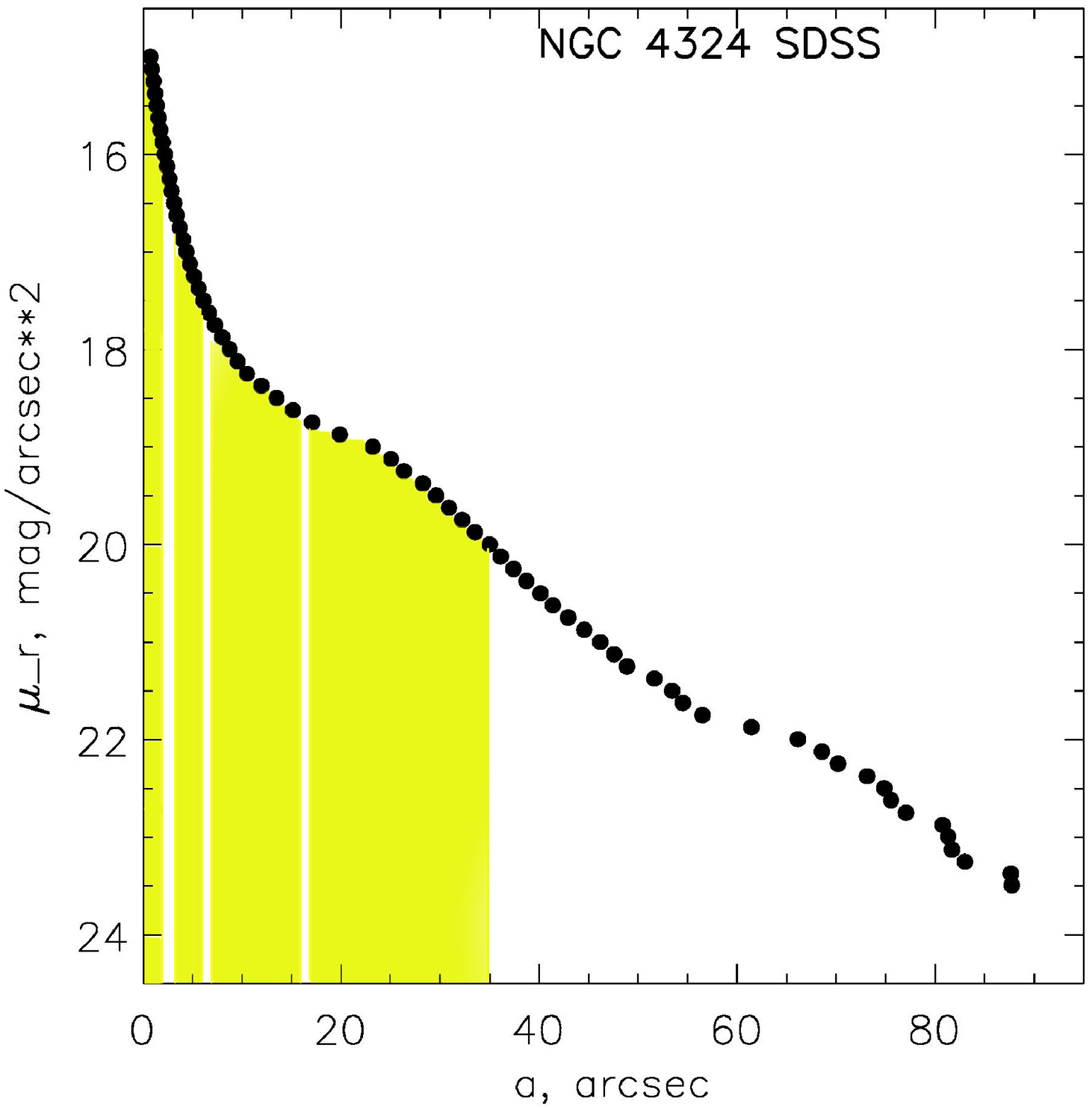} \\
\end{tabular} \\
\begin{tabular}{c}
\vspace{0.2cm}
 \includegraphics[width=3.5cm]{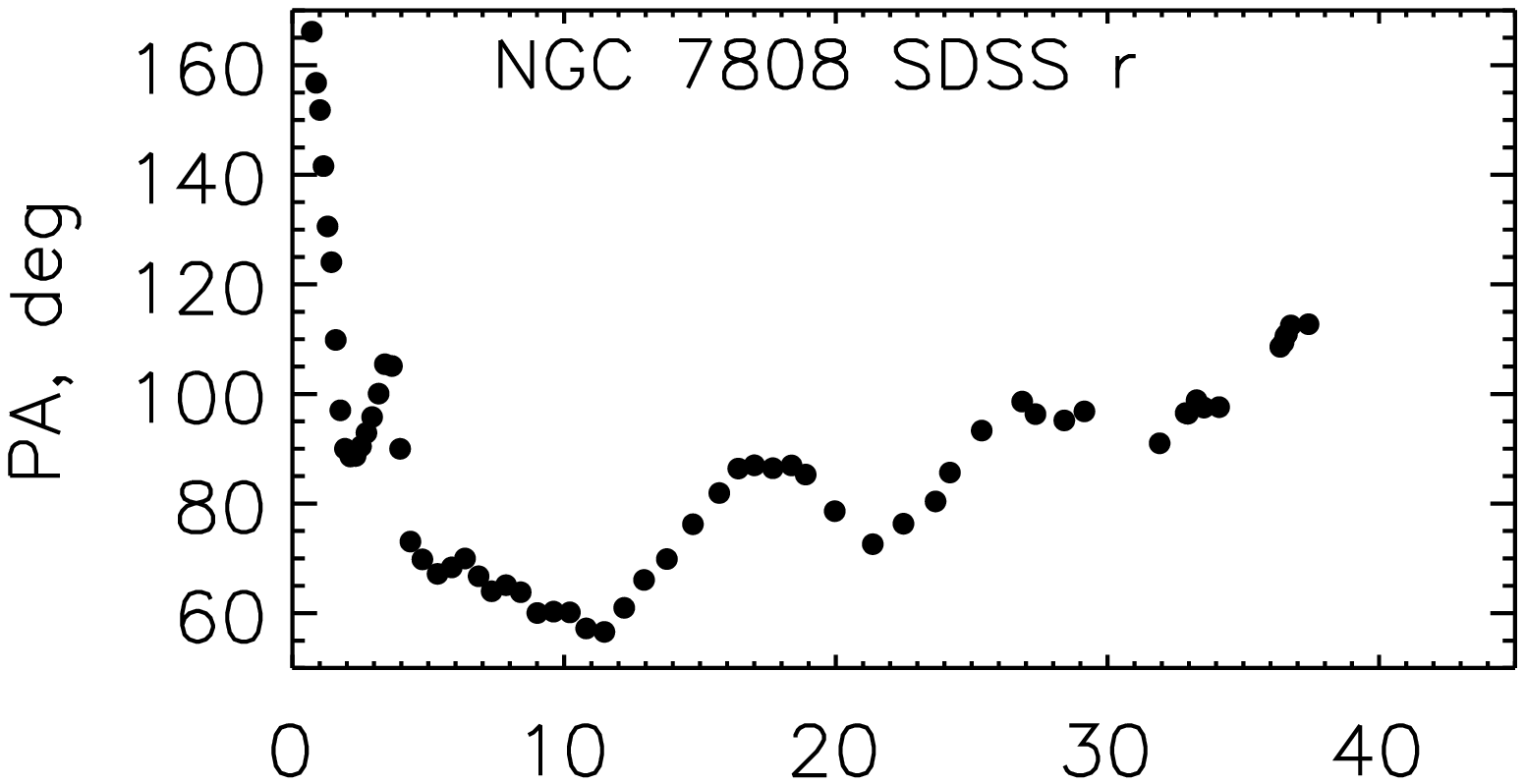} \\
 \includegraphics[width=3.5cm]{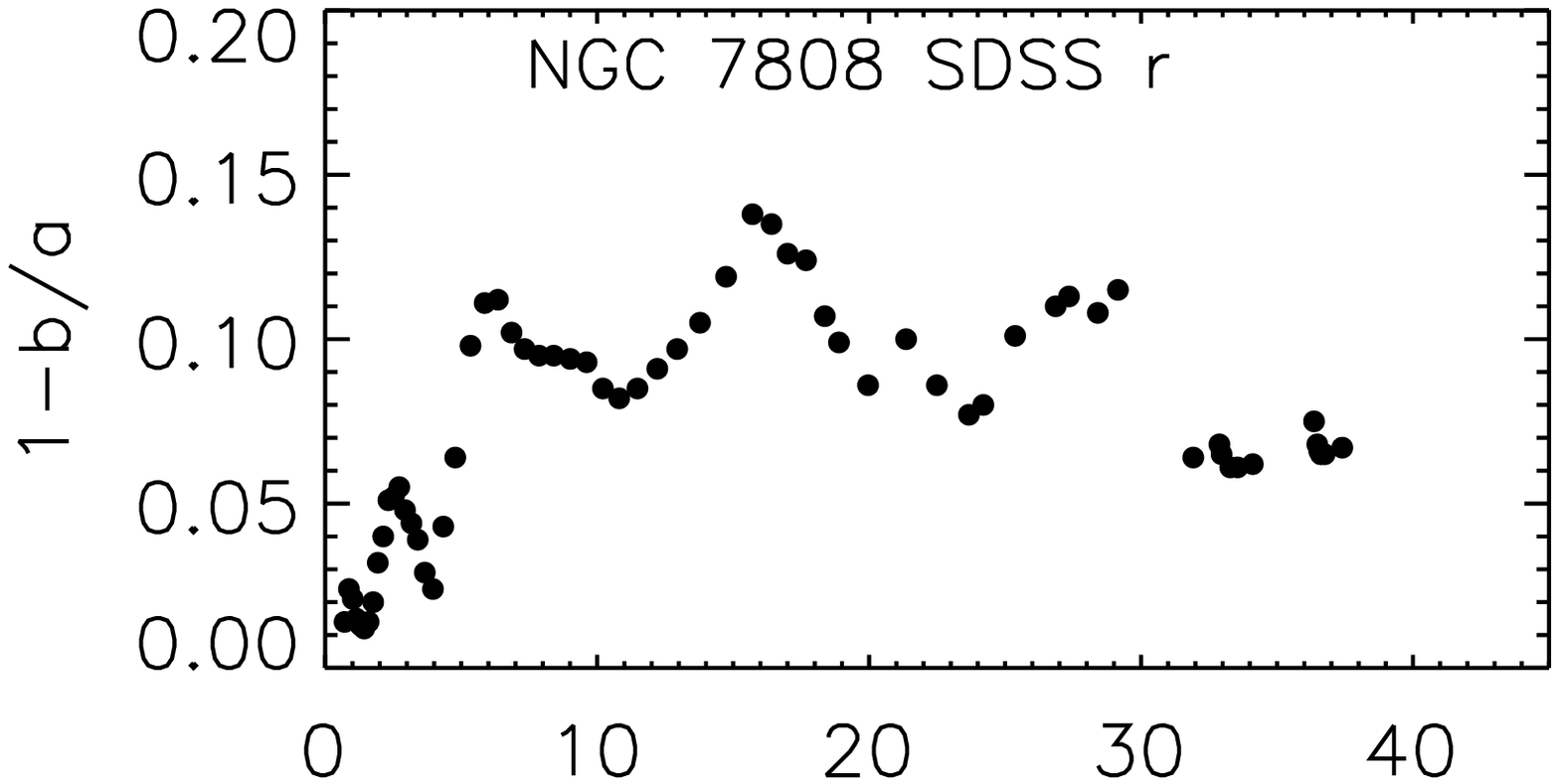} \\
\end{tabular}
 &
\begin{tabular}{c}
 \includegraphics[width=4.0cm]{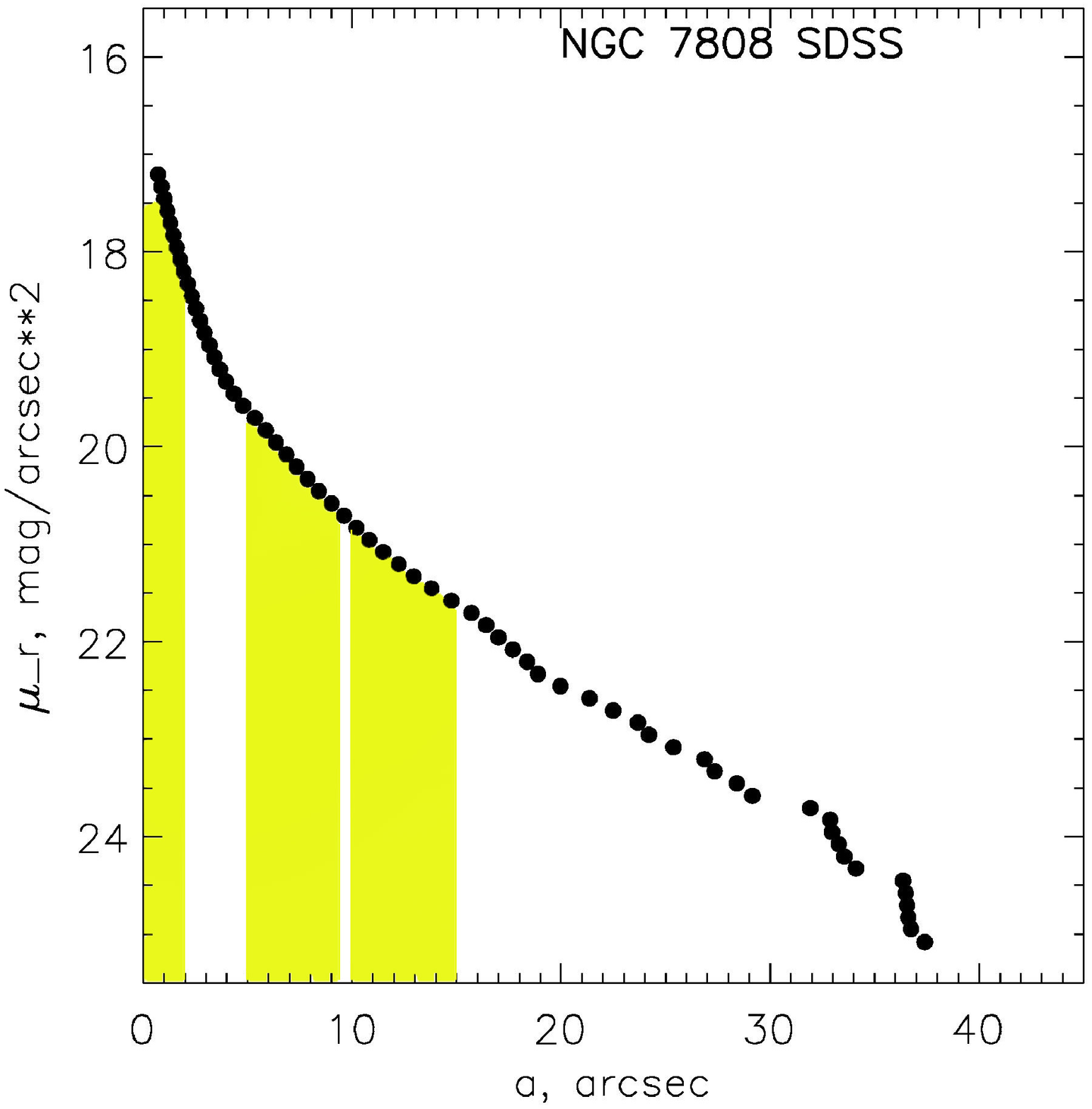} \\
\end{tabular} \\
\begin{tabular}{c}
\vspace{0.2cm}
 \includegraphics[width=3.5cm]{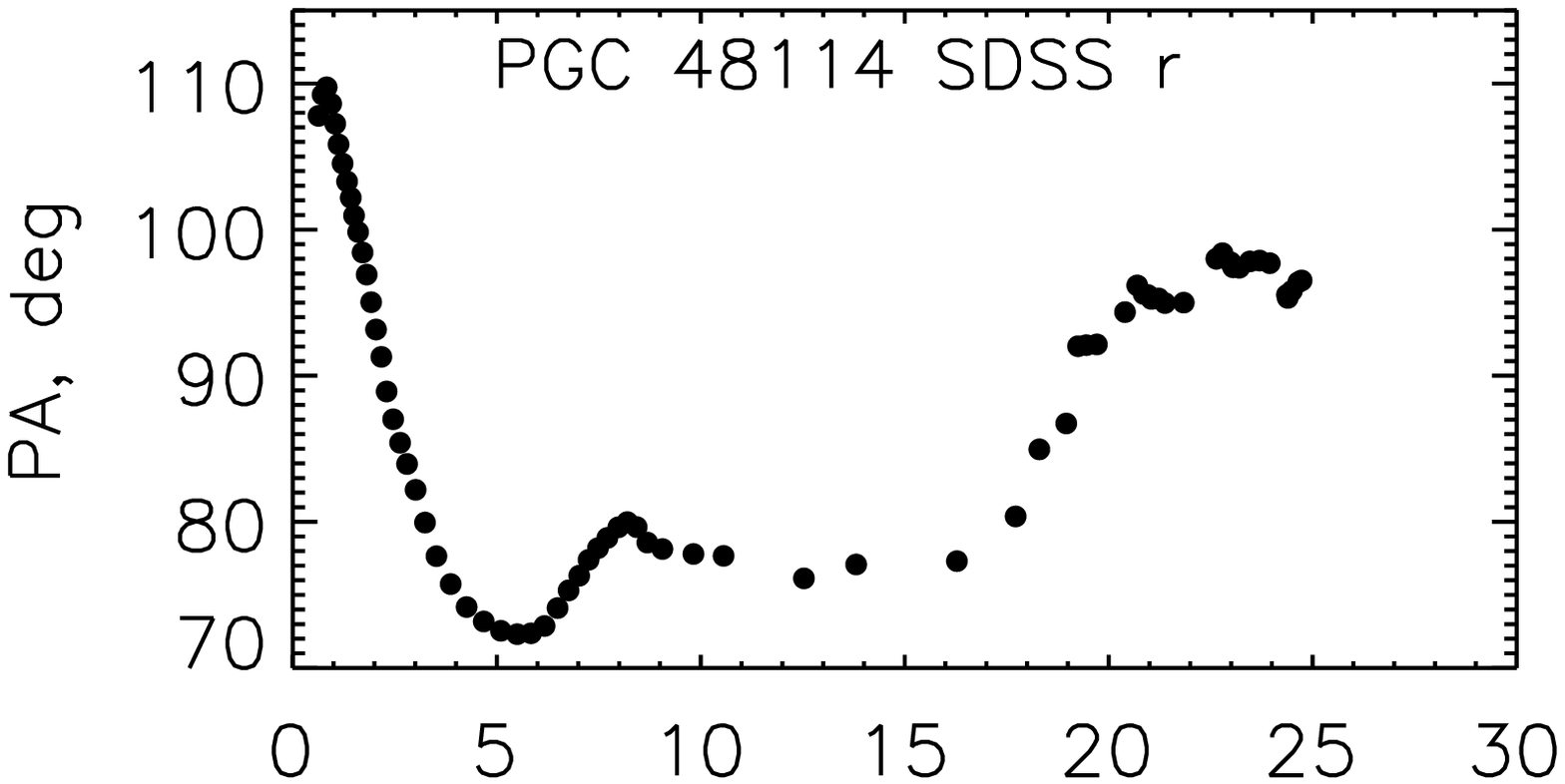} \\
 \includegraphics[width=3.5cm]{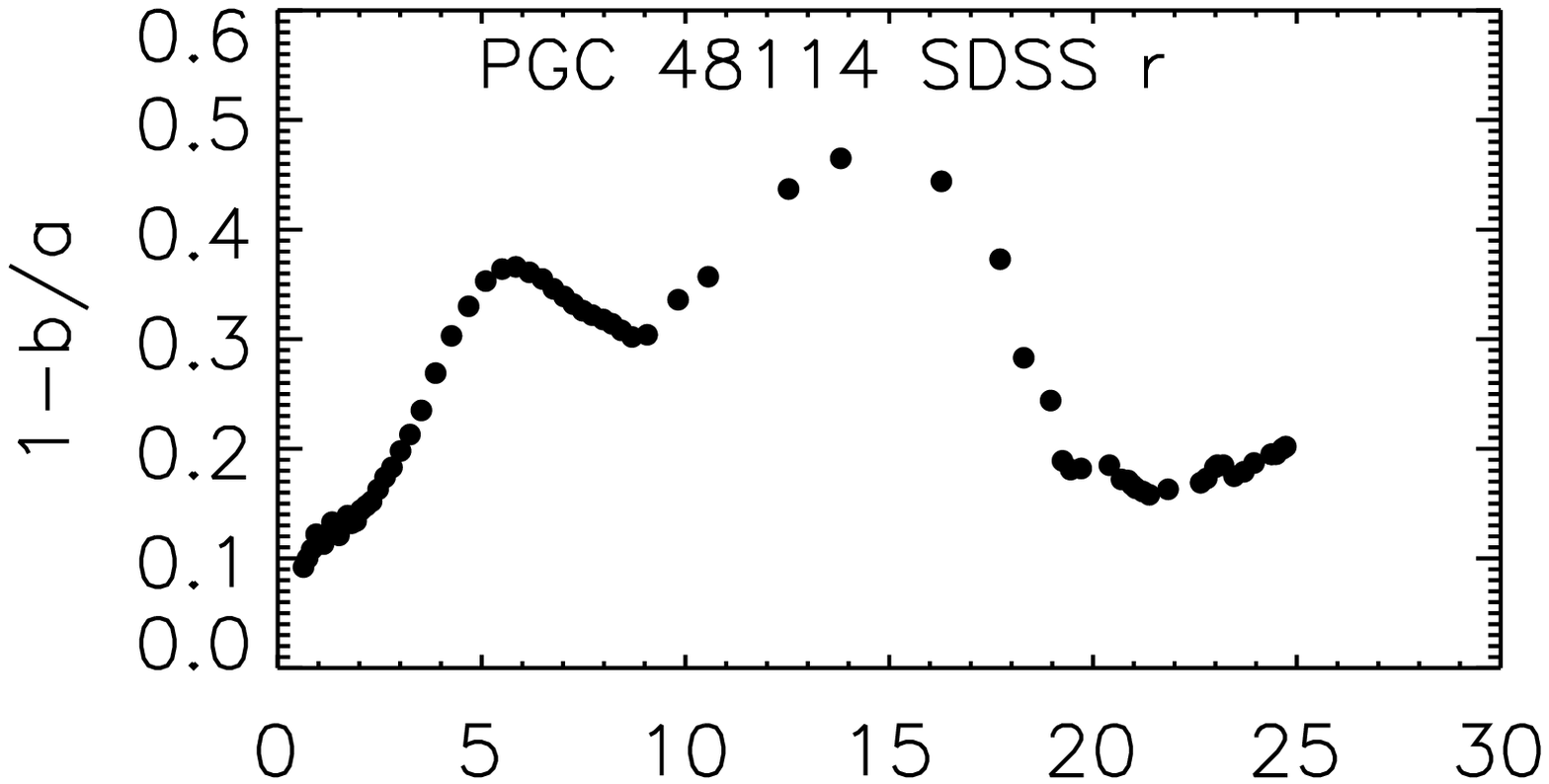} \\
\end{tabular}
 &
\begin{tabular} {c}
 \includegraphics[width=4.0cm]{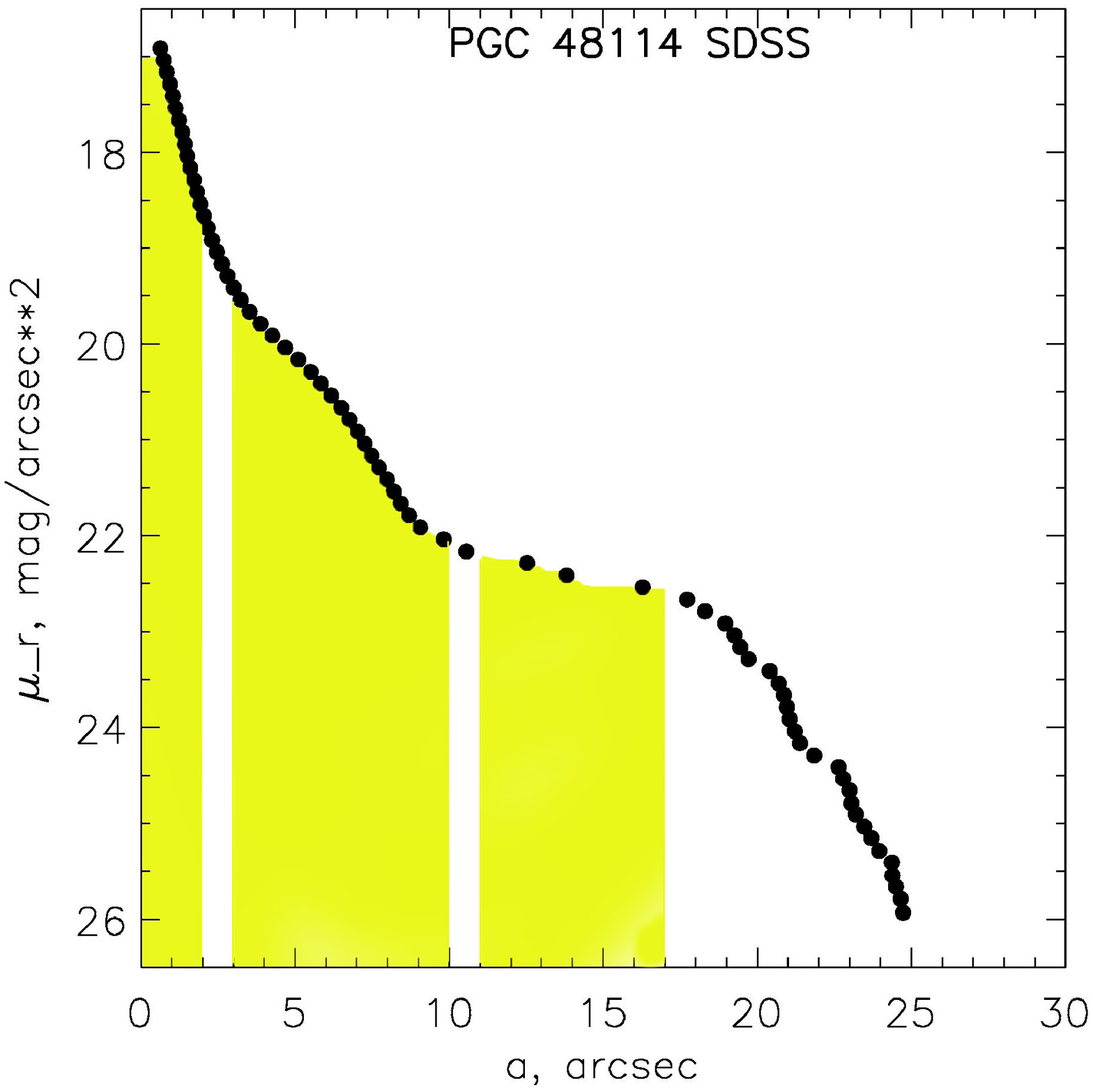} \\
\end{tabular} \\
\end{tabular}
\caption{Isophote major-axis position angle and ellipticity radial profiles (to the {\it left}) and $r$-band surface-brightness 
profiles (to the {\it right}). We have filled by yellow the radial zones for which the stellar population parameters are studied below. 
}
\label{iso}
\end{figure*}

Before deriving conclusions from our spectral results, we have studied the structure of the galaxies under
consideration, to be able to attribute the properties of radially resolved stellar populations to the bulges, or 
to the disks, or to the rings and the lenses if they are present. 
For 4 galaxies we have used the SDSS/DR9 imaging data \citep{dr9}; but
for NGC~2697 special observations were made with the telescope network LCO \citep{lco} since this galaxy is not
in the SDSS footprint. In all the cases the SDSS photometric system, $gri$, was used.
Figure~\ref{iso} shows the radial profiles of the isophote parameters and of the surface brightness
azimuthally averaged with the running values of the isophote ellipticities and orientation angles. At
the surface brightness profiles we mark in yellow the radius ranges for which the stellar population
characteristics will be analyzed. Initially, we suggested to measure the stellar population
properties in the nuclei, in the bulges, in the large-scale stellar disks, and in the rings. However,
we have failed to derive stellar population properties for the bulges in some cases, because NGC~2697
has no classical bulge at all, and in the distant objects NGC~7808 and PGC~48114 our spatial resolution is not
sufficient to separate the nuclei and the bulges, the bulges being rather compact. Instead, in some
cases we analyze {\it two} radial ranges containing rings, the inner ring being discovered only by inspecting 
the surface brightness profiles. We stress that here we define
the {\it stellar} rings as seen in the broad-band images. The {\it gaseous} rings marked by intense emission lines are
not necessarily exactly coincident with the stellar rings; this will be discussed further in a separate Section -- Section~5.

In general, four galaxies of the sample look axisymmetric, in accordance
with their nominal classification. Only in PGC~48114 the rings are both elliptical, demonstrating
local maxima of isophote ellipticity at the radii of 6\arcsec\ and 15\arcsec\ and the isophote major-axis position
angle turn at these radii by some 20--25 deg from the line-of-node orientation. NGC~7808 is seen
nearly face-on so the variations of the isophote major-axis PA in this galaxy could be attributed to the
low accuracy of its determination (the isophotes are nearly round). A quite curious feature is found in
NGC~4324: at $r=15$\arcsec, just inside the inner ring position at $r=23$\arcsec, the isophotes
demonstrate the major-axis turn and the ellipticity {\it minimum}. Such isophote behavior may be related to
a small bar orthogonal to the line of nodes. However we think that probably it is not a bar since no
kinematical disturbance of the gaseous disk is observed at this radius.

\section{Gaseous and stellar rotation}

\begin{figure*}[t]
\begin{tabular}{c c}
 \includegraphics[width=8cm]{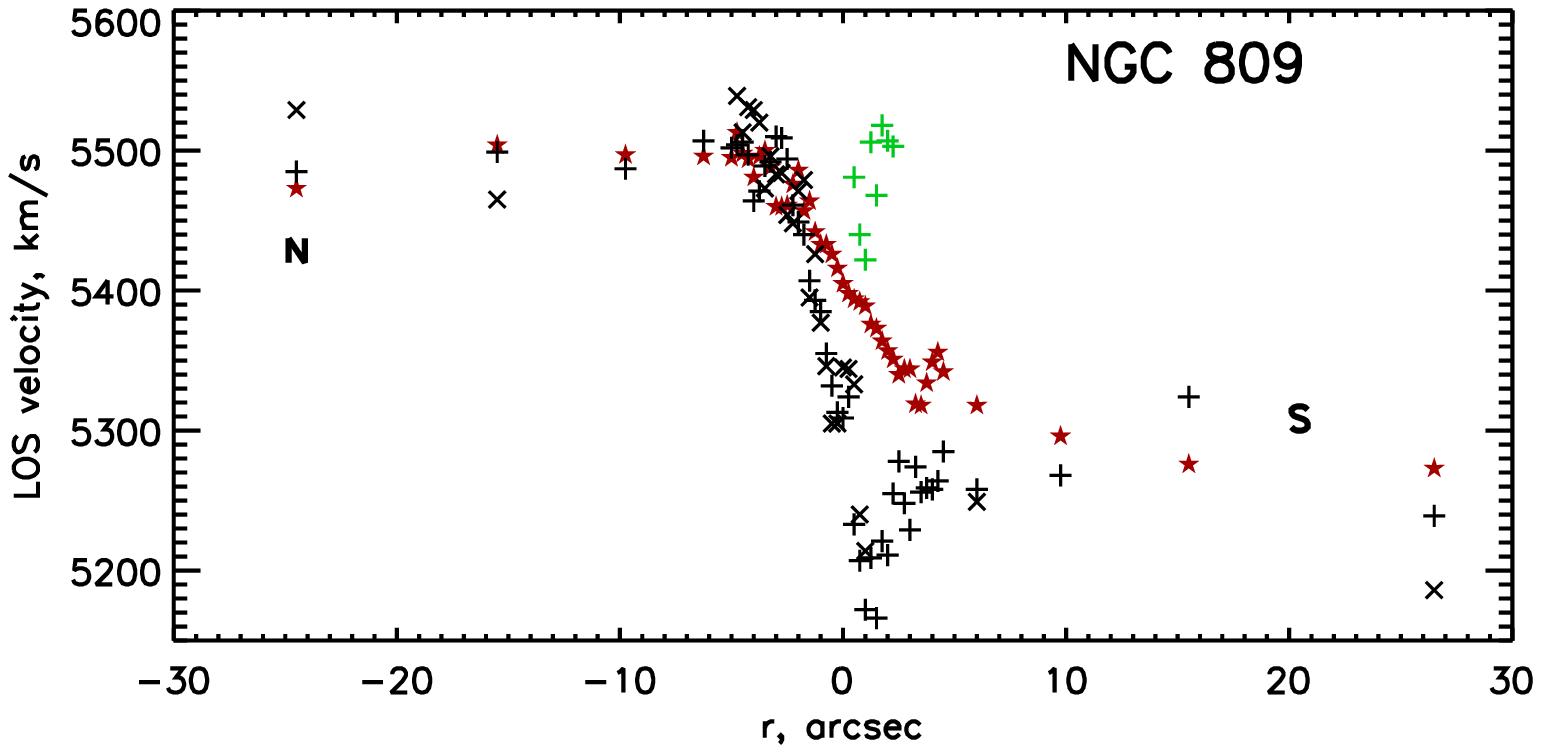} &
 \includegraphics[width=8cm]{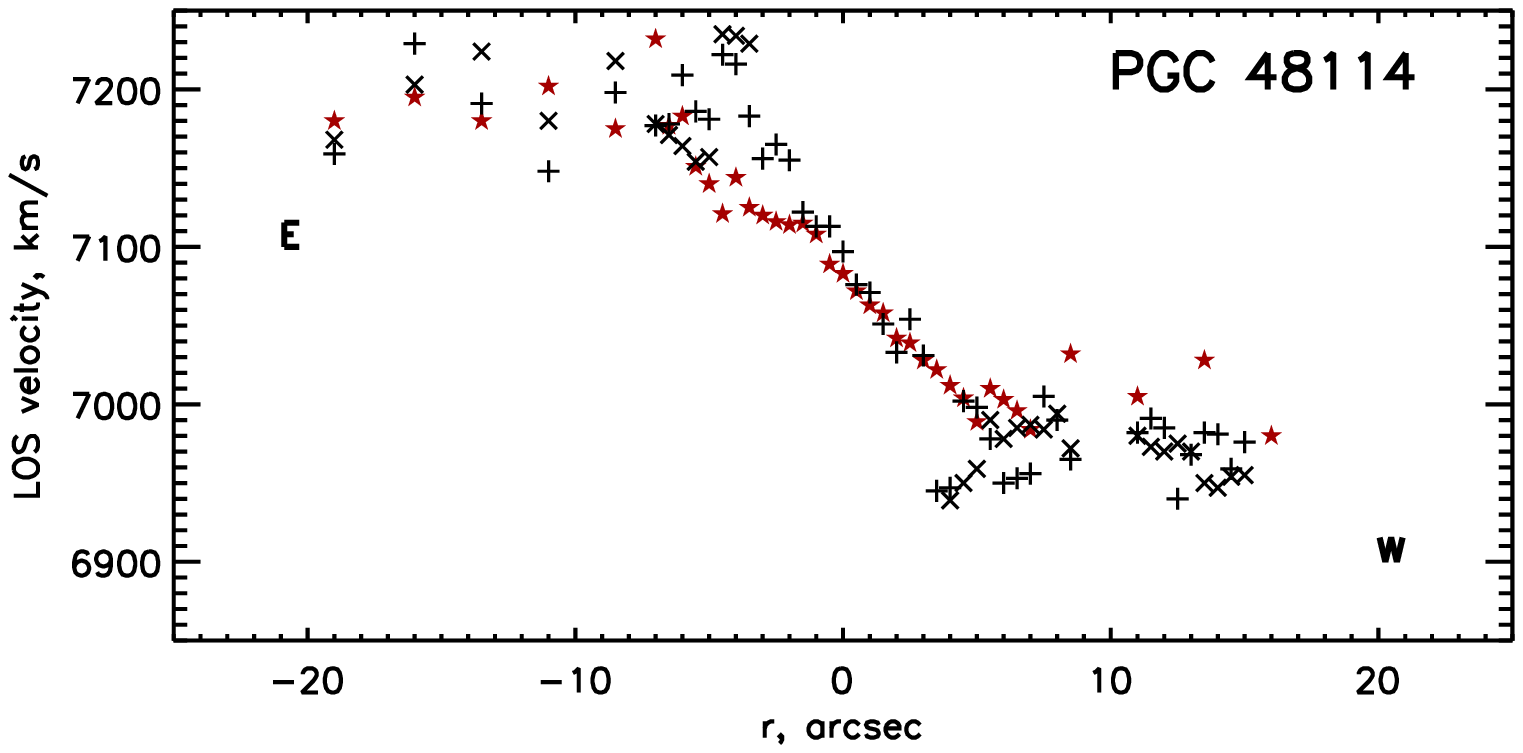} \\
 \includegraphics[width=8cm]{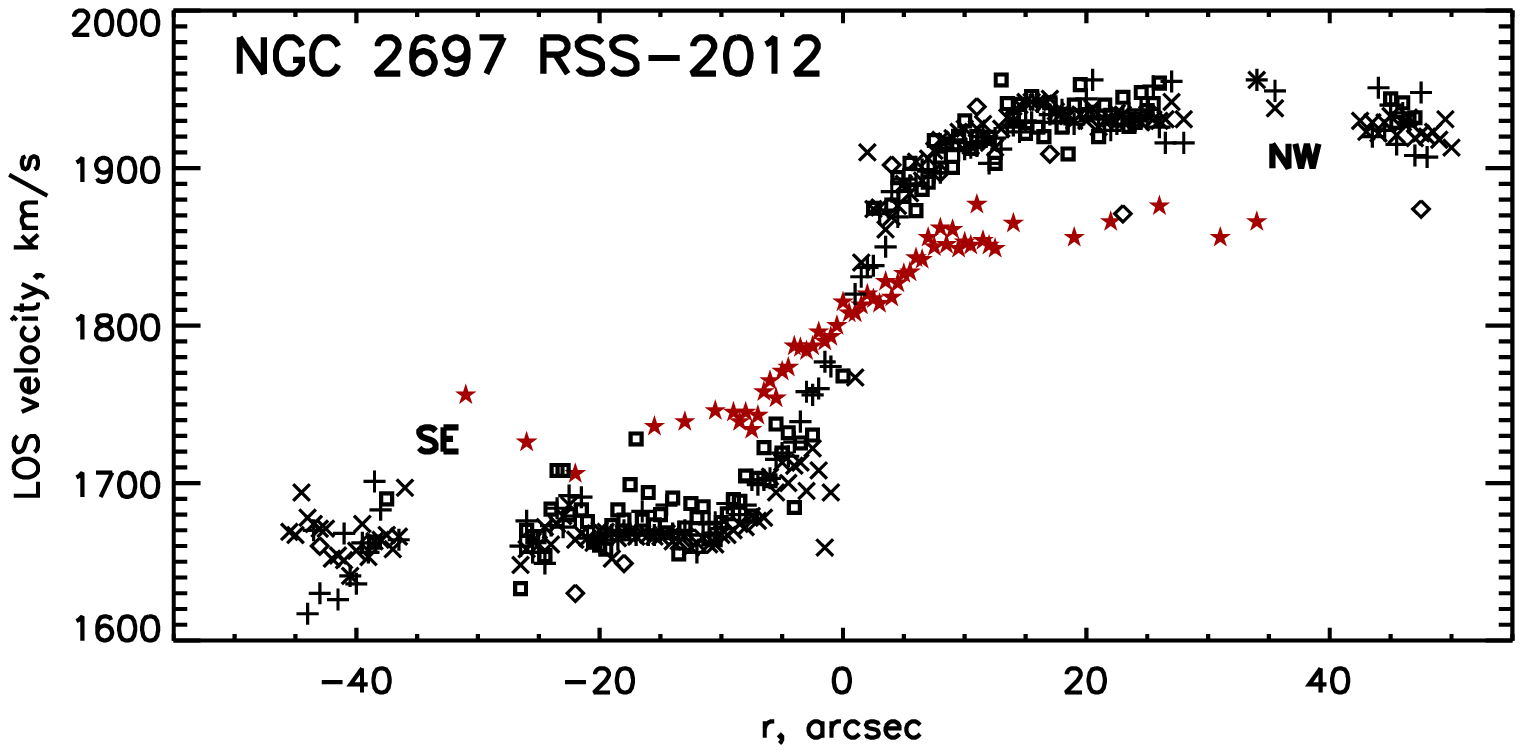} &
 \includegraphics[width=8cm]{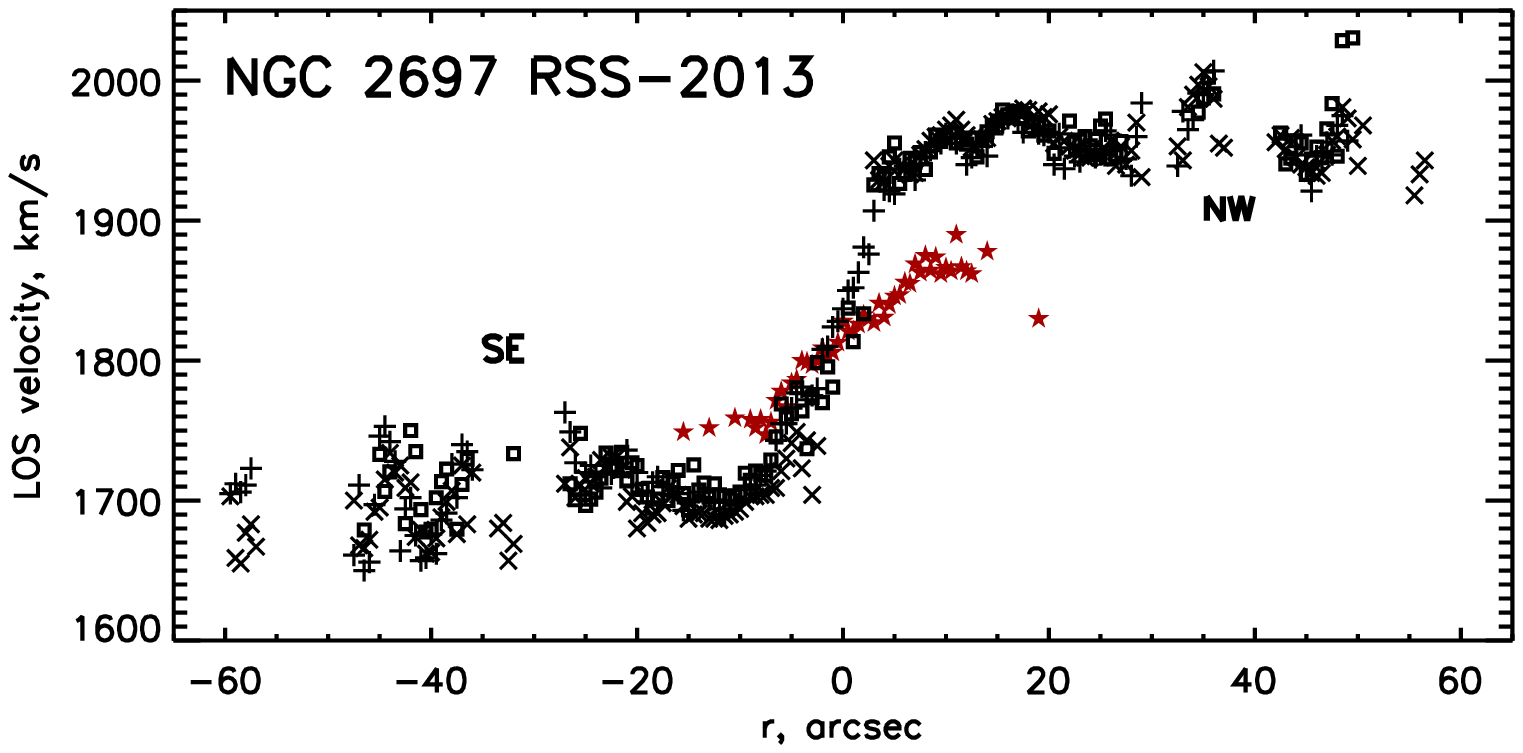} \\
 \includegraphics[width=8cm]{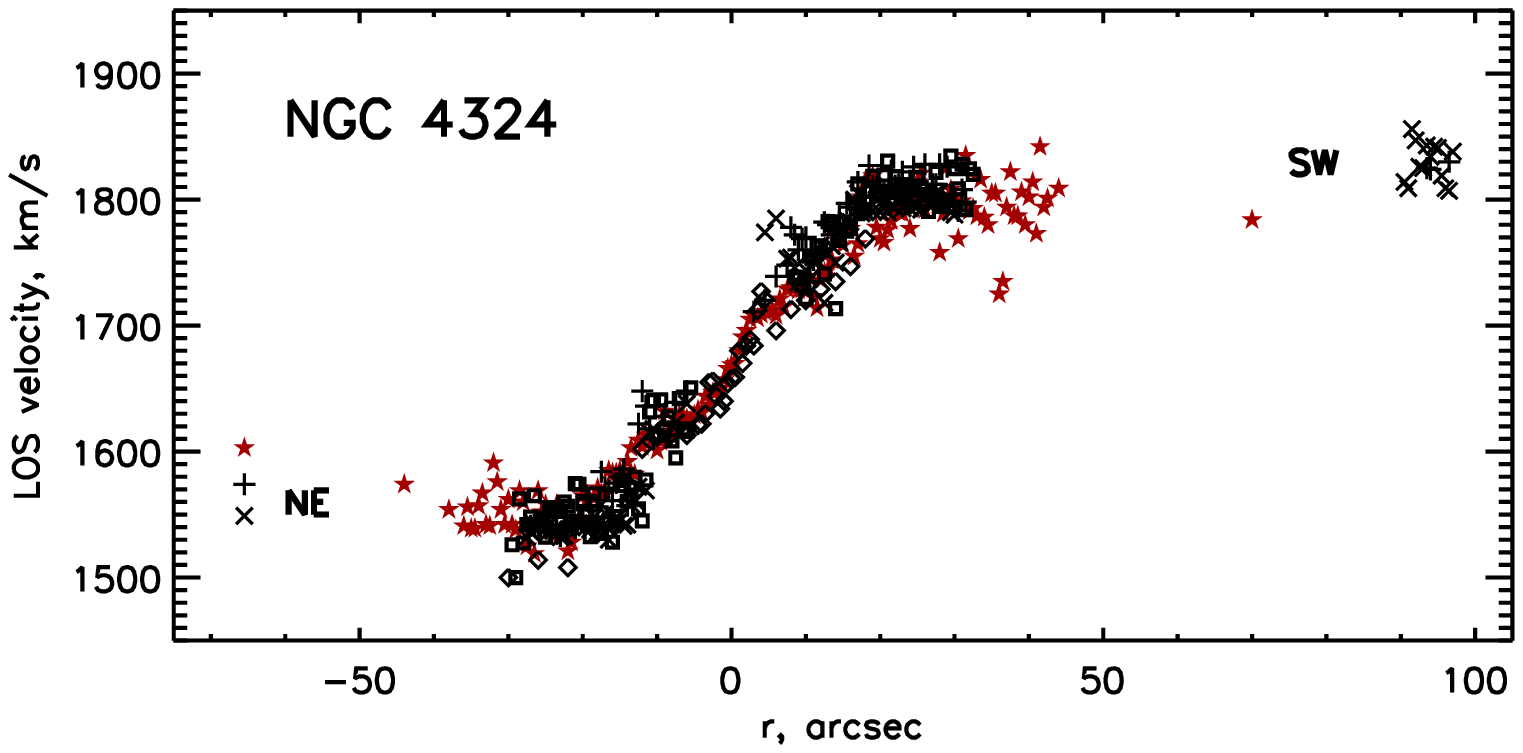} &
 \includegraphics[width=8cm]{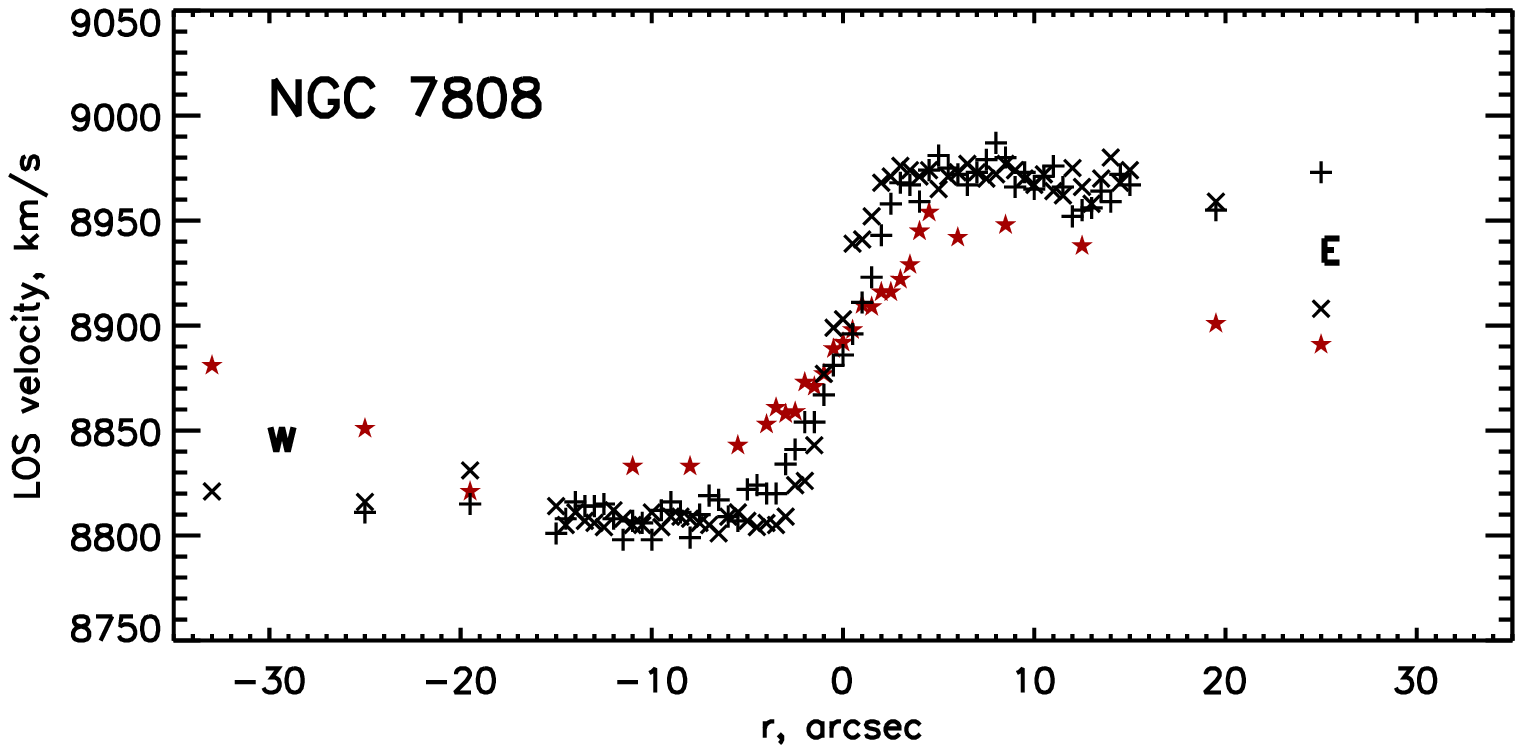} \\
\end{tabular}
\caption{Line-of-sight velocity profiles. Red stars signify stellar rotation,
black signs refer to the ionized-gas different emission lines: crosses -- H$\alpha$,
pluses -- [NII]$\lambda$6548+6583, squares -- [SII]$\lambda$6717+6731, diamonds --
[OIII]$\lambda$5007. For NGC~2697 two independent data sets are presented -- those obtained
in the same slit orientation in 2012 and in 2013. 
The emission line [NII]$\lambda$6583 in NGC~809 shows two components, well separated by their velocities, to the south from the nucleus.
}
\label{vels}
\end{figure*}

\begin{figure*}[tbp]
\centering
\includegraphics[width=5cm]{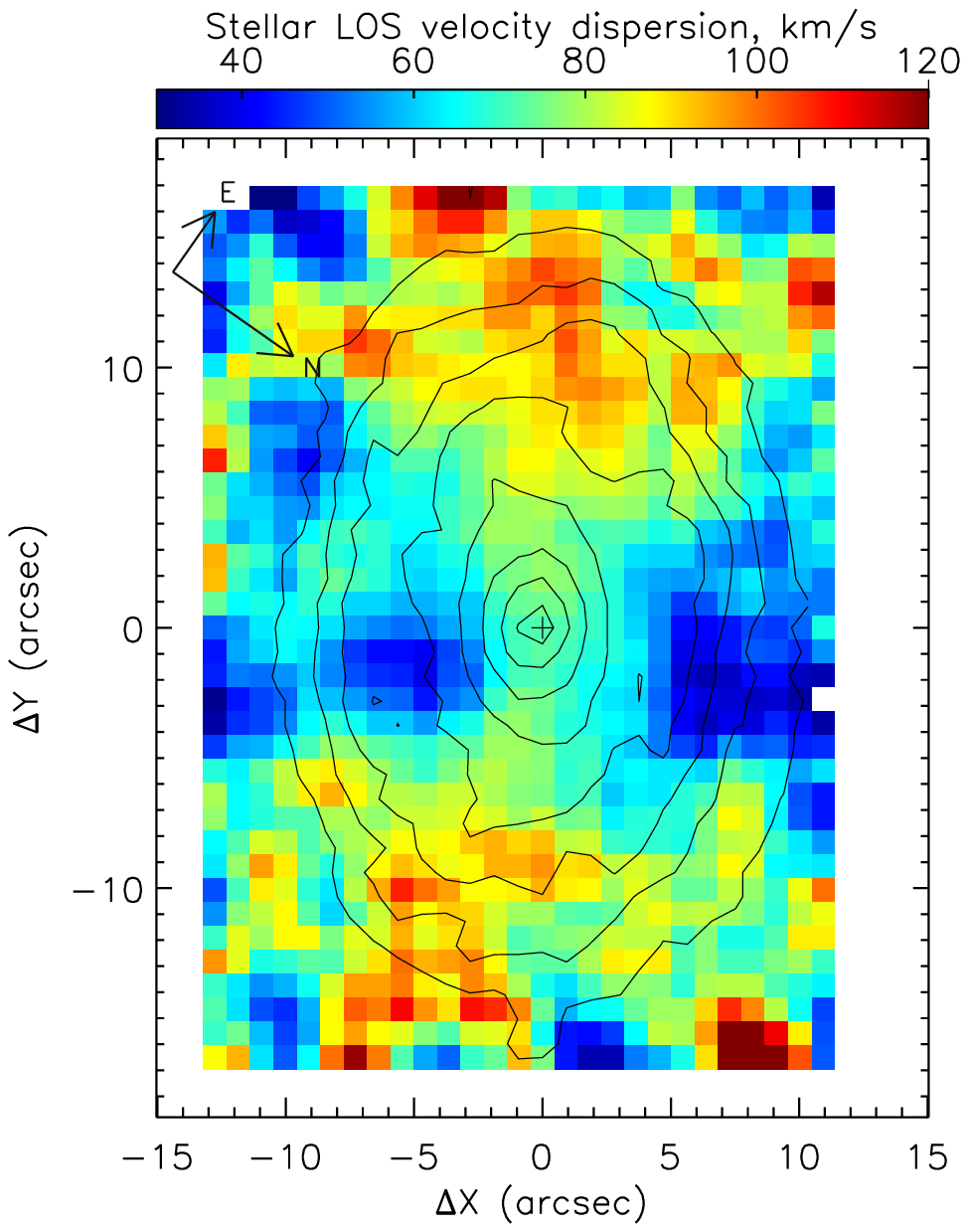}
\caption{The map of the stellar velocity dispersion for NGC~2697. Two quite untypical features can be
seen: the off-center stellar velocity dispersion maxima along the major axis of the disc and stellar 
velocity dispersion minima along the minor axis. The latter feature implies the absence of the bulge,
the former one betrays possible existence of two kinematically decoupled stellar subsystems.}
\label{n2697sig}
\end{figure*}

Figure~\ref{vels} presents line-of-sight velocity profiles both for the stellar and gaseous components in
our galaxies. NGC~2697 was observed twice in the same slit orientation, so we present two plots with independent
measurements. NGC~7808 was observed in two {\it different} slit orientations. But accidentally these two
cross-sections have appeared to deviate symmetrically from the line of nodes, and the measured velocity
profiles, both for the gas and stars, have coincided perfectly; so we present a united plot for this galaxy.
In dynamically cold systems, the agreement between the velocity profiles (rotation curves) derived
by measuring emission lines and absorption lines implies indirectly the coincidence of the rotation planes 
of the stars and ionized gas. The gaseous disks are likely coplanar with the stellar disks in NGC~4324 and PGC~48114.
In NGC~2697 and NGC~7808 the stellar component rotates more slowly than the ionized gas that may
indicate that the gaseous disks are tilted with respect to the stellar disks. In the latter galaxy the difference
may be alternatively an effect of asymmetric drift because NGC~7808 is the most luminous galaxy in our sample, and 
we may expect high velocity dispersion in its stellar disk.
As for NGC~2697, according to the SAURON data which we have retrieved from the open ING Archive,
the stellar and ionized-gas components demonstrate very different visible rotation velocities: we see the rotation
with the speed less than $80\, \mbox{km} \cdot \mbox{s}^{-1}$ in the stellar component and about $200\,\mbox{km} \cdot \mbox{s}^{-1}$
for the ionized gas. At the distance of 10\arcsec--15\arcsec\ from the center a sharp increase of the stellar velocity
dispersion, up to more than 100 km s$^{-1}$, is observed just at the line of nodes (Fig.~\ref{n2697sig}).
It is a disk-dominated area (Fig.~\ref{iso}) which is expected to be dynamically cold; therefore we may
suspect an overlapping of two kinematically decoupled stellar subsystems along our line of sight -- two stellar disks
with slightly different inclinations. One of them may be formed currently from freshly accreted gas. But 
the low spectral resolution  of the observations does not allow us to state surely the presence of two
kinematically different stellar disks. Also because of the low spectral resolution we cannot measure asymmetric
drift for the stellar disk in NGC~7808.

In NGC~809 there are apparently strong perturbations in the dynamics of the gaseous disk -- we see double-horn
profiles of the [NII] emission line just to the south from the nucleus. It may be explained by the possible presence
of a compact bar in the central part of this galaxy which was implied also by the photometric data: the isophote 
major axis is turned by some 70\degr\ within the radius of 5\arcsec. The edges of bars often host shocks betrayed
by the dust lanes. If our slit has caught simultaneously shock-excited decelerated gas at the bar edge {\it and}
freely rotating starforming gas, we may expect a double-peak emission-line profile reflecting the presence of
two kinematically different gaseous subsystems on our line of sight.

\section{Excitation and chemistry of the ionized gas}

\begin{figure*}[!h]
\centering
\vspace{1.5cm}
\begin{tabular}{c c}
 \includegraphics[width=6cm]{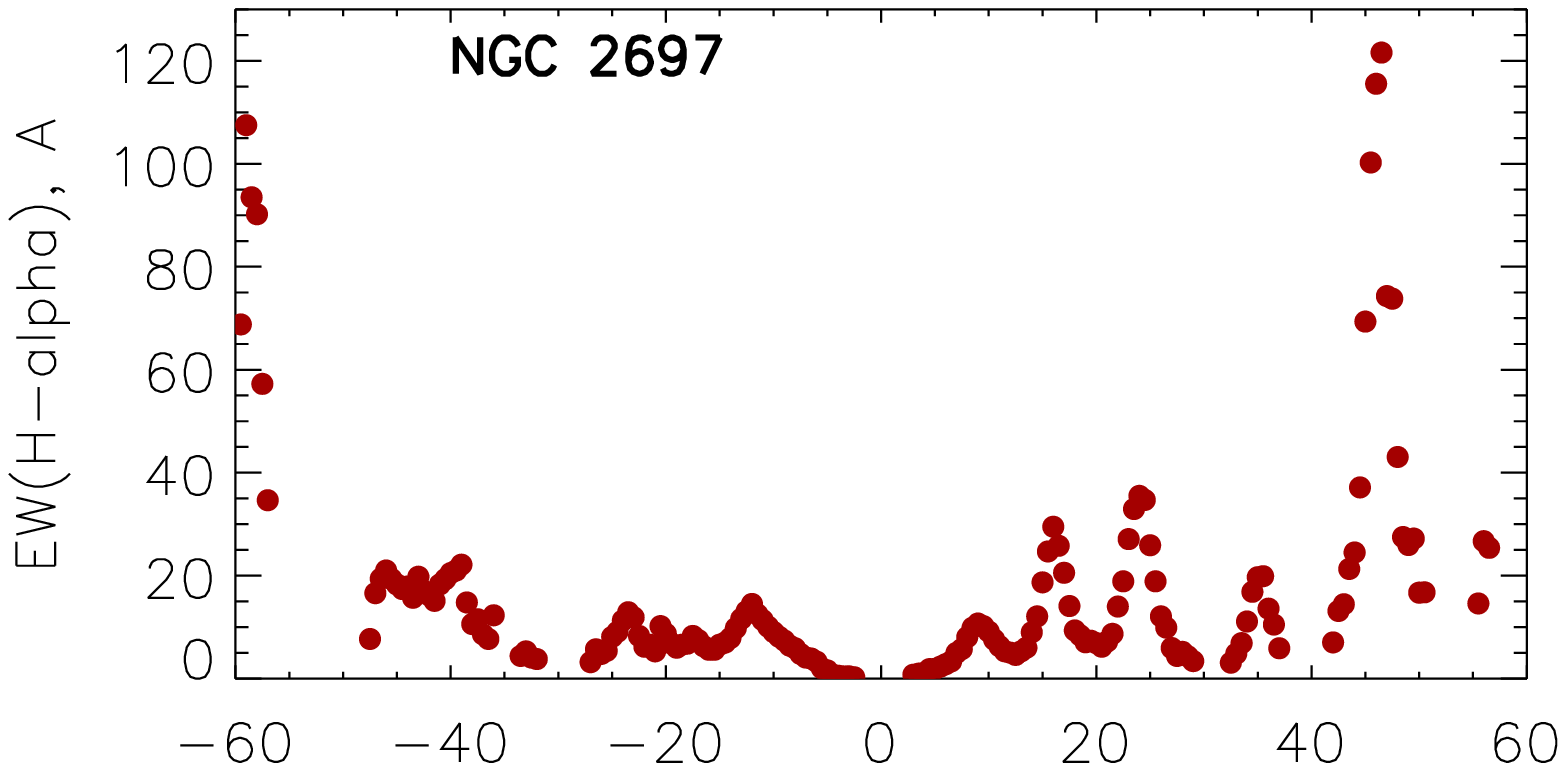} &
 \includegraphics[width=6cm]{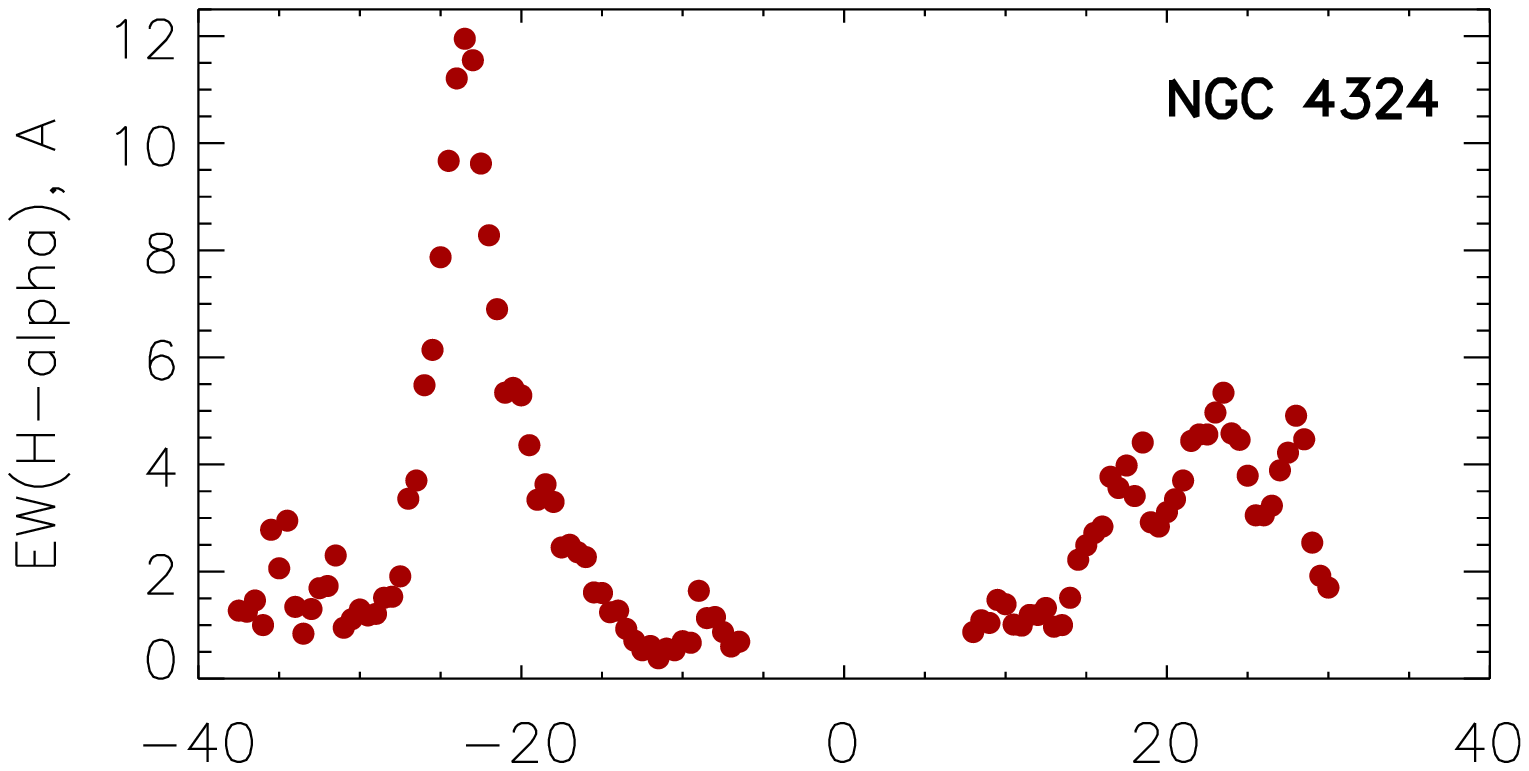} \\
\end{tabular}
%\vspace{1.5cm}
\begin{tabular}{c c}
 \includegraphics[width=6cm]{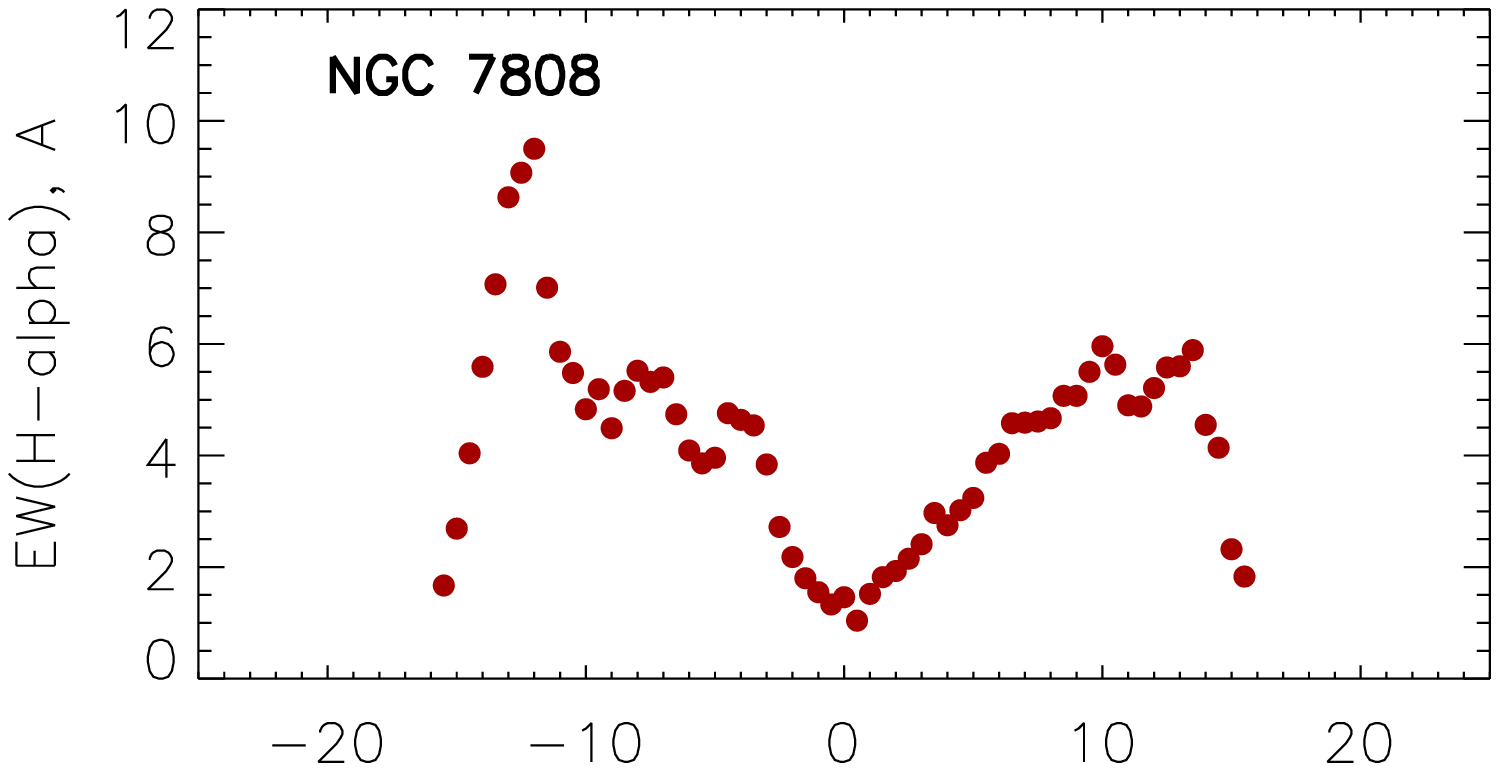} &
 \includegraphics[width=6cm]{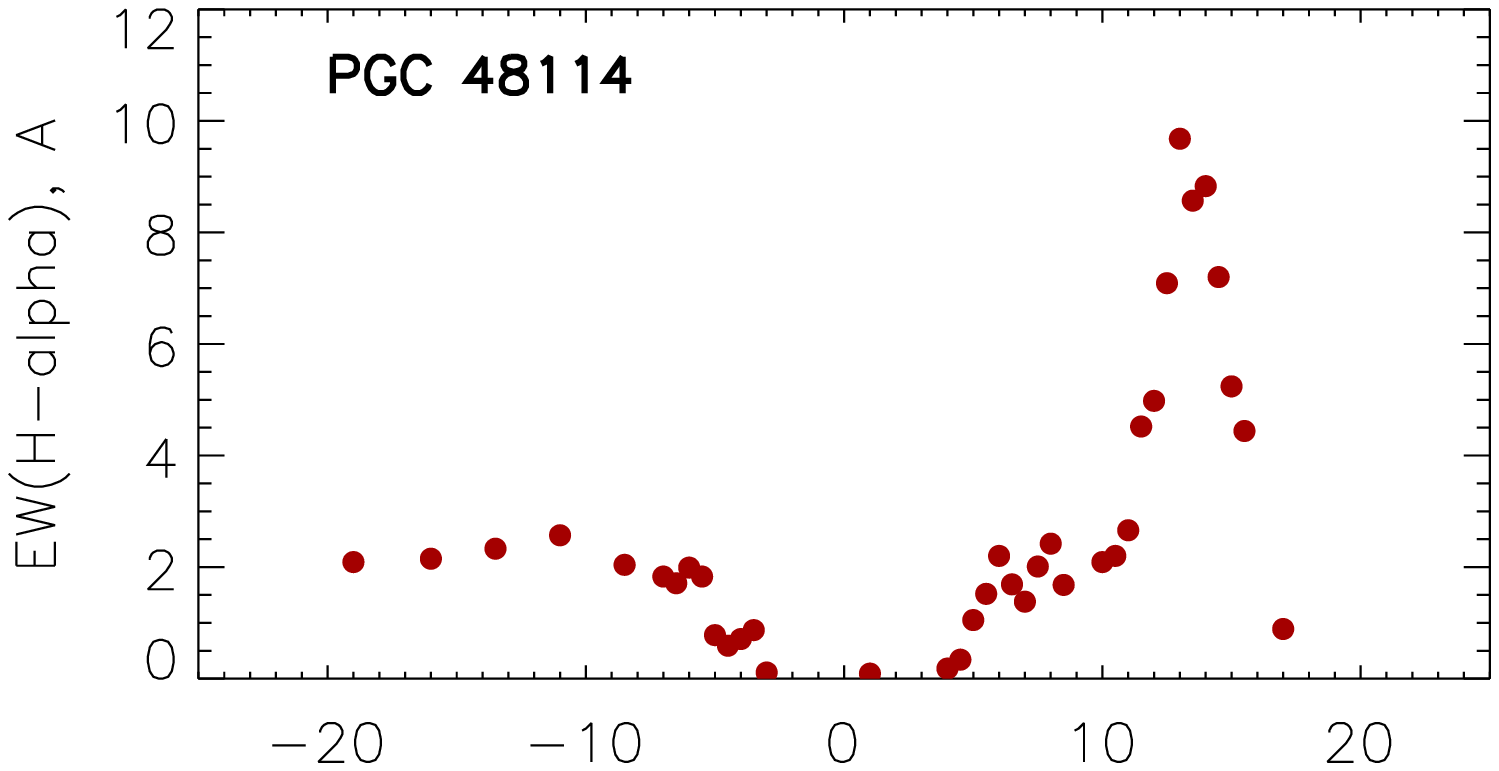} \\
\end{tabular}
\caption{Emission-line equivalent width radial cut-off: Starforming rings seen in the H$\alpha$ emission line.
}
\label{rings}
\end{figure*}

\begin{figure}[tbp]
\centering
\includegraphics[width=6cm]{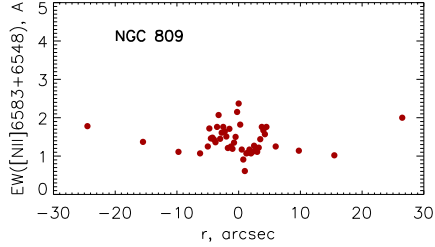}
\caption{[NII] emission-line intensity profile in NGC~809.}
\label{n809n2}
\end{figure}

The existence of prominent emission lines H$\alpha$ and [NII] in the long-slit spectra of the galaxies under
consideration, and especially, their good visibility well off the center, are common for all of them. 
Figs.~\ref{rings} and \ref{n809n2} show the radial profiles of the equivalent widths of the emission lines: H$\alpha$ 
for four galaxies and [NII]$\lambda$6583 for NGC~809 where it is the strongest one. 
One can see that the equivalent widths of the emission lines vary along the radius of the 
galaxies and show one or more prominent peaks. Presumably this indicates rings or spirals containing
emission-line regions in the gaseous disks of the galaxies. In the spectra of the nearby 
galaxies NGC~2697 and NGC~4324 we have also detected strong emission lines due to [SII]$\lambda$6716.4, 6730.6.
For the more distant galaxies NGC~809, NGC~7808, and PGC~48114 these spectral lines are unfortunately beyond the spectral
range of the spectrograph configuration used by us. In the long-slit spectra of NGC~4324, NGC~2697, and NGC~7808
we have also detected strong emission lines H$\beta$ and [OIII]$\lambda$5007 in the green range seen at the same distances 
from the center of the galaxies as the emission lines H$\alpha$ and [NII] that makes possible a confrontation between the strong
emission-line ratios used for the gas excitation mechanism diagnostics. We have analyzed the relations of the intensities
of these emission lines to reveal the excitation mechanism by exploiting BPT-diagrams \citep{BPTdiag}.
 
\begin{figure*}[!t]
\centering
\begin{tabular}{c c c}
 \includegraphics[width=5cm]{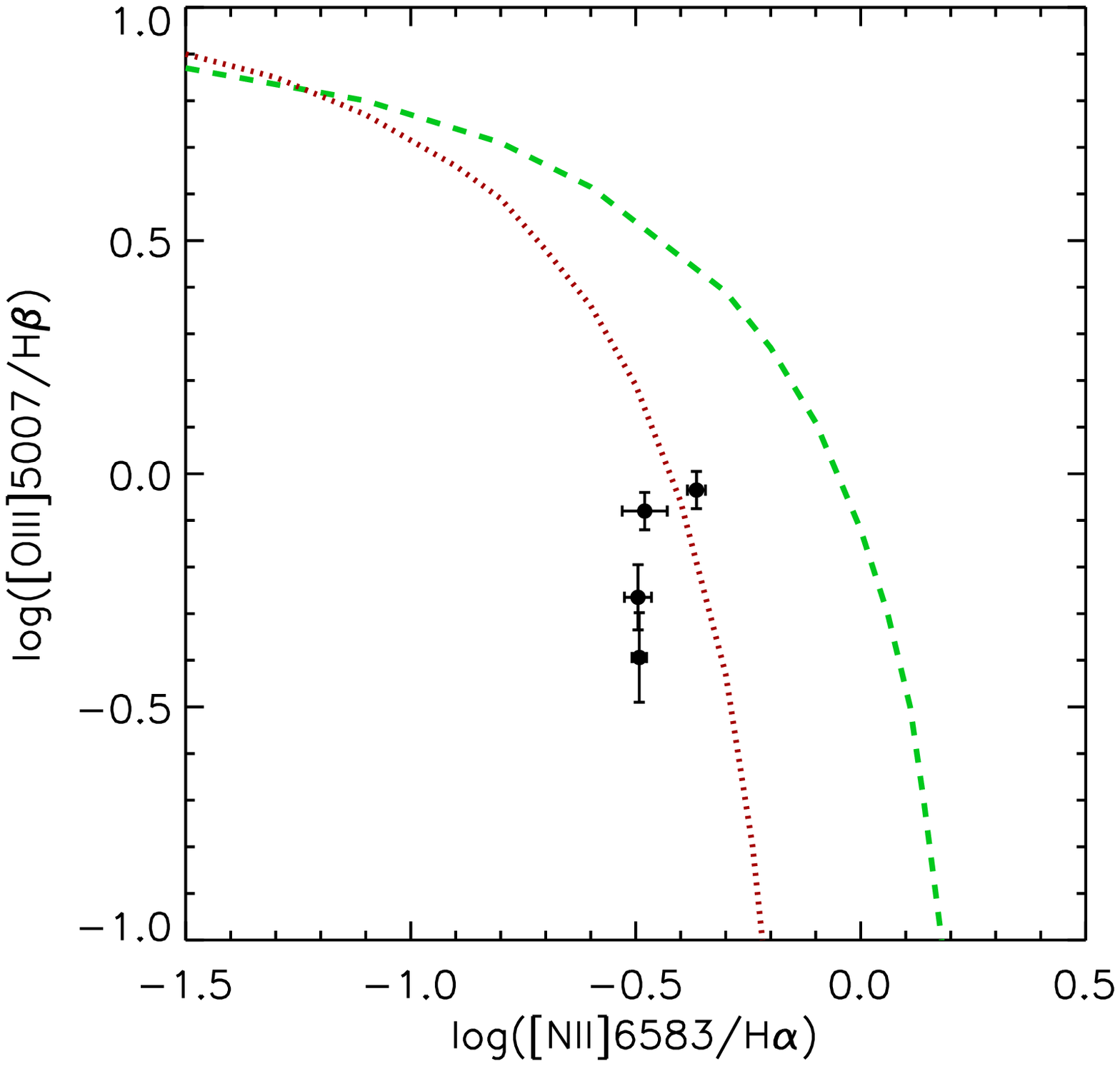} &
 \includegraphics[width=5cm]{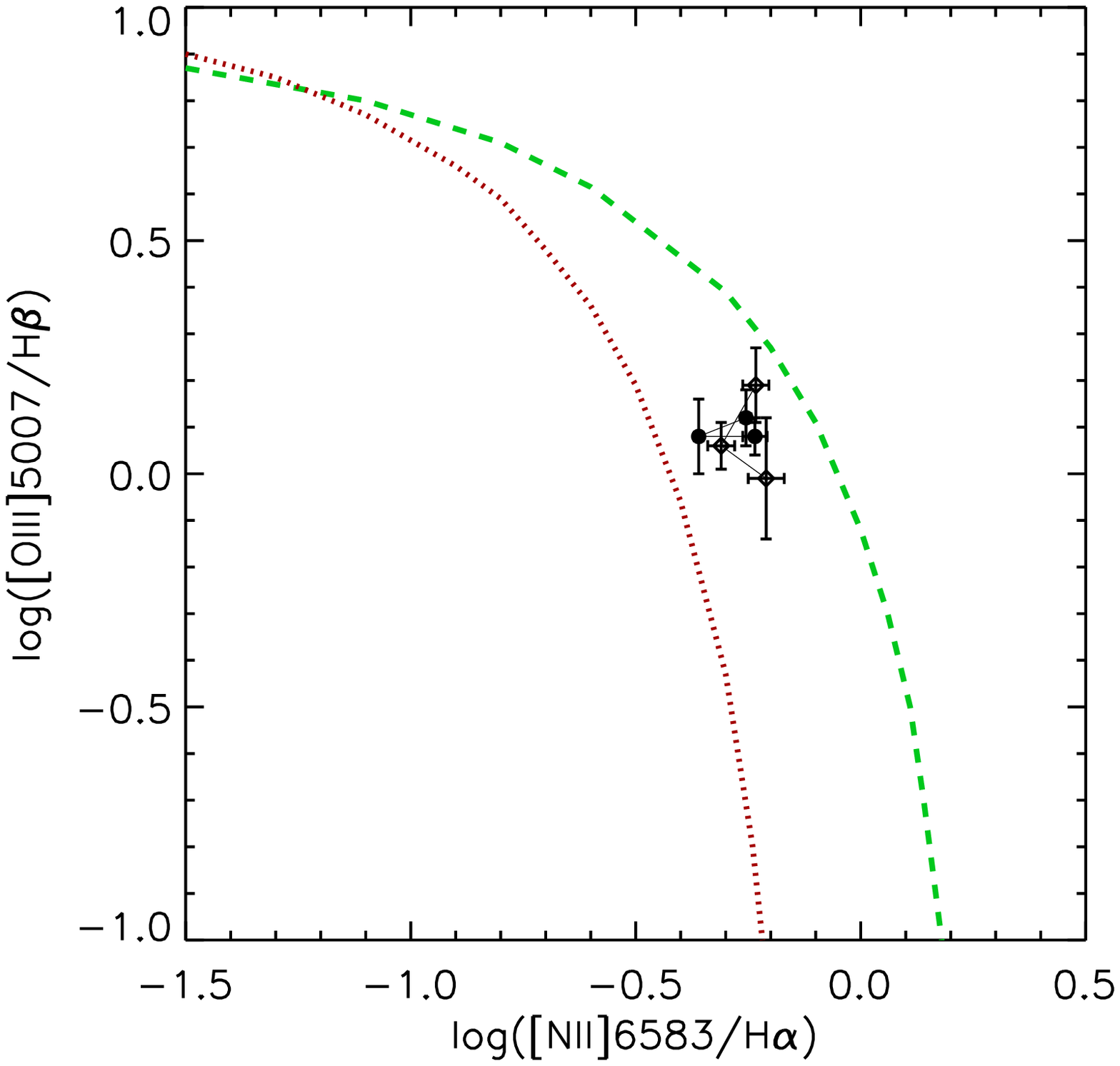} &
 \includegraphics[width=5cm]{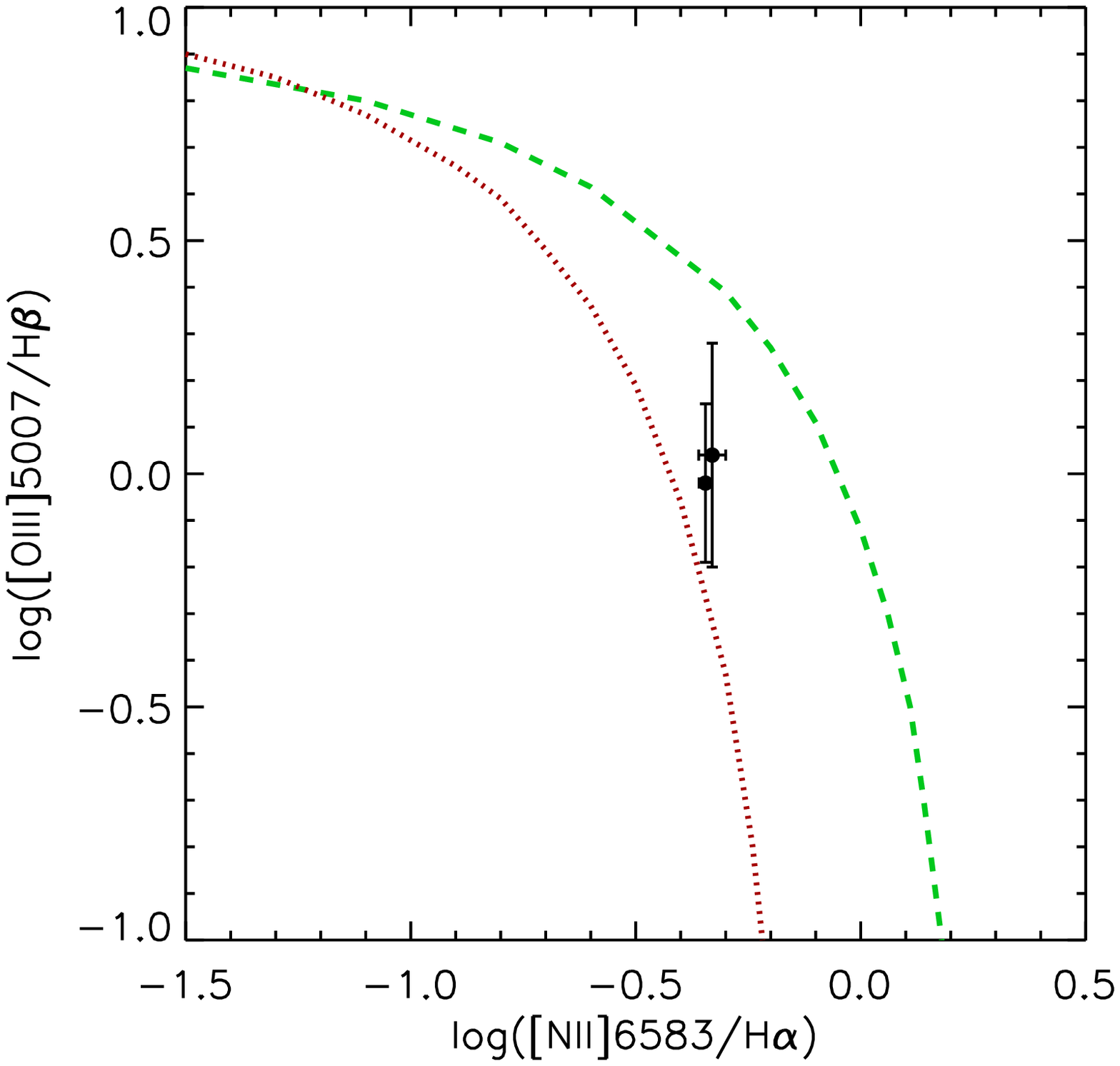}\\
\end{tabular}
\caption{BPT-diagrams for the emission-line rings in NGC~2697 ({\it left}), NGC~4324 ({\it middle}), and NGC~7808 ({\it right}).
The dividing curved lines are: the green dashed one marks the theoretical border of star formation excitation, from \citet{kewley01}, 
the red dotted one is the empirical boundary between starforming galactic nuclei in SDSS and other types of nuclear emission
excitation from \citet{kauffman_2003}. For NGC~4324 we show the radial cuts through the ring (points connected
by solid lines): the middle of the ring demonstrates the excitation the most close to the HII-region one.
}
\label{bpt}
\end{figure*}

To determine the nature of the gas excitation -- by young stars or by other mechanisms -- in the ring-like area
of strong emission lines, we have used BPT-diagnostic diagrams with the models by \citet{kewley01} and the empirical
boundaries found by \citet{kauffman_2003}. Figure~\ref{bpt} presents
$\log(\mbox{[OIII]/H}\beta)$ vs $\log(\mbox{[NII]6583/H}\alpha)$ for the galaxies with the full collection
of the measured emission lines -- NGC~2697 ($r=16\arcsec-46\arcsec$), NGC~4324 ($r=20\arcsec-26\arcsec$), and NGC~7808 ($r=10\arcsec-14\arcsec$).
The point clouds are within the area characterized by the gas excitation by young stars (NGC~2697) and within
the transition zone characterized by the mixing excitation by young stars {\it and} shock waves (NGC~4324 and NGC~7808).
In NGC~4324 we see systematic trends of the line intensity ratios between the inner and outer edges of the
ring, on one hand, and its middle line on the other hand: both for the northern and southern cross-points we have connected 
three measurements, those at $r=20\arcsec$, $r=22\farcs 5$, and $r=25\arcsec$, and plotted them at the BPT-diagram. 
In Fig.~\ref{bpt} one can see that the middle measurements both at the northern and southern tip of the ring 
are the closest to the line dividing HII-region excitation and the transition zone. So we may suspect that the ring
in NGC~4324 is dynamically compressed by the ambient medium producing shock fronts at its edges.

\begin{figure*}[!h]
\centering
\begin{tabular}{c c }
 \includegraphics[width=7cm]{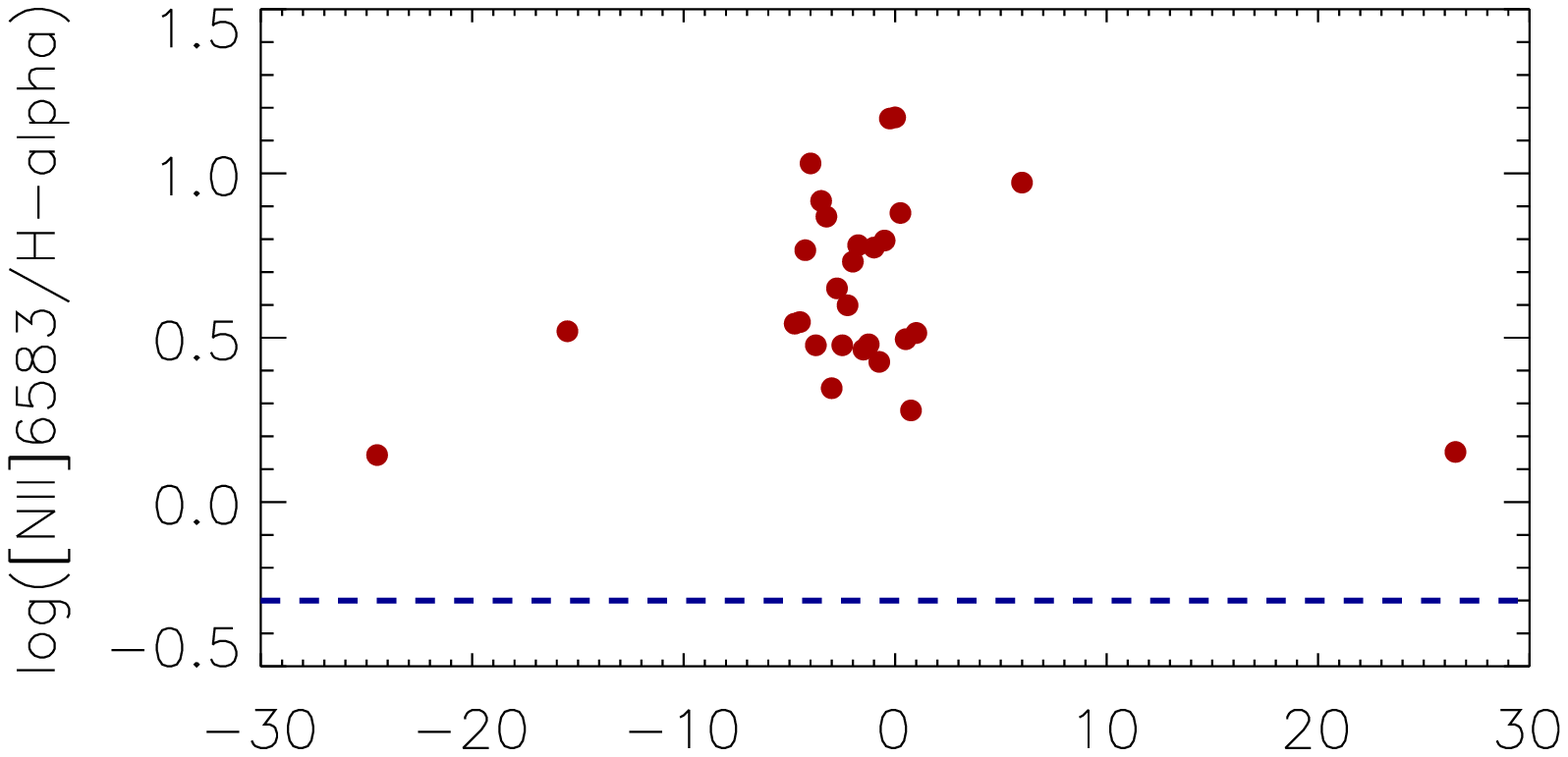} &
 \includegraphics[width=7cm]{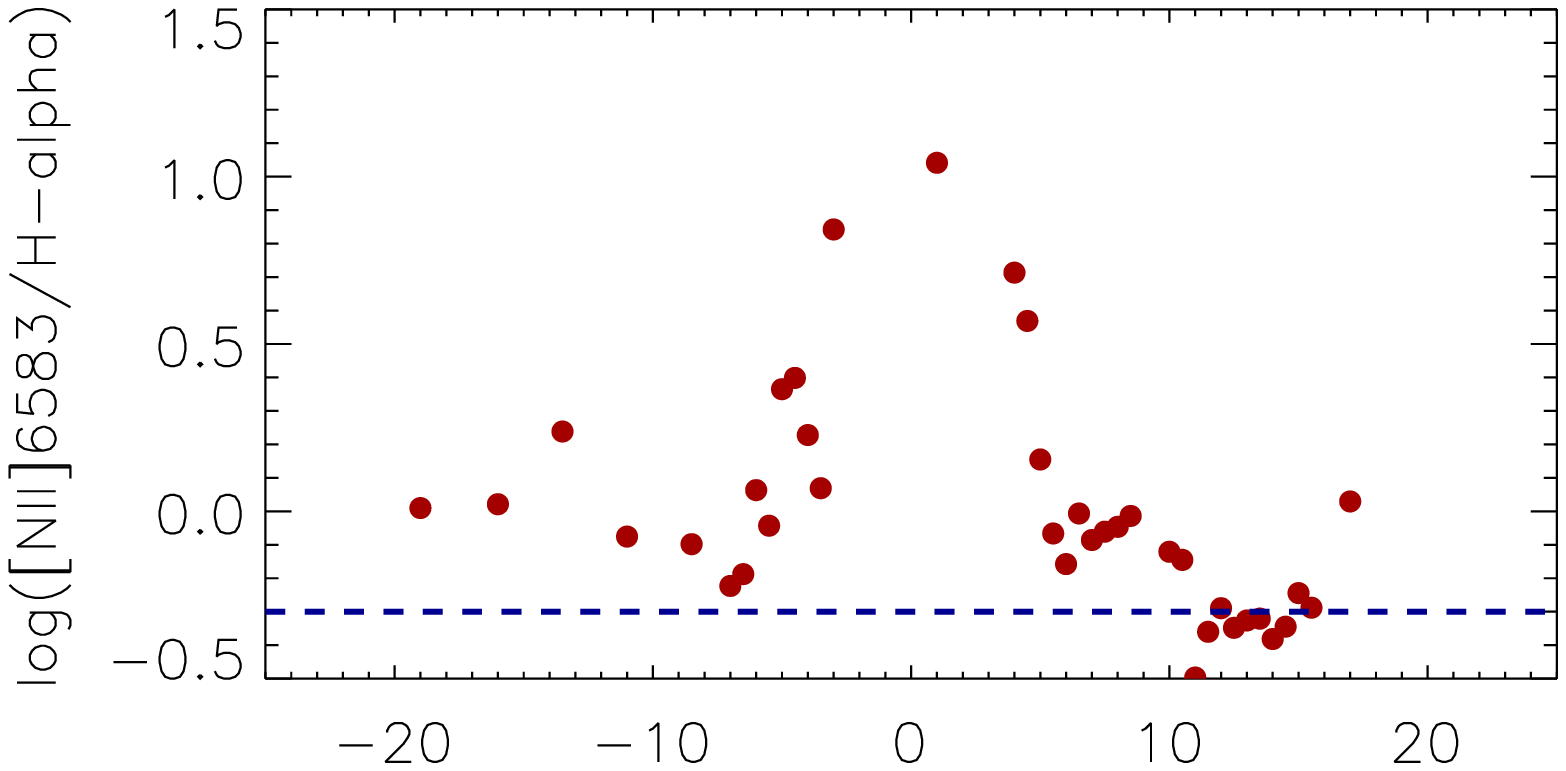} \\
\end{tabular}
\caption{Profiles of emission-line intensity ratios for NGC~809  ({\it left}) and
and PGC~48114 ({\it right}). The horizontal dashed line divides the area occupied by the spectral
characteristics of the ionized gas excited by young stars (below the line) and by other mechanisms of
the gas excitation (above the line).
}
\label{proflog}
\end{figure*}

By adopting the restrictions onto the line intensity ratios from \citet{kewley06}, we suggest that
the N2-ratio $\log(\mbox{[NII]6583/H}\alpha) \le -0.3$ signifies the gas excitation by young stars in any case.
Fig.~\ref{proflog} traces the ratios, $\log(\mbox{[NII]6583/H}\alpha)$, along the radius of the galaxies.
In PGC~48114, only the western half of the disk demonstrates
the coincidence of the sharp peak of the emission-line intensity and of the $\log(\mbox{[NII]6583/H}\alpha)$ ratio typical
for star formation, at $r\approx +13\arcsec$. In NGC~809 the emission-line ratio $\mbox{[NII]6583/H}\alpha$ is everywhere
larger than 1, so in this galaxy, despite the gas presence and the bright UV-ring (Fig.~\ref{sdssview}),
we do not see any signs of current, $T<30$~Myr, star formation in the ring.

The strong-line intensity ratios are used by us to estimate the physical and chemical characteristics of the gas.
The relation of two lines of the sulfur doublet, [SII]6716.4/[SII]6730.6, allows us to estimate electron density 
in the ionized gas by using the calibration of \citet{oster06}.  To derive oxygen abundances for the regions where
the gas excitation is dominated by the star formation process, we have used quite recent strong-line calibrations
by \citet{marino}; two different indicators, O3N2 and N2, are exploited. The results are presented in Table~\ref{tbl:Gas}
(the radius ranges correspond to the intensity peaks of the emission lines in the disks).
The gas metallicity in the star-forming rings of the galaxies under consideration looks homogeneous and only
slightly subsolar, being  everywhere higher than the stellar metallicity of the underlying galactic disks
which is commonly less than half solar (see the next Section).

\begin{table*}
\centering
\caption{Ionized-gas and star properties of the rings in the studied galaxies}
%\begin{flushleft}
\label{tbl:Gas}
\begin{tabular}{lccccc}
\hline\noalign{\smallskip}
Galaxy & Radius & $<N_e>$,cm$^{-3}$ & Gas [O/H] (O3N2) & Gas [O/H] (N2) & [Z/H](stars) \\
\hline
NGC~2697 & $+16^{\prime \prime}$ & 100--250 & $-0.145\pm 0.04$ & $-0.19\pm 0.05$ & --0.4 \\
         & $+24^{\prime \prime}$ & 100--250 & $-0.15\pm 0.03$ & $-0.18\pm 0.05$ & \\
         & $+46^{\prime \prime}$ & $<60$ & $-0.21\pm 0.03$ & $-0.19\pm 0.05$ & \\
\hline
NGC~4324 &  $-23^{\prime \prime}$ & 100 & $-0.26\pm 0.03$ & $-0.11\pm 0.04$& --0.5 \\
         &  $+94^{\prime \prime}$ & & & $-0.18\pm 0.06$ & \\
\hline
NGC~7808 & $-24^{\prime \prime}$ &  &  & $-0.15\pm 0.03$ &      \\
           & $-12^{\prime \prime}$ & & $-0.24\pm 0.07$ & $-0.11\pm 0.04$ & --0.4 \\
           & $+12^{\prime \prime}$ & & $-0.24\pm 0.02$ &  $-0.12\pm 0.01$ & --0.4 \\
\hline
PGC~48114 & $+13^{\prime \prime}$ & & & $-0.105\pm 0.04$ & --1.0 \\
\hline
\end{tabular} 
%\end{flushleft}
\end{table*}

\section{Stellar populations}

To estimate stellar population properties in different parts of the galaxies we have measured the Lick indices 
(H$\beta$, Mgb, Fe5270, Fe5335) along the slit which is mostly aligned with the major axes.
Since the galaxies reveal emission lines in their spectra at some radii, their stellar Lick indices H$\beta$ are
contaminated by the Balmer emission line of the ionized hydrogen. The equivalent widths of the H$\alpha$ emission line 
derived by us from the red-range spectra were used to calculate the correction for the H$\beta$ index. The H$\beta$ emission-line 
intensities are related to the H$\alpha$ emission-line intensities through Balmer decrement which depends on ionization model 
under the assumption of the hydrogen excitation mechanism. We have used 
BPT diagnostic diagrams (see the previous Section) to determine nature of the excitation mechanism
(by young stars or due to other reasons), to accept some model-based Balmer decrement, and to calculate the H$\beta$ index correction. 
In any case, $\Delta \mbox{H}\beta \le 0.35\,EW(\mbox{H}\alpha \, emission)$; the lowest Balmer decrement, $\mbox{H}\alpha /\mbox{H}\beta = 2.85$,
characterizes the HII-region-like excitation.

After correcting the H$\beta$, we have compared the measured Lick indices to the models of Simple Stellar Populations (SSP)
by \citet{thomod} to determine mean stellar population properties such as SSP-equivalent ages, metallicities, and 
magnesium-to-iron abundance ratios.  The models by \citet{thomod} are calculated for several values of [Mg/Fe]. 
It allows us to determine 
magnesium-to-iron abundance ratios for the stellar populations over the different parts of the galaxies and 
to estimate the duration of the last major star-forming episode. Chemical evolution models suggest the difference 
in the timescales of iron and magnesium production by a single stellar generation. Brief star formation, $\Delta T <1$~Gyr, 
would give significant magnesium overabundance up to [Mg/Fe]$=+0.3 - +0.5$, and only continuous star formation during more 
than 2-3~Gyr provides solar Mg/Fe abundance ratio \citep{chem86}.

\begin{figure*}[p]
\centering
\begin{tabular}{c c c}
 \includegraphics[width=4.2cm]{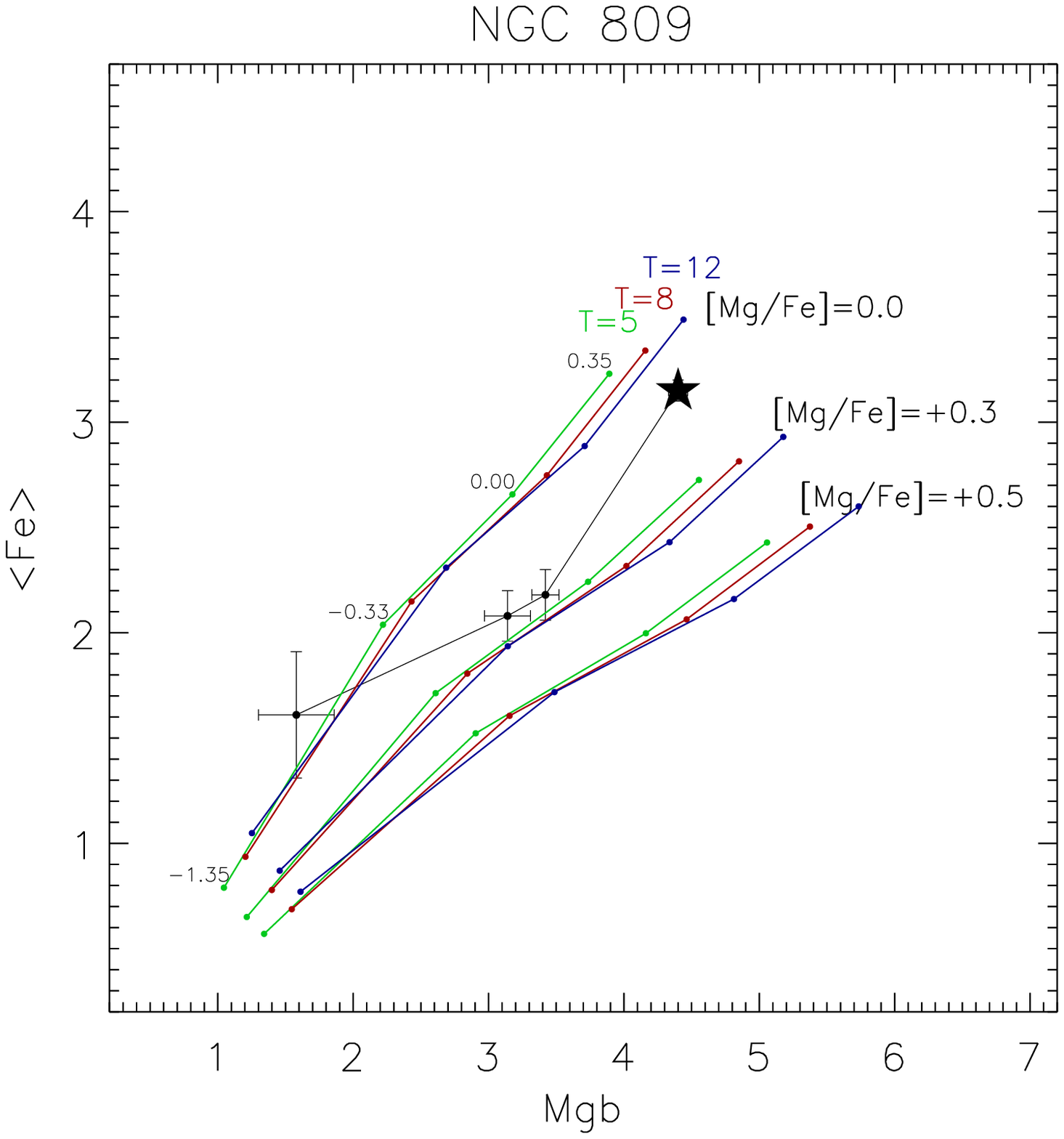} &
 \includegraphics[width=4.2cm]{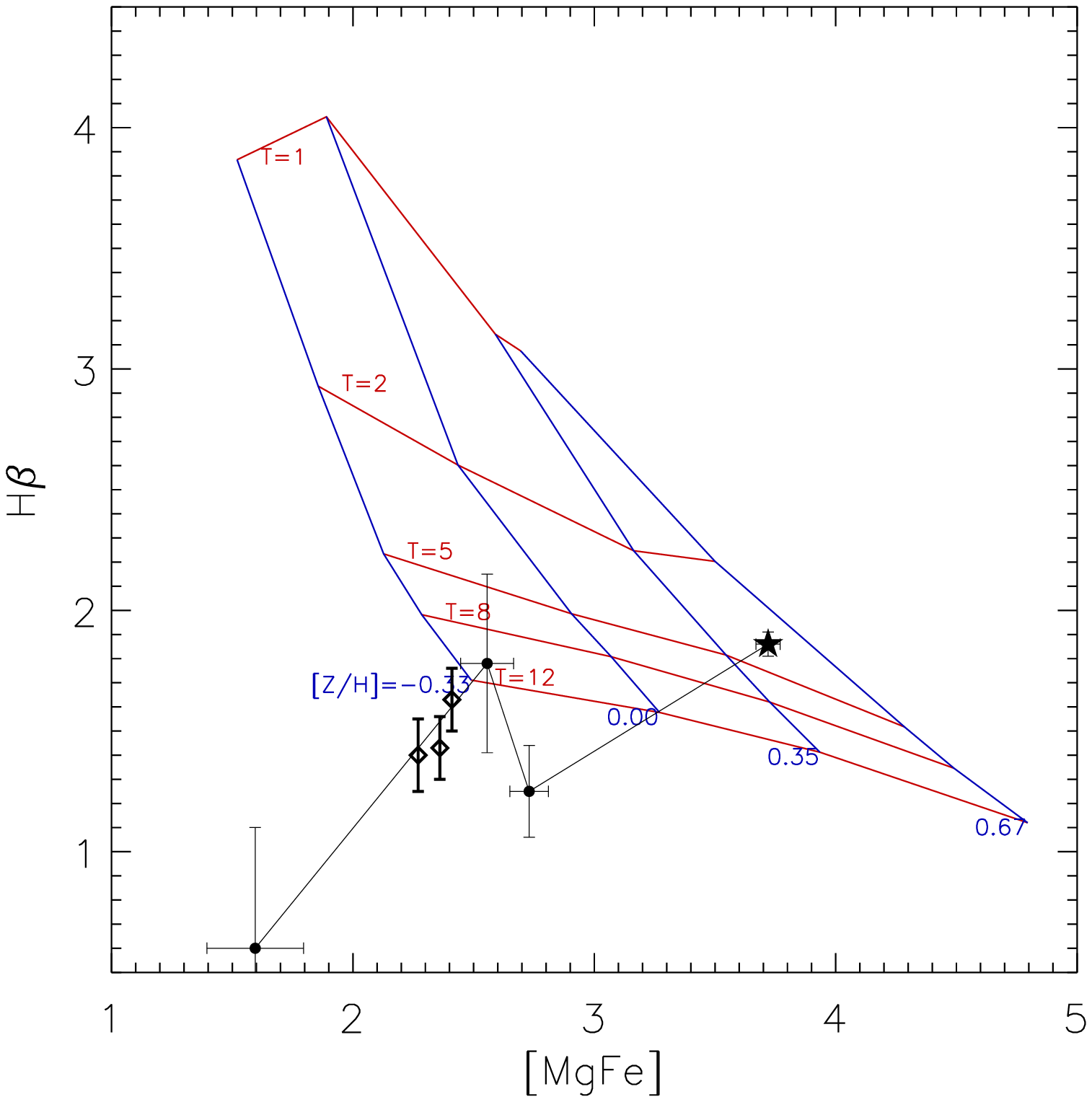} &
 \includegraphics[width=4.2cm]{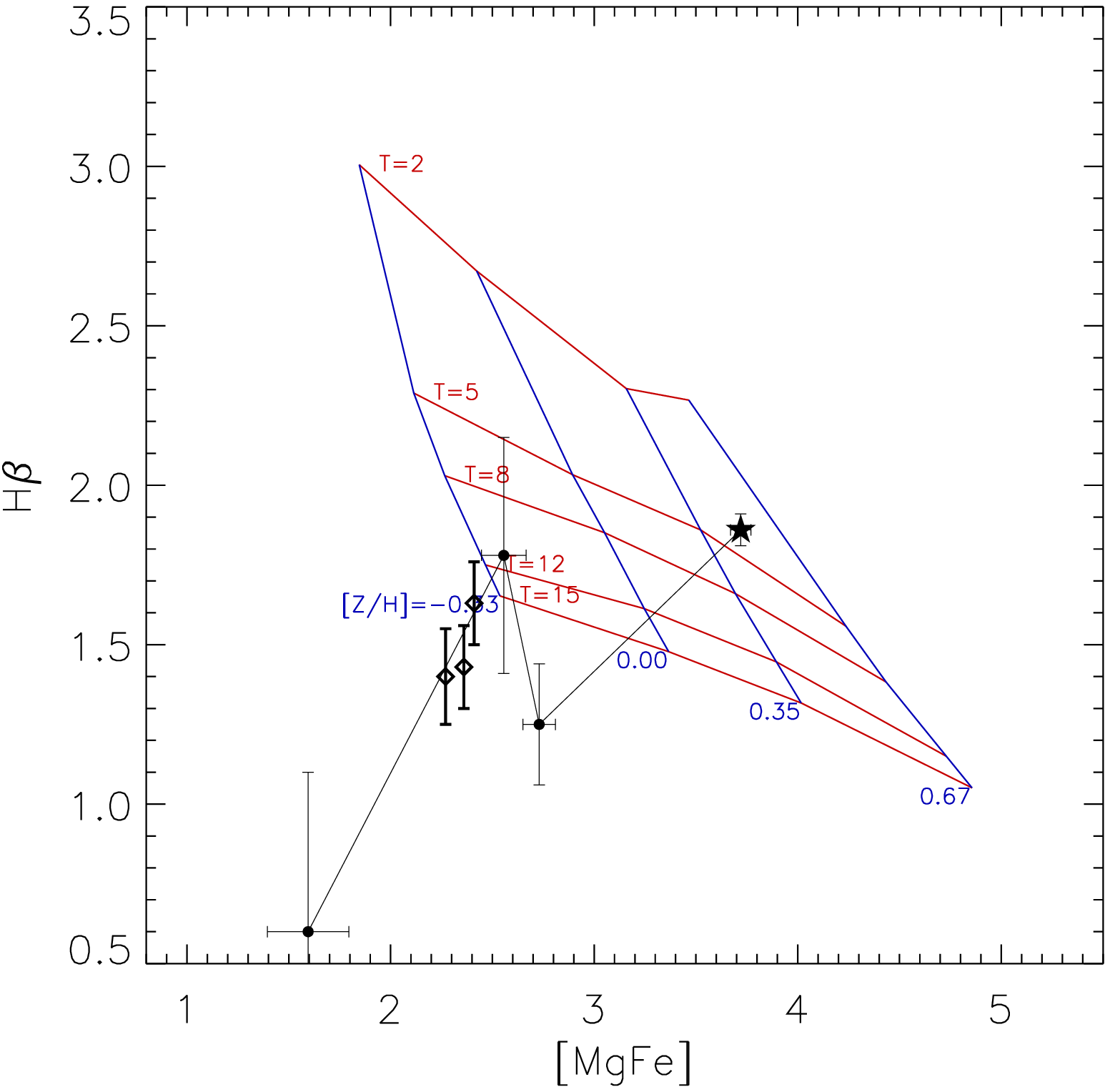} \\
 \includegraphics[width=4.2cm]{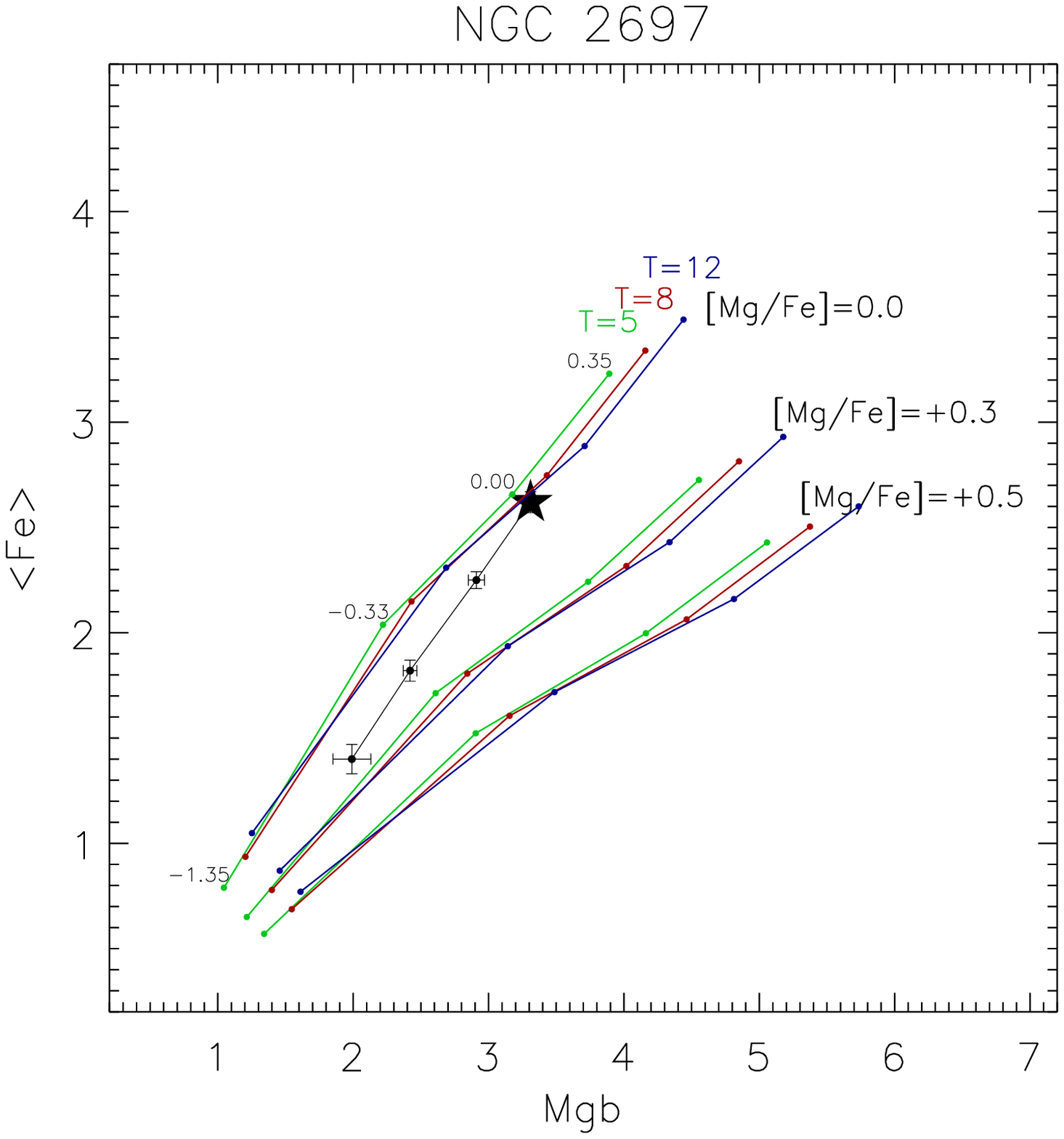} &
 \includegraphics[width=4.2cm]{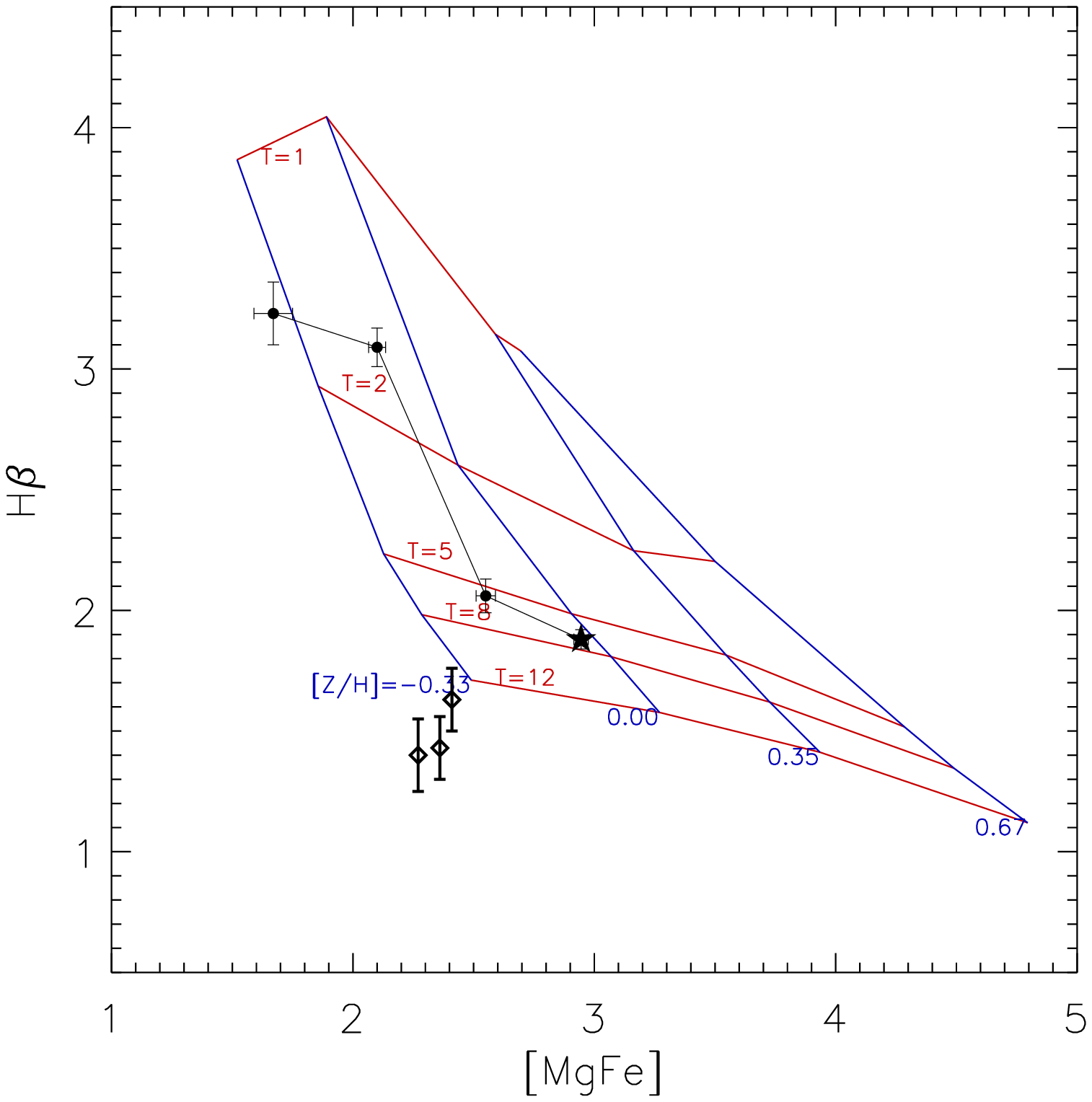} &
 \includegraphics[width=4.2cm]{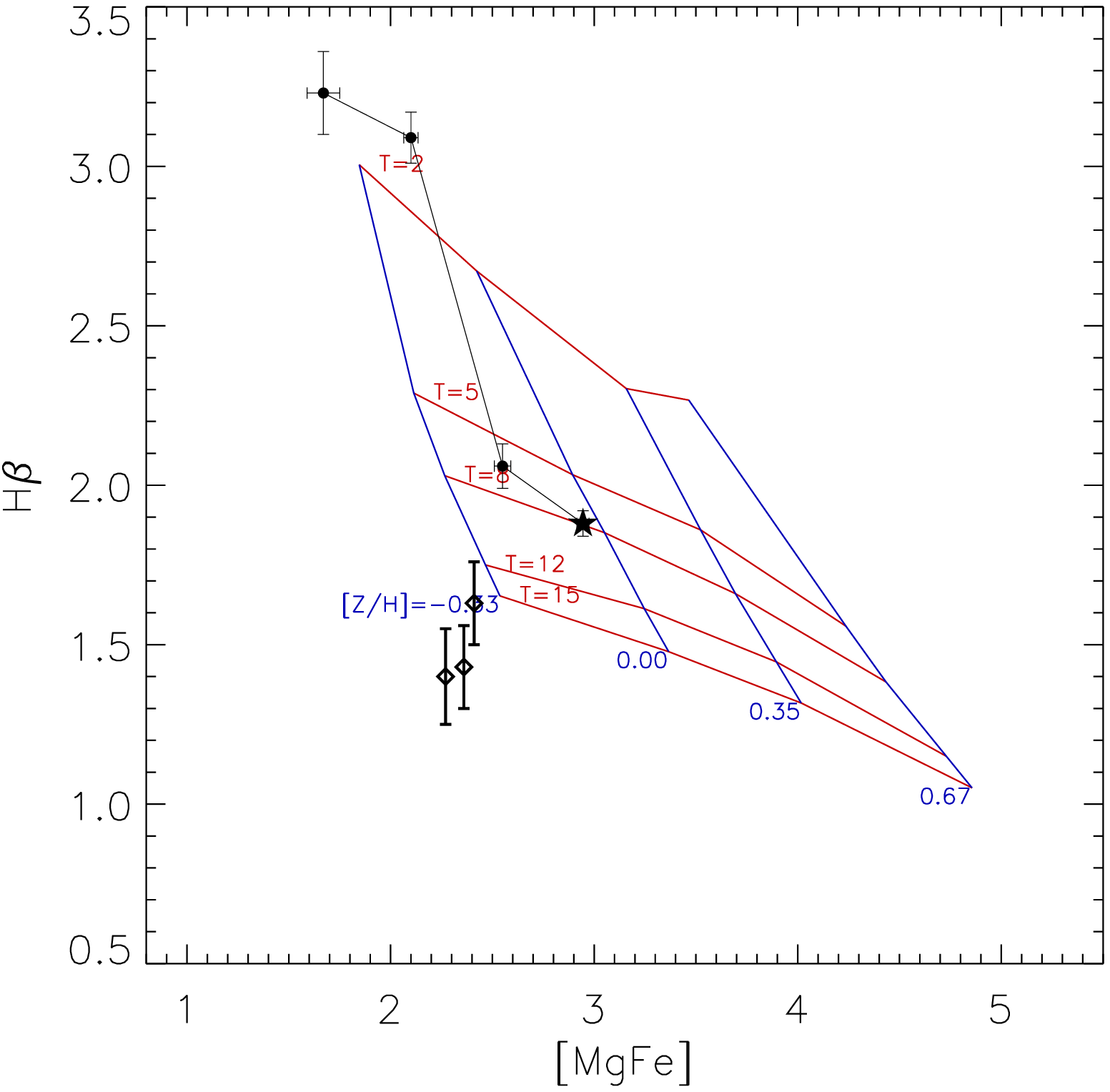} \\
 \includegraphics[width=4.2cm]{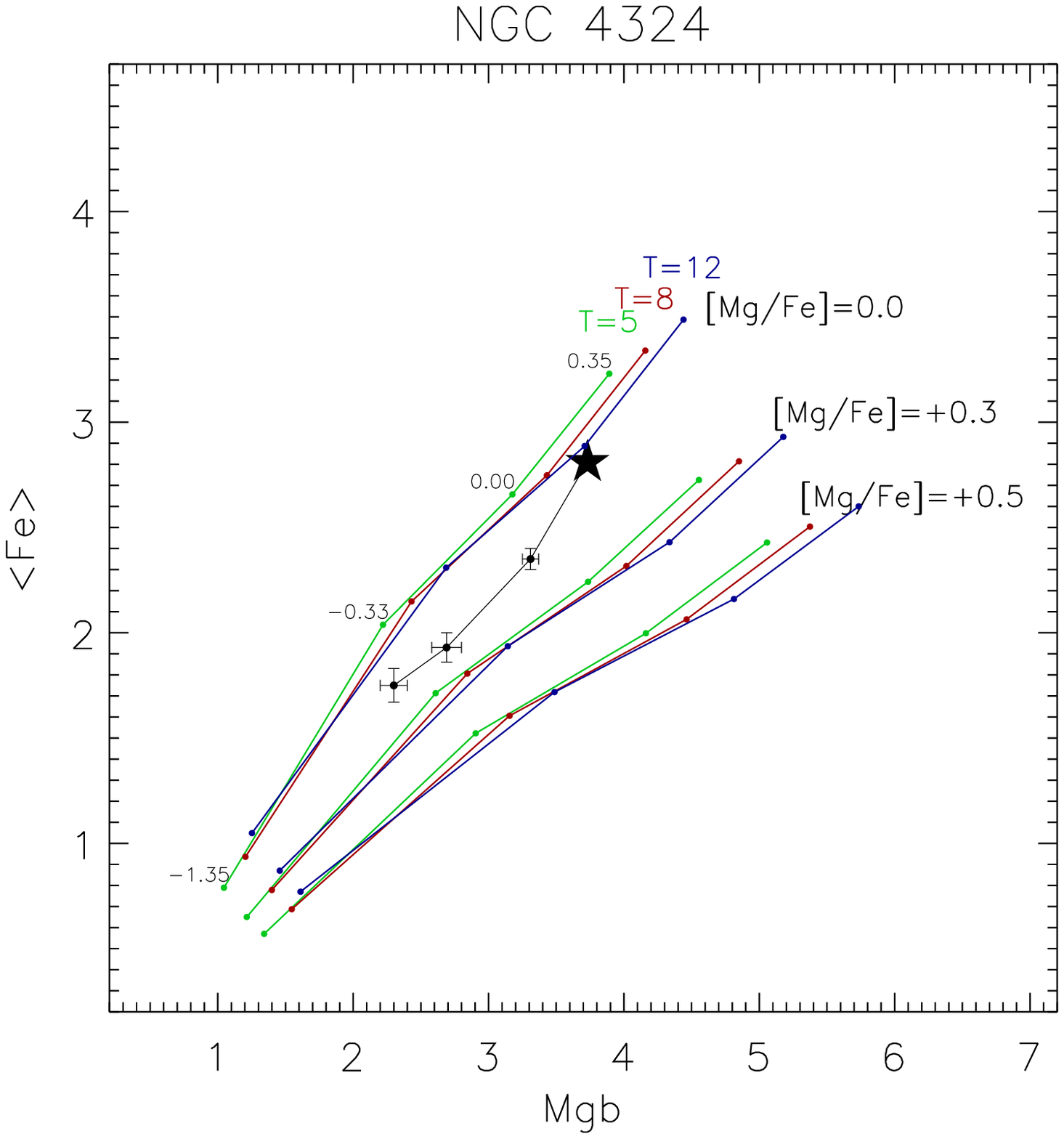} &
 \includegraphics[width=4.2cm]{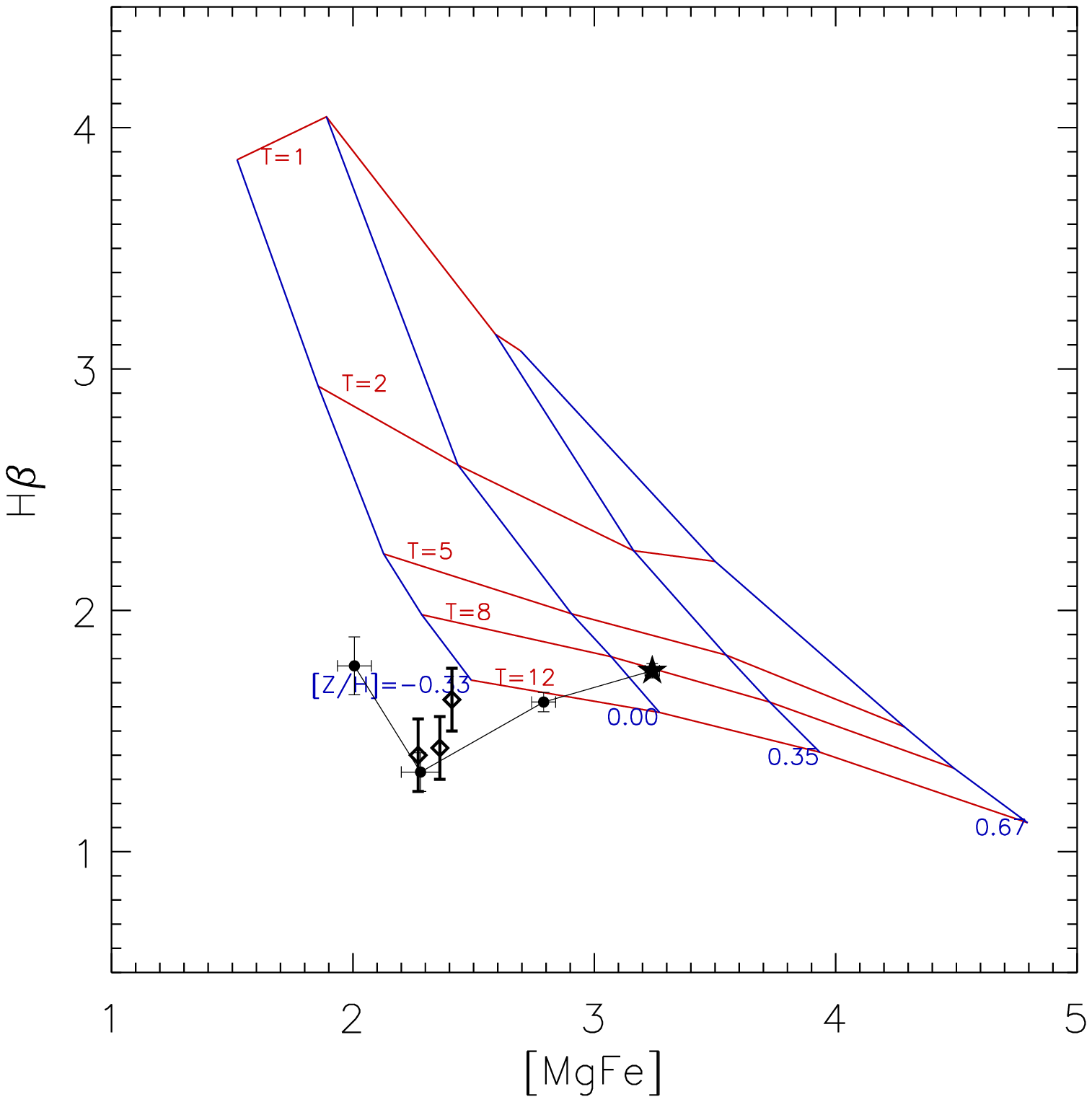} &
 \includegraphics[width=4.2cm]{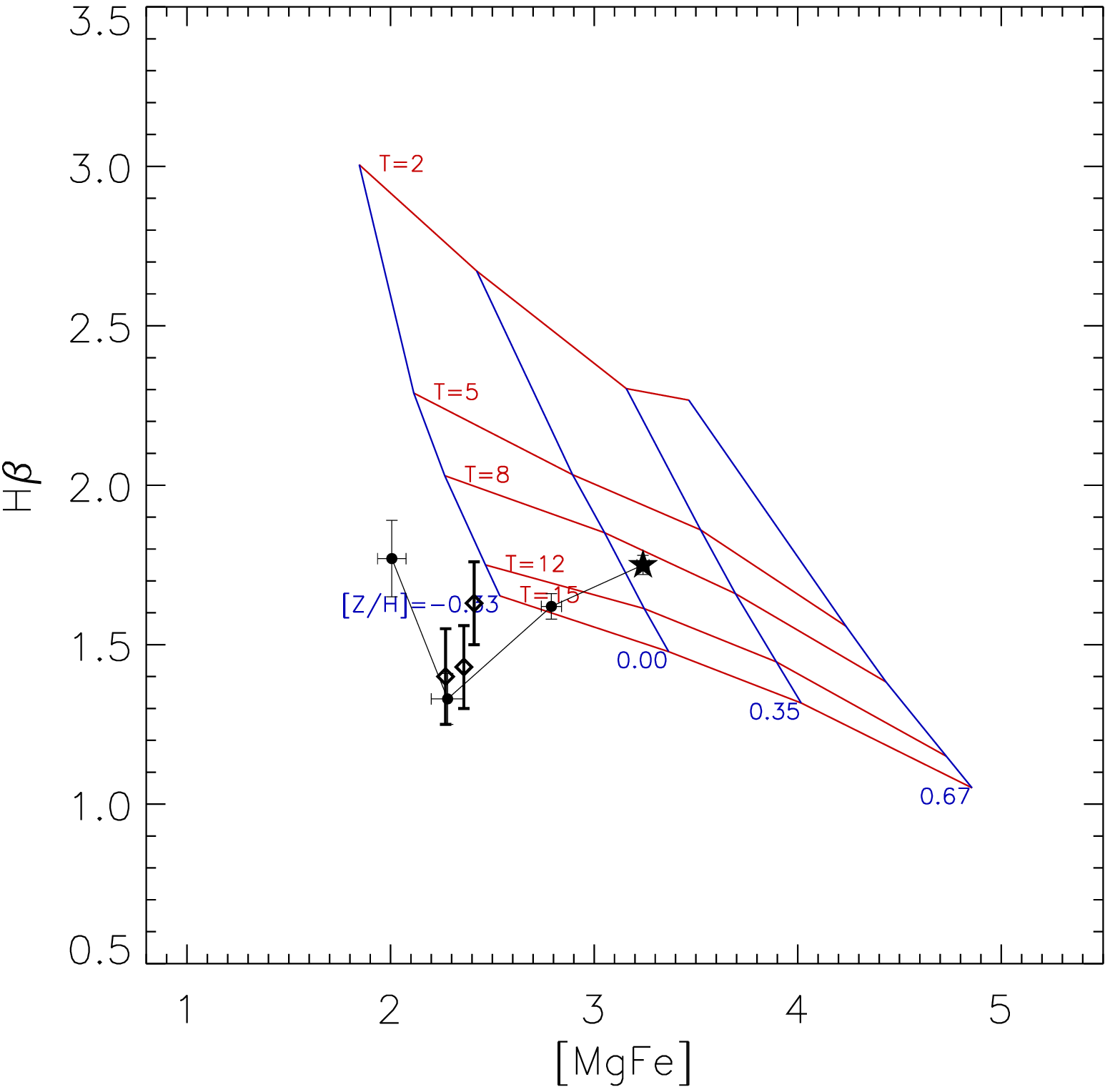} \\
 \includegraphics[width=4.2cm]{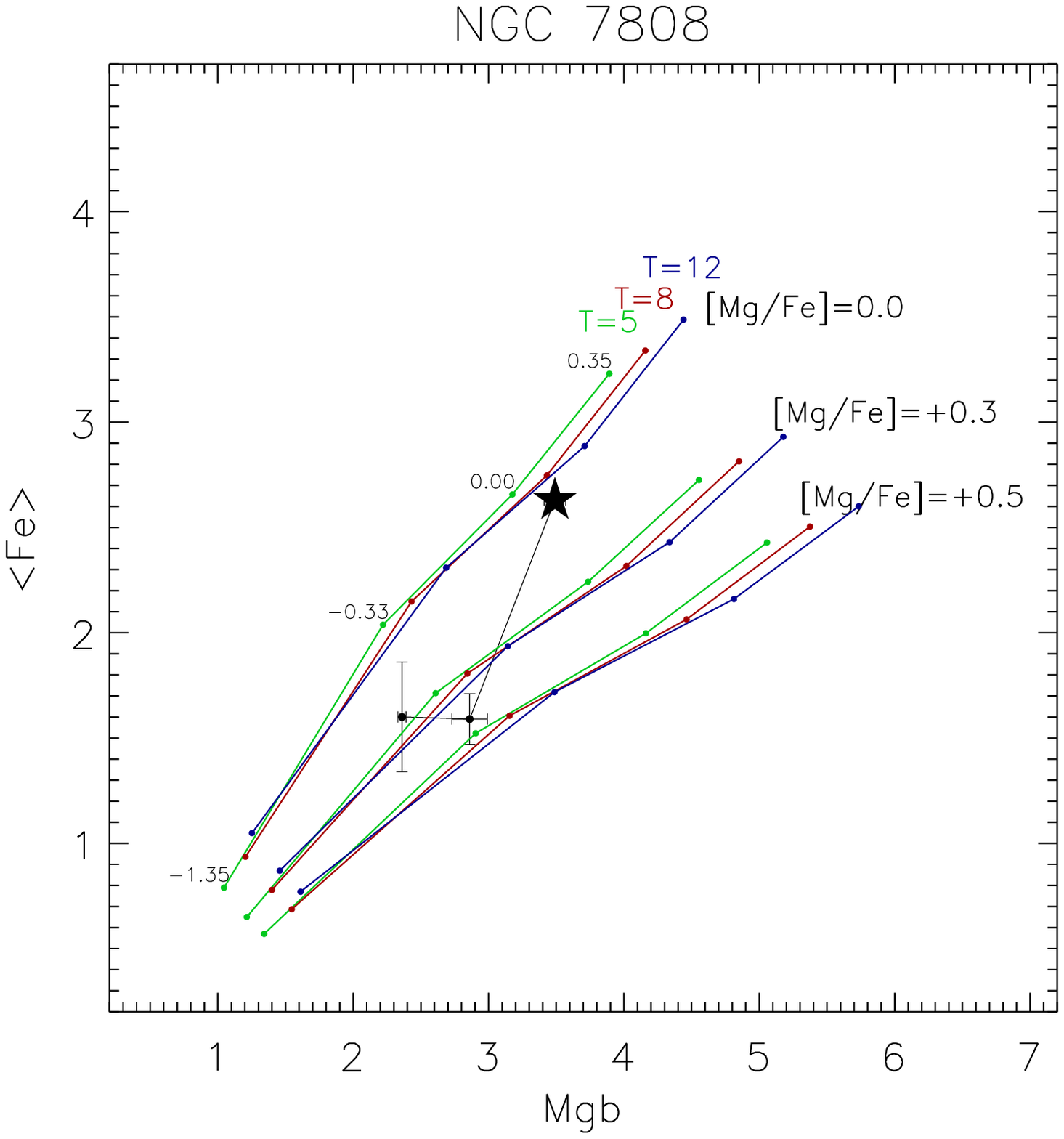} &
 \includegraphics[width=4.2cm]{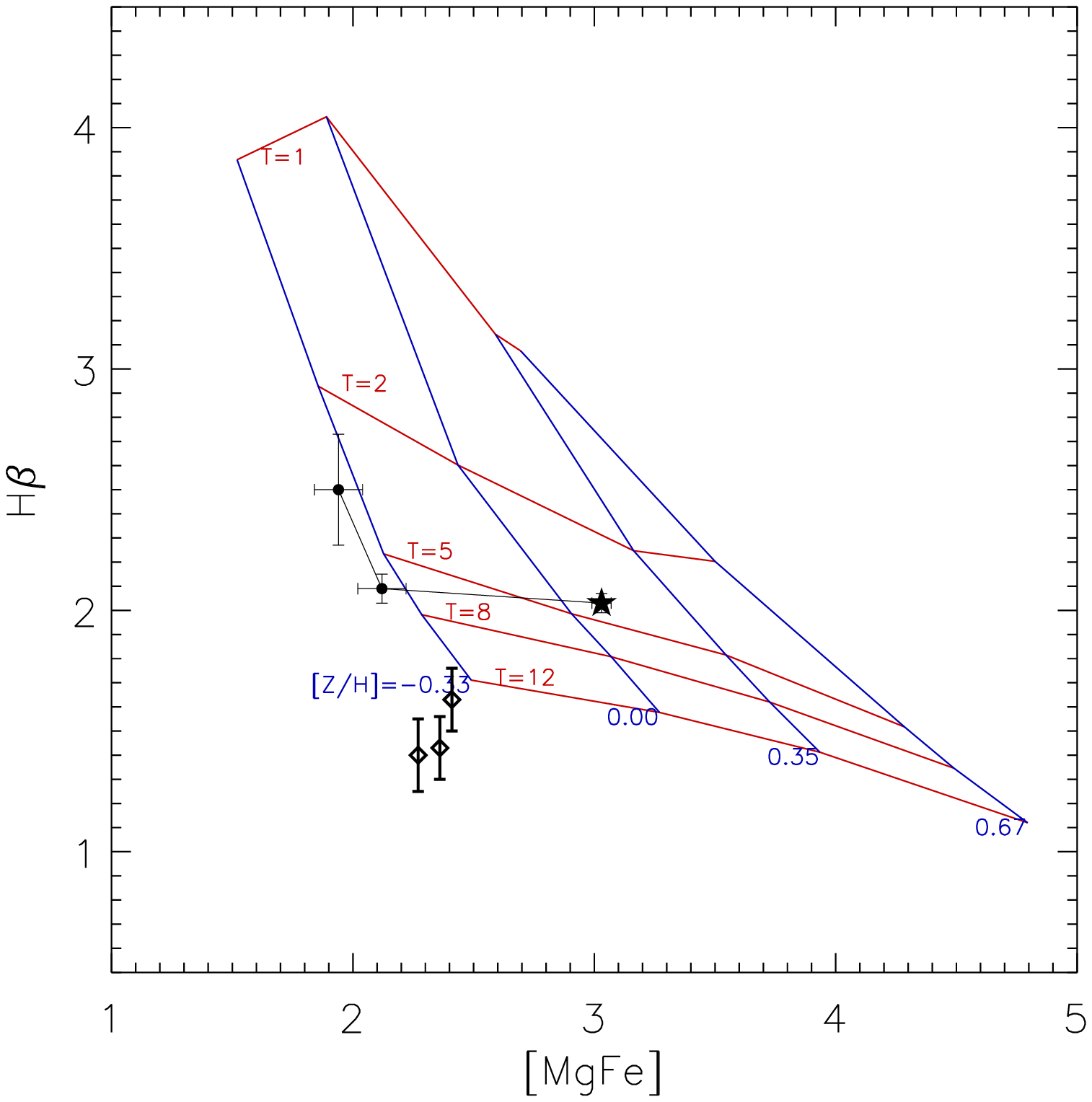} &
 \includegraphics[width=4.2cm]{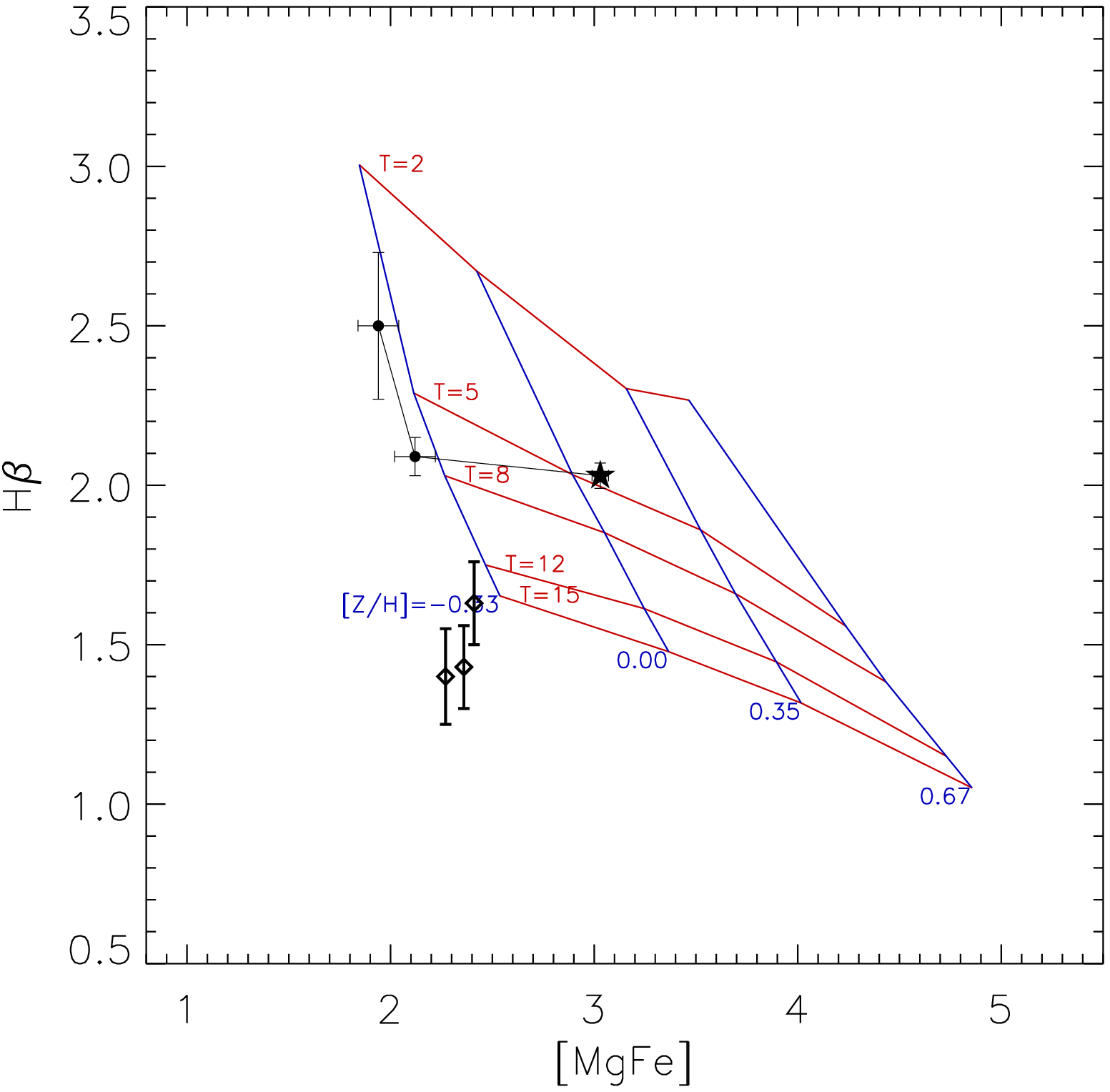} \\
 \includegraphics[width=4.2cm]{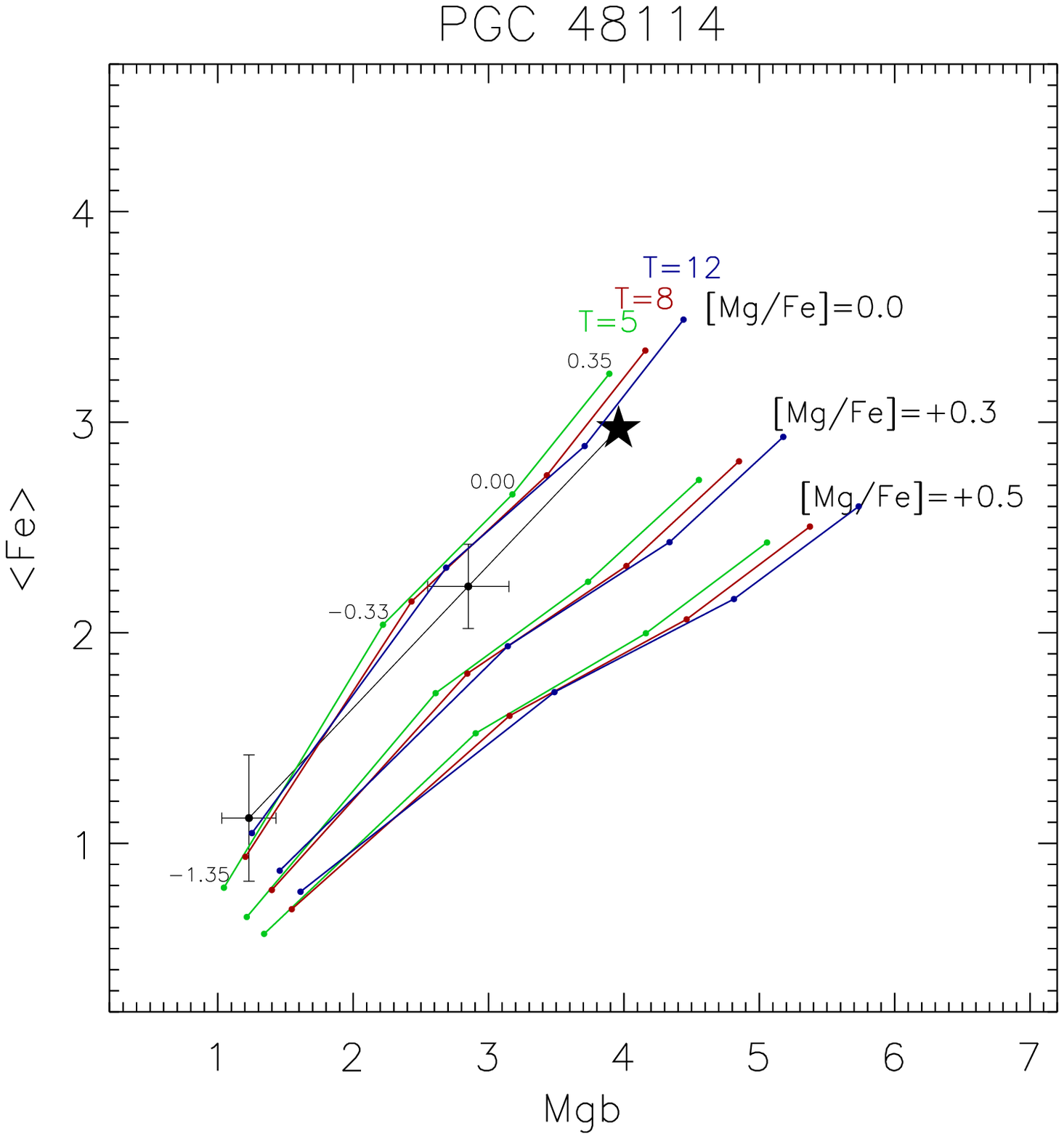} &
 \includegraphics[width=4.2cm]{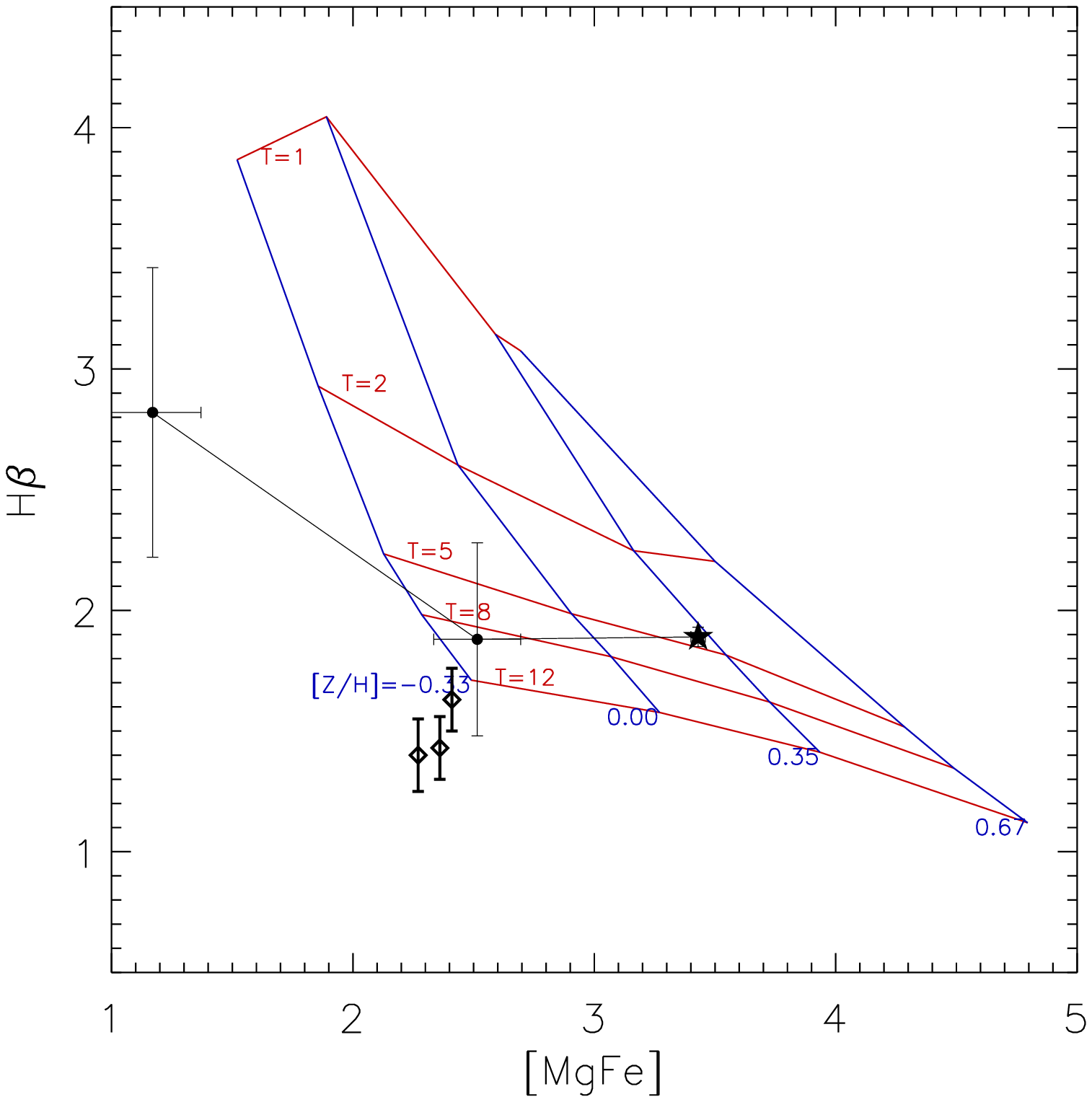} &
  \\
\end{tabular}
\caption{Diagnostic diagrams to determine SSP-equivalent parameters of the stellar populations: the left column
presents the $\langle \mbox{Fe} \rangle =$(Fe5270+Fe5335)/2 versus Mgb diagram, other two columns -- the
H$\beta$ versus [MgFe]$=(\mbox{Mgb}\langle \mbox{Fe} \rangle)^{1/2}$ diagrams, for [Mg/Fe]$=0$ in the center and
for [Mg/Fe]$=+0.3$ to the right. Large black stars mark the nucleus for every galaxy, and then we go along the
radius through the zones described in the Section devoted to the structure of the galaxies. The diamonds present a few
globular clusters from \citet{beasley} belonging to the Galactic bulge, as the reference frame.
}
\label{iidiag}
\end{figure*}

In Fig.~\ref{iidiag} we show the Lick indices profiles along the major axes of the galaxies
plotted onto diagnostic diagrams $\langle \mbox{Fe} \rangle =$(Fe5270+Fe5335)/2 versus Mgb and H$\beta$ versus 
[MgFe]$\equiv \sqrt{\mbox{Mgb}\langle \mbox{Fe} \rangle}$. With these diagrams, we can trace the mean ages,
metallicities, and magnesium-to-iron ratios along the radius in every galaxies studied. In general, we can note 
that in all five galaxies the central stellar populations are of intermediate age, of 5--8~Gyr, have solar or
supersolar metallicity, and no current
or even recent star formation are detected in the nuclei and in the bulges. The star formation
histories in their stellar disks are rather different, but everywhere the mean stellar metallicities in the disks
including the stellar rings are below [Z/H]$\sim -0.3$.

\section{Discussion}

The current paradigm of the evolution of disk galaxies implies persistent accretion of outer cold gas
onto disks that provides, e.g., the flat relation between the ages and metallicities of disk stars in our Galaxy
and also continuous star formation within galactic thin disks of other spiral galaxies during the last 8--10~Gyr
despite the current supply of the gas in the disks being sufficient only for $\sim 2$~Gyr of star formation with the
rate observed. Interestingly, the timescale of gas consumption, 2~Gyr, is nearly constant for the disks of nearby
spirals \citep{bigiel}, and this fact has inspired suggestions that galactic starforming disks are now
in equilibrium \citep{dave}: the rate of outer gas accretion is nearly equal to the star formation 
rate plus gas outflow in so called `galactic fountains' due to star formation feedback. In the frame of this
paradigm {\it all the gas} observed currently in the disks of spiral galaxies must be recently accreted from
outside; no gas can be considered as a primordial aspect. If correct, there must be no general difference between
the well-settled gaseous disks of spiral galaxies and the (often) inclined gaseous disks of lenticulars. We may expect
to find star formation in lenticular galaxies with noticeable gas content. \citet{pe87,pogge_eskridge} searched 
for HII-regions in gas-rich S0s, and have found them only in half of the sample studied. Two subsamples,
with and without star formation, showed similar mean HI content contradicting, at first glance,
the Kennicutt-Schmidt prescription that the star formation rate has to be proportional to the gas mass.
The cause of the star formation suppression in the half of gas-rich S0s remained unclear. However, the galaxies which
we have studied in this work were selected through the UV-ring visibility criterion, and indeed, just
four of five galaxies reveal currently starforming regions detected through the intense H$\alpha$ emission line. 
Here we consider the S0s from the first half, in the terms by Pogge and Eskridge.

\subsection{Star formation rate in the S0 rings}

\begin{figure}[tbp]
\centering
\includegraphics[width=8cm]{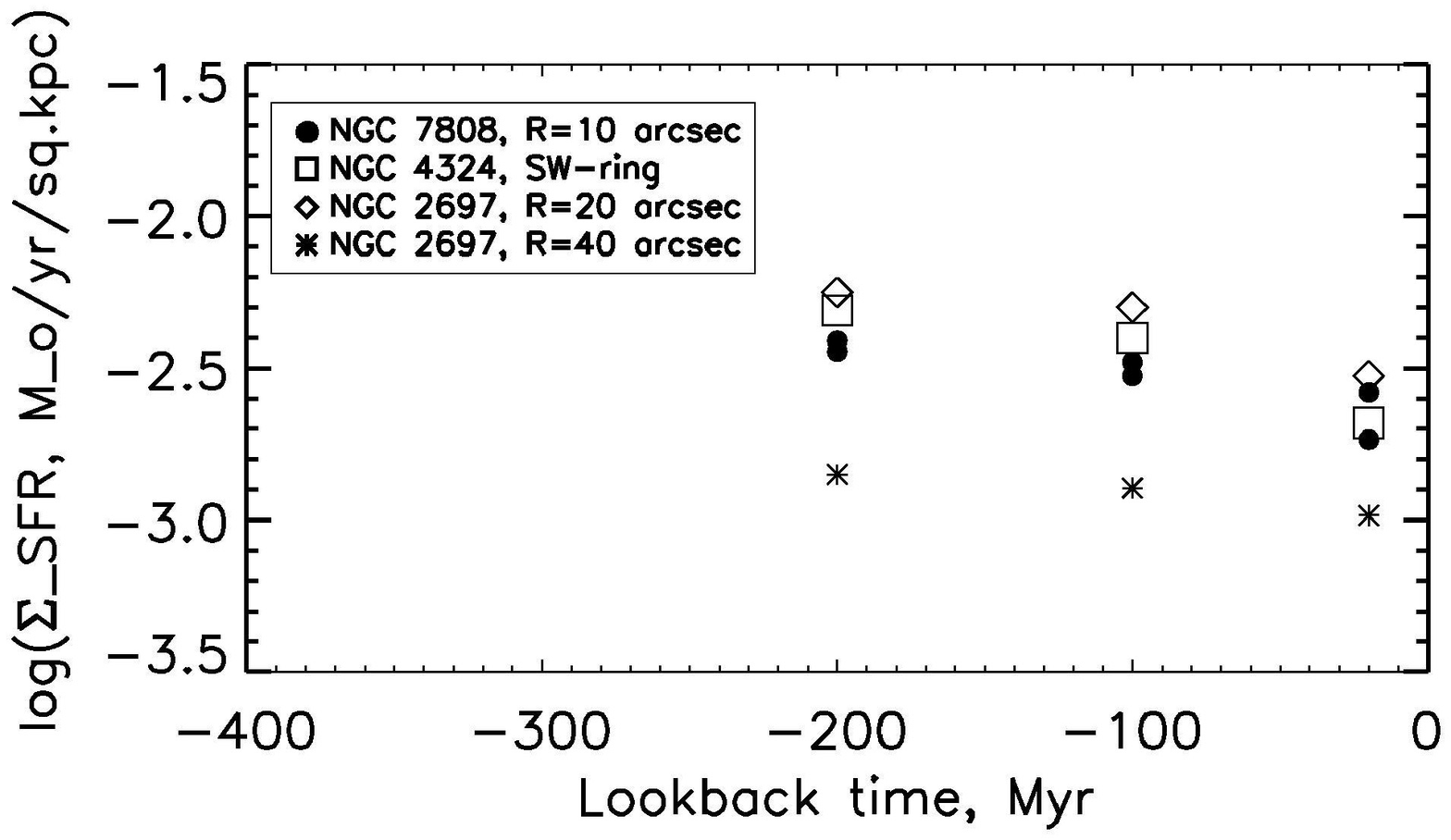}
\caption{Recent star formation histories in the rings of NGC~7808, NGC~2697, and NGC~4324. For NGC~7808
both its spectral cross-sections, in $PA=50\deg$ and in $PA=123\deg$, are plotted separately.}
\label{sfhistory}
\end{figure}

We can determine integrated star formation rates (SFRs) related evidently to the rings in these particular galaxies
by using the FUV-magnitudes measured by the GALEX and listed in the NED 
(NASA/IPAC Extragalactic Database, http://ned.ipac.caltech.edu). We have taken the calibration of
the star formation rate (SFR) through the GALEX FUV-magnitudes from the work by \citet{lee11}. The integrated
FUV-magnitudes listed in the NED were corrected for the foreground extinction but not for the intrinsic
dust of the lenticular galaxies under consideration. For NGC~809 and NGC~7808 we have taken `asymptotic' FUV-magnitudes
from the recent compilation by \citet{bai_galex}; these FUV-magnitudes are more luminous by 0.1--0.3~mag than the
data in the NED. The calculated estimates of the star formation rate (over the timescale of 100~Myr) for 3 galaxies 
are between 0.1~\Ms\ per year (PGC~48114 and NGC~4324) to 0.2~\Ms\ per year (NGC~2697); NGC~7808 appears to be rather
intensely forming stars, with SFR$=1$~\Ms\ per year. We have found a few similar luminous S0 galaxies
with a SFR of about $1$~\Ms\ per year in the SDSS sample by \citet{nair_abraham} -- for example, an unbarred
central group S0 PGC~36834 resembles NGC~7808 very much. In general, such level of star formation puts all four S0s
close to the `main sequence' -- in particular within the scaling relation between the integrated SFR and blue
absolute magnitude for starforming galaxies of Sa-Sc type by \citet{bothwell}. NGC~809, with its SFR$=0.066$~\Ms\ per year,
falls slightly below the boundaries of the `main sequence'; and it is the only galaxy of our sample where the UV-bright
ring shows no emission-line spectrum excited by young stars. For NGC~2697 and NGC~4324 there are measurements of the
gas content, both of the neutral hydrogen (the data in Table~1 are taken from the Extragalactic Distance Database,
http://edd.ifa.hawaii.edu) and of the molecular gas \citep{atlas3d_18}. With the SFR estimates obtained above, the
molecular gas consumption times are 2.1~Gyr for NGC~2697 and 800~Myr for NGC~4324 the latter being shorter 
than typical 2~Gyr for spiral galaxies \citep{bigiel}. The HI consumption times, 3~Gyr for NGC~2697 and 14~Gyr
for NGC~4324, are quite typical \citep{bothwell}. The general impression is that NGC~2697 and NGC~4324
have normal (for their masses) HI content and star formation rate \citep{bothwell}. Nevertheless they
are not spirals, they are lenticulars, and their star formation is assembled into the UV-bright rings.

In fact, we have in hand three independent indicators of the star formation rate: two GALEX-measured fluxes, those
in the Near-Ultraviolet (NUV) band and in the Far-Ultraviolet (FUV) band, and also the H$\alpha$ emission-line
intensity. For the latter, we have converted the fluxes into the absolute energy units by normalizing the
red continuum measured over the galactic spectra along the slit to the known photometric $r$-band surface-brightness
profiles (Fig.~\ref{iso}). These three indicators probe the star formation rate over different timescales: the NUV-flux -- over the
last 200~Myr, the FUV-flux -- over the last 100~Myr, and the H$\alpha$ emission produced by Stromgren zones around
massive stars provides the current star formation rate since stars with masses larger than 10~\Ms\ possessing Stromgren zones 
live no more than 10--20~Myr \citep{kennrev}. By comparing the SFR estimates made by using three different indicators we
can determine the qualitative trends of the star formation histories. We have cut the GALEX-images of our galaxies with digital
slits applying the slit characteristics given in the Table~\ref{tbl_logobs} for every galaxy, have converted the
GALEX counts into the energy units by using the recommendations from \citet{galex_calib}, have determined the
NUV-based and FUV-based star formation rates by using the formulae from \citet{kennrev}, and have corrected them
for the intrinsic dust by using WISE 22$\mu$m images. The current SFRs have been determined from the
H$\alpha$ emission-line fluxes by using also the formulae from \citet{kennrev}. After that we have divided the
SFRs obtained by the areas covered by the slit and in such a way have derived the local SFR surface density in the rings.
Just these estimates are plotted in Fig.~\ref{sfhistory} for four rings in three galaxies. Almost all the star formation
histories behave similarly: they fall with time over the timescale of some $10^8$ yrs. Only the NE tip of the NGC~4324 ring
demonstrates quite recent start of the local star formation (the H$\alpha$-estimated SFR exceeds those
derived from the UV fluxes) -- in this galaxy the star formation is very inhomogeneous along the ring.
By fitting three SFR ticks for every ring with a straight line, we have roughly estimated the e-folding
times of the star formation rate decline; these timescales have appeared to be mostly in the range of 200 to 350~Myr,
with the standing alone e-folding time of 600~Myr in the outer ring of NGC~2697 -- that is
much shorter than typical SFR e-folding times in the disks of spiral galaxies. Note that it is an upper limit of
the e-folding times, because if star formation has started more recently than 200~Myr ago, the NUV-based estimate
of the SFR must be increased, and the points at $T=200$~Myr in Fig.~\ref{sfhistory} must be shifted to later times -- both
corrections making the star formation rate decline steeper.

\subsection{Chemical evolution of the starforming rings in S0s}

By using strong-line calibrations, we have determined the gas metallicities in the starforming
rings and have compared them with the stellar metallicities of the underlying disks. One can see
in the Table~\ref{tbl:Gas} that the gas metallicities in all 4 galaxies with intense star
formation in the rings are tightly grouped around [O/H]$=-0.1\cdots -0.2$ dex.
This is based on the recent calibrations by \citet{marino}. If we use the most popular calibration by
\citet{pettini_pagel}, we would obtain strictly solar oxygen abundance. This is not
the first hint that star-forming gas in S0s is of homogeneously solar metallicity. We have measured solar
gas metallicities in the rings of NGC~6534 and MCG~11-22-015 \citep{we_aap}. \citet{ilyina14}
obtained just such metallicities for the starforming rings of S0 galaxies NGC~252 and NGC~4513.
Much earlier \citet{pe99}  in their survey of HII-regions
in gas-rich S0s noted that in all six galaxies with spectra of sufficient quality the oxygen
abundance of the ionized gas was solar, including two galaxies with counterrotating (and so
evidently accreted!) gaseous disks. Now we know also that the underlying stellar disks have lower
metallicity (Table~\ref{tbl:Gas}).

Some recent models of gas chemical evolution in disks of spiral galaxies, which include persistent gas inflow
and outflow \citep[e.g.][]{kudritzki15}, show that the gas oxygen abundance reaches rather quickly some 
asymptotic enrichment level close to the solar metallicity. The timescale of this process is below but
close to the gas consumption time \citep{lilly13}. Our finding of the homogeneously solar oxygen
abundance in the gas of starforming rings of lenticular galaxies is consistent with their accretion origin:
pure gas accretion, over period of a few hundred Myr, with a rate of order, say, a half of the SFR, provides
the asymptotic oxygen abundance of the ionized gas within starforming rings just after depletion of
half of the gas supply \citep{kudritzki15}. With our e-folding times of 200--350~Myr, it means that the
current gas oxygen abundance observed is reached after only 150--200~Myr of continuous star formation.

\subsection{Gas origin}

The origin of the accreted gas remains vague. Its metallicity is close to solar, though we observe
{\it the outer} parts of the galactic disks. There is a well-elaborated idea that the gas feeding star formation
in the disks of spiral galaxies represents cooling gas from the halo coming into the disks with galactic fountains
returning after blowing out from the starforming sites \citep{marinacci10}. Since in this case the gas clouds
leaving central parts of the disks fall to the outer disk regions following ballistic orbits \citep{fra_binney06}, we can
expect that in the outer rings we can meet the gas pre-enriched close to the galactic centers; commonly known
negative metallicity gradients in the disks of spiral galaxies serve well for this. However, the estimated timescale of
galactic fountain return into the disks is less than $10^8$~yr \citep[e.g.][]{houck_bregman} so to involve such a mechanism
of accretion we are needing recent star formation in the central parts of our galaxies. In contrast, all of
our galaxies have central stellar populations older than 4~Gyr. In our particular case of the S0 galaxies without
recent star formation in the centers, the mechanism of the own gas re-accretion through galactic fountains seems improbable.

Our galaxies belong to the `field' -- their environments are not very dense. Only NGC~4324 may be ascribed
to the outskirts of the Virgo cluster. None of those demonstrate
any signs of interaction.  In view of this, possible sources of the gas accretion may be filaments of the large-scale
Universe structure or minor merging with gas-rich satellites. In fact, these two mechanisms may play
concurrently. In NGC~809 and PGC~48114 where the rings are well seen in the UV and are detached from the main
bodies of the galaxies in the optical bands, we have found a significant difference of the stellar population properties
between the rings and the inner parts of the galaxies: the rings demonstrate solar magnesium-to-iron ratios
and very low stellar metallicity, [Z/H]$\approx -1$, unlike supersolar Mg/Fe and metallicity at the level of
a half-solar in the usual disks of lenticular galaxies \citep{s0disks}. We may suggest that both the gas and a major part of
stars in the rings of these two galaxies have an external origin, and the most probable source of the accretion seems
to be a late-type satellite merging from a high-momentum orbit. In the other three galaxies, NGC~2697, NGC~4324, and NGC~7808,
the UV-rings are embedded into their large-scale stellar disks, and the only certain difference between the stellar
population of their rings and the stellar population of the surrounding stellar disks is a systematically lower mean stellar 
age in the formers that is not astonishing taking into account current star formation localized
just within the rings. In these cases outer cold gas accretion from filaments is a reasonable interpretation
because the rather intense star formation in the rings for a few hundred Myr is able to increase the gas
metallicity toward the nearly solar value even from a very low initial level. However, the accretion events
might occur only less than 1~Gyr ago, and it is a special challenge to invent a scenario of a galaxy group dynamical
evolution that would provide a sporadic gas accretion from a cosmological filament onto the early-type group members at particular
moments long after the group assembled.

\section{Conclusions}

We have studied spectrally 5 lenticular galaxies with UV-bright outer rings. Four of the five rings
are also bright in the H$\alpha$ emission line, and the spectra of the gaseous rings extracted around the
maxima of the H$\alpha$ equivalent width reveal excitation by young stars betraying current star
formation in the rings. The integrated level of this star formation is 0.1--0.2~\Ms\ per year, with
the outstanding value of 1~\Ms\ per year in NGC~7808, that is within the `main sequence' of nearby
starforming galaxies. However, the difference of chemical composition between the ionized gas of the
rings and the underlying stellar disks which are metal poor implies star formation burst within
the lookback time of less than 1~Gyr induced by recent accretion of gas or another possible impact
event (e.g. infall of a small gas-rich satellite).

\acknowledgments

The study of the outer rings in disk galaxies has been supported by the grant no.18-02-00094a of the
Russian Foundation for Basic Reseaches.
The study is based on the observations
made with the Southern African Large Telescope (SALT), programs \mbox{2011-3-RSA\_OTH-001},
\mbox{2012-1-RSA\_OTH-002}, and \mbox{2012-2-RSA\_OTH-002}.
AYK acknowledges the support from the National
Research Foundation (NRF) of South Africa.
This research has made use of the NASA/IPAC Extragalactic Database (NED) which is
operated by the Jet Propulsion Laboratory, California Institute of Technology,
under contract with the National Aeronautics and Space Administration, and of
the Lyon Extragalactic Database (HyperLEDA, http://leda.univ-lyon1.fr). In this study, we used the SDSS DR9
data (http://www.sdss3.org). Funding for the SDSS and SDSS-II has been provided by the Alfred P. Sloan
Foundation, the Participating Institutions, the National Science Foundation, the
U.S. Department of Energy, the National Aeronautics and Space Administration,
the Japanese Monbukagakusho, the Max Planck Society, and the Higher Education
Funding Council for England. To estimate the star formation rates in the galactic rings,
we have used public archive data of the space experiments GALEX and WISE.
The NASA GALEX mission data were taken from the Mikulski Archive for Space
Telescopes (MAST). STScI is operated by the Association of Universities for Research in Astronomy,
Inc., under NASA contract NAS5-26555.
The WISE data exploited by us were retrieved from the NASA/ IPAC Infrared Science Archive,
which is operated by the Jet Propulsion Laboratory, California Institute of Technology,
under contract with the National Aeronautics and Space Administration.
This paper makes use of data obtained from the Isaac Newton Group Archive which is maintained as part 
of the CASU Astronomical Data Centre at the Institute of Astronomy, Cambridge.

\vspace{5mm}
\facilities{SALT(RSS), LCO, SDSS, GALEX, WISE, NED, HyperLEDA} 

\appendix

\section{Notes on the individual galaxies}

\noindent
{\bf NGC~809}.
In NGC~809 we inspect the nucleus, the bulge, and two rings -- the inner one, purely stellar
and red, at $r=10\arcsec$ (3.5~kpc), and the outer one, at $r=19\arcsec -32\arcsec$ \citep{ilyina_sil}, 
demonstrating the prominent UV signal (Fig.~\ref{sdssview}) which may be also classified as a pseudoring.
Meantime, in the outer ring the stellar population is older than 15 Gyr (Fig.~\ref{iidiag}); 
the relative abundance of $\alpha$-elements is close to the solar value, and the stellar metallicity falls to [Z/H]$\sim -1$, 
both latter facts pointing at continuous ineffective star formation which is typical for example for dwarf galaxies.
Stellar population of the inner galaxy in the area between $3\arcsec < r < 19\arcsec$ demonstrates the following properties: 
the mean age is 15 Gyr, the metallicity [Z/H]$\sim -0.3$, and the abundance ratio [Mg/Fe]$\approx +0.3$, 
that implies a brief ancient star formation epoch in the main disk of the galaxy.
The strong difference between the chemical properties of the outer stellar ring and the rest of the (inner)
galaxy, together with the detached outer ring seen in all photometric bands (Fig.~\ref{sdssview}), 
imply possible accretion of a small satellite with high orbital momentum onto the outer part of NGC~809.  
Another minor merging with a gas-rich satellite could lead to prolonged effective star formation 
episode in the central part of the galaxy ($r< 3\arcsec$) which had stopped about 3.5--5~Gyr ago and had then 
rejuvenated the stellar population of the galactic center. As a result, in the nucleus of NGC~809 
the mean age of the stellar population is 4--5~Gyr, the metallicity [Z/H]$\sim +0.5$, and the ratio [Mg/Fe]$=+0.1$. 
One can see that in the center the relative abundance of $\alpha$-elements is lower than the one in the disk of the 
galaxy but exceeds slightly the solar value because all the iron from the last starburst didn't have time 
to come into the last generation of the newly born stars before the star formation ceased.

\begin{figure*}[!t]
\centering
\begin{tabular}{c c}
 \includegraphics[width=6cm]{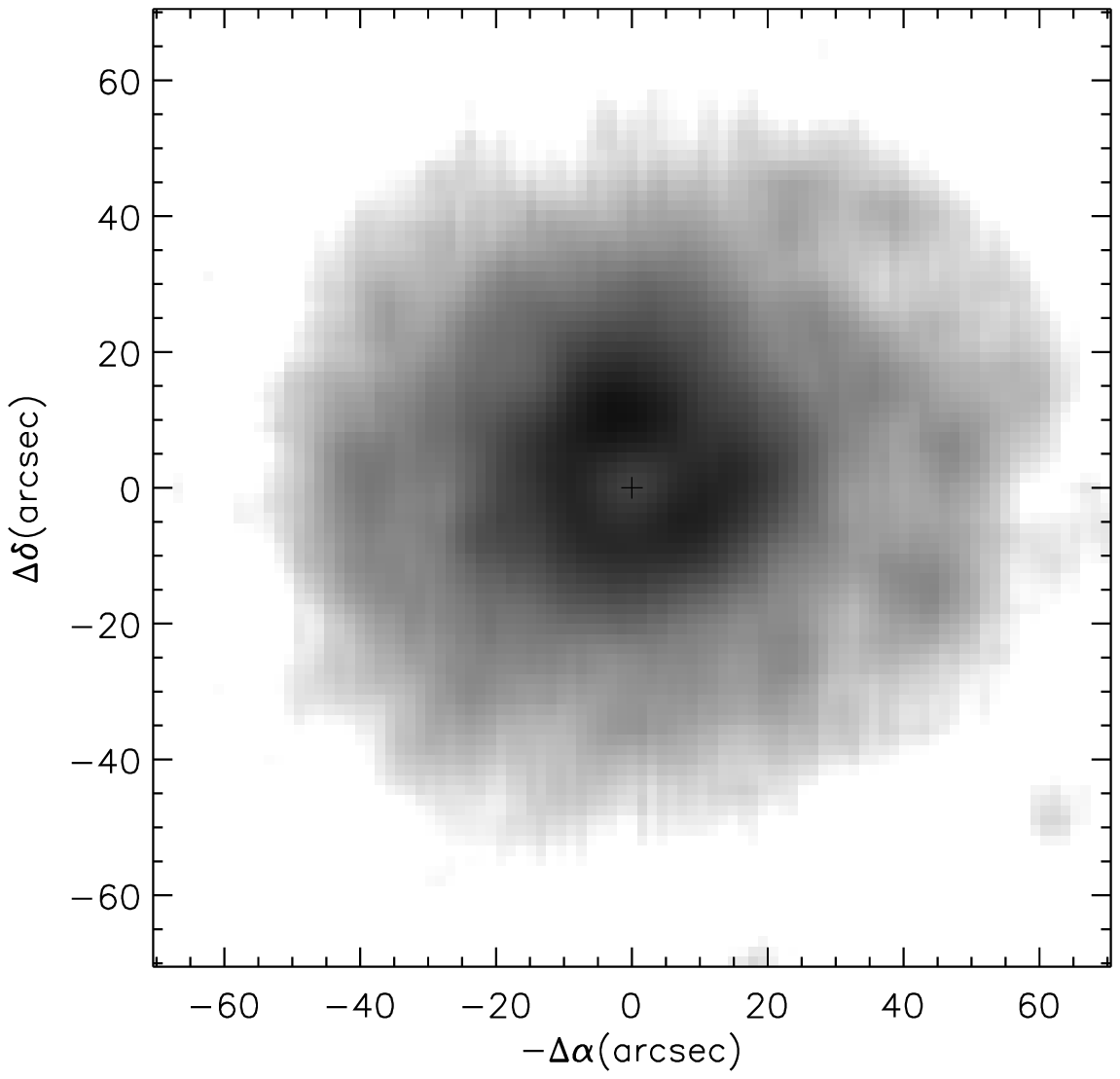} &
 \includegraphics[width=6cm]{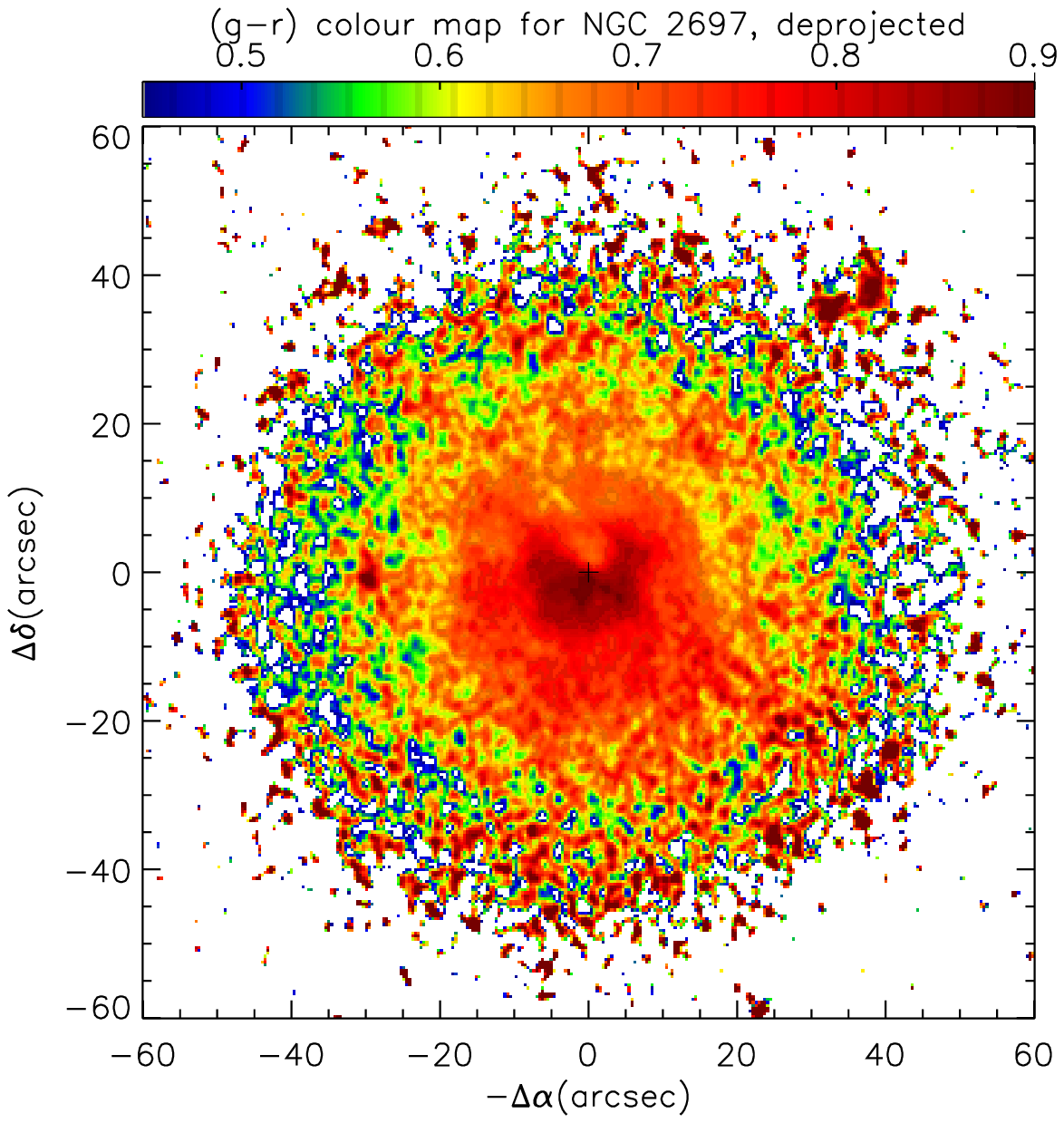} \\
\end{tabular}
\caption{The deprojected maps for NGC~2697: we have turned the initial images by $\sim 30\degr$ to put the line of nodes
horizontally, and then have stretched them in vertical direction to take into account the disc inclination of 60\degr\ implied
by the continuum isophote elllipticity. The GALEX FUV-image ({\it left}) reveals bright round ring with $r\approx
9\arcsec$ and a clumpy elliptical de-centered ring at $r\approx 40\arcsec$. The $(g-r)$ deprojected map derived from our
LCO observations demonstrates a circumnuclear dust ring with a radius of about 5\arcsec\ ({\it right}).
}
\label{n2697fuvgr}
\end{figure*}

\noindent
{\bf NGC~2697}.
In NGC~2697 we recognize the nucleus, the circumnuclear dust (red)
ring at $r=3\arcsec -7\arcsec$ (Fig.~\ref{n2697fuvgr}), the inner UV-ring at $r=7\arcsec -11\arcsec$ (Fig.~\ref{n2697fuvgr}),
and a regular stellar disk between $r=11\arcsec$ and $r=21\arcsec$.
The galaxy has no classical bulge (see Fig.~\ref{iso}, the surface brightness profile of NGC~2697), but
at the distance of 7\arcsec--10\arcsec\ from the center the low-contrast inner stellar ring can be detected
in the $r$-band. It is also the location of the UV-ring in this galaxy, and just starting from this radius
we observe strong emission lines excited by young stars.
At the distances of $r>7$\arcsec\ from the center the stellar population is rather young, with the mean age of 1.5~Gyr 
and the relative abundance of $\alpha$-elements [Mg/Fe]$\approx +0.2-+0.3$. A prominent metallicity
gradient is observed in the stellar component, from $\mbox{[Z/H]}\sim -0.1$ at the radii of $7\arcsec <r<11\arcsec$ 
till [Z/H]$<-0.33$ at the radii of $11\arcsec \le r < 21\arcsec$. In the inner part of the galaxy
($2\arcsec <r \le 7\arcsec$) the mean (SSP-equivalent) age of the stellar population is 5--7~Gyr, 
the abundance ratio [Mg/Fe]$\approx +0.1$, and the metallicity $\mbox{[Z/H]}\approx -0.2$. 
In the center ($r\le 2\arcsec$) the mean age of the stellar population is 7~Gyr, 
and the abundance ratio, as well as the total metallicity, is close to the solar value  
that points at continuous effective star formation in the central part of the galaxy.
With its negative metallicity and age gradients, NGC~2697 supports the idea of 'inside-out' disk formation in accordance
with the classical paradigm \citep{CDM_disk,two_infall}. This implied that star formation began in
the galactic center and ceased there about 7~Gyr ago, then star formation proceeded farther from the center, in the region
$2\arcsec <r \le 7\arcsec$ where it was inefficient and ceased about 2.5-4~Gyr ago (for the connection
between the SSP-equivalent age and the age of star formation quenching -- see e.g. \citet{rsmith}). 
After that there was a pause in the starforming process. Presumably, about 1.5~Gyr ago, a dwarf galaxy merged
with NGC~2697, a process which had added metal-poor stars to the outer parts of the galaxy, had ignited intense star formation
in the inner disk of the galaxy, and caused a series of brief starbursts organized like rings
which we are still observing presently as clumpy pattern of the disk. Furthermore, the gas has nearly solar metallicity
just where the emission lines are excited by star-forming process.

\noindent
{\bf NGC~4324}.
In NGC~4324 we investigate the nucleus,
the bulge, the roundish lens between $r=7\arcsec$ and $r=16\arcsec$, and the UV-ring area.
In Fig.~\ref{iidiag} one can see that in the center ($r\le 3\arcsec$) of NGC~4324 the mean age of the stellar population 
is about 8~Gyr, the abundance ratio is close to the solar value, [Mg/Fe]$\ge 0$, and the metallicity is slightly supersolar, 
[Z/H]$\sim +0.1$ that points at continuous effective star formation in the nucleus of the galaxy. At the radii of 
$3\farcs 5 \le r \le 6\arcsec$ from the center, in the bulge-dominated area, the mean age of the stellar population 
is about 13~Gyr, the abundance ratio rises to [Mg/Fe]$=+0.15$, and the metallicity decreases to $\mbox{[Z/H]}=-0.2 \cdots -0.3$.
In the inner part of the disk ($r= 7\arcsec - 16\arcsec$) the stellar population is old, the abundance ratio reaches  
[Mg/Fe]$=+0.2$, and the metallicity decreases to [Z/H]$< -0.33$ -- demonstrating the properties that match exactly the
stellar population characteristics of the globular clusters in the bulge of our Galaxy which are also plotted in our
age-diagnostic diagrams for comparison. Such characteristics imply a brief single starburst
which took place more than 10~Gyr ago and had formed the large-scale stellar disk of NGC~4324.
In the ring-dominated area of the disk ($r= 17\arcsec - 35\arcsec$) the dominant stellar population is also old though
slightly younger than in the inner disk, and the mean chemical properties of the stars are nearly the same. 

\noindent
{\bf NGC~7808}.
In NGC~7808 we have succeeded to distinguish the nucleus($+$bulge), the disk, and the inner starforming ring (the outer UV-bright
ring at $r>23\arcsec$ \citep{ilyina_sil} is too faint in continuum, and we cannot derive its stellar population properties).
In this galaxy we had two long-slit cross-sections in two different slit orientations, but because the velocity profiles
appear to be identical, we have added both spectra before analyzing the stellar population properties.
As a result, the stellar population characteristics are taken under some 36\degr\ to the major axis.
In Fig.~\ref{iidiag} one can see that in the center ($r<2\arcsec$) the mean age of the stellar population is 4--5~Gyr,
the abundance ratio [Mg/Fe]$\approx +0.1$, and the stellar metallicity is close to the solar one -- all these
characteristics point at prolonged star formation which stopped there about 2.5~Gyr ago. The history of the nucleus
in NGC~7808 resembles that of the nucleus of NGC~809. In the inner disk, at $r= 5\arcsec - 9\farcs 5$, the mean age of
the stellar population is 8~Gyr, the abundance ratio increases to [Mg/Fe]$\ge +0.3$, and the metallicity decreases
to [Z/H]$\le -0.33$. And at the radii $10\arcsec \le r< 15\arcsec$ we observe a ring-like region, this time with star formation,
where the mean age of the stellar population drops to 3~Gyr, but the abundances are the same as in the inner stellar disk. 
Presumably, the ring-like star formation burst is brief and has happened at least twice at the same radius,
about 2.5 Gyr ago and quite recently (since we observe the H$\alpha$ emission line excited by young stars in the ring).
As for NGC~2697, we conclude that the disk formation in NGC~7808 has evolved as inside-out.

\noindent
{\bf PGC~48114}.
In PGC~48114 we inspect the nucleus and two stellar rings -- a red one at $r=3\arcsec -10\arcsec$ (1.5--4.8~kpc) and 
that at $r=11\arcsec -17\arcsec$, the latter corresponding also to the UV-bright area.
The double ring in PGC~48114 may be classified as $R_1R_2$; this time a weak bar presence can be
suspected. In the outer bright inhomogeneous ring starting from $r >12\arcsec$ the strong emission lines have been
detected partially excited by star-forming process (along our slit -- only to the west from the nucleus). 
In the center ($r<2\arcsec$) of PGC~48114 the mean age of the stellar population is about 5~Gyr, 
abundance ratio [Mg/Fe]$\approx +0.1$, and the metallicity $\mbox{[Z/H]}=+0.3$ -- typical
values for the nuclei of lenticular galaxies \citep{sil06,sil16}. However, the low [Mg/Fe],
within the range of $0\cdots +0.1$, kept all along the galaxy is quite untypical for the contemporary lenticular 
galaxies \citep{s0disks}; it implies an extended star formation epoch in the disk. The inner elliptical ring
is rather old, $T= 9\pm 5$~Gyr, with the typical low stellar metallicity, $\mbox{[Z/H]}=-0.3$. However, in the area
of the outer, starforming ring the mean stellar age drops to 3~Gyr, and the metallicity drops also, to $\mbox{[Z/H]}< -1$.

\bibliographystyle{aj}
\bibliography{saltring} % if your bibtex file is called example.bib

\end{document}